\documentclass[aps,prd,nofootinbib,twocolumn,showpacs,preprintnumbers,amsmath,amssymb,superscriptaddress,groupedaddress,floatfix]{revtex4-1}

\usepackage[pdftex,
       colorlinks=true,
       pdfpagemode=UseNone,
       bookmarksopen=true]{hyperref}
\usepackage[utf8]{inputenc}
\usepackage{amsmath,amssymb}
\usepackage{braket}
\usepackage{color}
\usepackage{dsfont}
\usepackage{graphicx}
\usepackage{slashed}
\usepackage{soul}

\newif\ifdocomment
\docommentfalse
\ifdocomment
\newcommand{\comment}[1]{\textcolor{red}{[#1]}}
\newcommand{\str}[1]{}

\else
\newcommand{\comment}[1]{\PackageError{comment}{unresolved comment}{}}
\newcommand{\str}[1]{}

\fi

\newcommand{\tmin}{\ensuremath{{t_{\text{min}}}}}
\newcommand{\tmax}{\ensuremath{{t_{\text{max}}}}}
\newcommand{\sutwo}{\ensuremath{\text{SU}(2)}}

\begin{document}
\title{
Isospin 0 and 2 two-pion scattering at physical pion mass using distillation with periodic boundary conditions in lattice QCD
}

\newcommand{\etal}{\emph{et al.}}
\newcommand\bnl{Physics Department, Brookhaven National Laboratory, Upton, NY 11973, USA}
\newcommand\bnlccs{Computational Science Initiative, Brookhaven National Laboratory, Upton, NY 11973, USA}
\newcommand\bern{Albert Einstein Center, Institute for Theoretical Physics, University of Bern, Switzerland}
\newcommand\cu{Physics Department, Columbia University, New York, NY 10027, USA}
\newcommand\pu{School of Computing \& Mathematics, Plymouth University, Plymouth PL4 8AA, UK}
\newcommand\riken{RIKEN-BNL Research Center, Brookhaven National Laboratory, Upton, NY 11973, USA}
\newcommand\regensburg{Fakult\"at f\"ur Physik, Universit\"at Regensburg, Universit\"atsstra{\ss}e 31, 93040 Regensburg, Germany}
\newcommand\edinb{School of Physics and Astronomy, The University of Edinburgh, Edinburgh EH9 3FD, UK}
\newcommand\uconn{Physics Department, University of Connecticut, Storrs, CT 06269-3046, USA}
\newcommand\soton{School of Physics and Astronomy, University of Southampton,  Southampton SO17 1BJ, UK}
\newcommand\york{Mathematics \& Statistics, York University, Toronto, ON, M3J 1P3, Canada}
\newcommand\cssm{CSSM, University of Adelaide, Adelaide 5005 SA, Australia}
\newcommand\cern{CERN, Physics Department, 1211 Geneva 23, Switzerland}
\newcommand\mib{Dipartimento di Fisica, Universit\'a di Milano-Bicocca, Piazza della Scienza 3, I-20126 Milano, Italy}
\newcommand\infn{INFN, Sezione di Milano-Bicocca, Piazza della Scienza 3, I-20126 Milano, Italy}
\newcommand{\ucb}{Physics Department, University of California, Berkeley, CA 94720, USA}
\newcommand{\lbnl}{Nuclear Science Division, Lawrence Berkeley National Laboratory, Berkeley, CA 94720, USA}
\newcommand{\llnl}{Nuclear \& Chemical Sciences Division, Lawrence Livermore National Laboratory, Livermore, CA 94550, USA}

\author{M.~Bruno}\affiliation{\mib}\affiliation{\infn}
\author{D.~Hoying}\affiliation{\bern}
\author{T.~Izubuchi}\affiliation{\bnl}\affiliation{\riken}
\author{C.~Lehner}\affiliation{\regensburg}
\author{A.~S.~\surname{Meyer}}\email{asmeyer.physics@gmail.com}
\thanks{Present address: \llnl}
\affiliation{\ucb}\affiliation{\lbnl}
\author{M.~Tomii}\affiliation{\uconn}

\date{\today}

\begin{abstract}
 The two pion channel in Lattice QCD has long been a primary testing ground
  for studying multiparticle scattering in finite volume QCD.
 With the development of sophisticated techniques such as distillation,
  it is possible to carefully study two-pion scattering
  in order to constrain associated low-energy constants.
 In this work, correlation functions with multiparticle interpolating
  operators are constructed to compute pion scattering phase shifts
  and scattering lengths in the isospin 0 and 2 channels
  with both sea and valence quarks at physical mass.
 Contamination from vacuum and thermal contributions are explicitly quantified
  with dedicated calculations and the results obtained after subtracting these
  nuisance terms are compared with the traditional correlator
  time series subtraction method.
 Two physical point ensembles with different lattice actions are used,
  and our finest ensemble gives results for scattering lengths and phase shifts
  consistent with phenomenology to within the reported statistical uncertainty.
\end{abstract}

\maketitle

\def\figsize{0.4}

\section{Introduction}
 \label{sec:introduction}

With advances in computational techniques
 and increases in computing power have come
 ensembles with finer lattice spacings and
 physical pion masses.
Paired with computational techniques such as distillation,
 Lattice QCD (LQCD) can now probe calculations of spectra
 and matrix elements involving multiparticle states.
These advances are vital to obtain precise measurements
 of processes involving multiparticle scattering
 that are relevant for future and ongoing experiments,
 including predictions of CP violation in the
 kaon sector~\cite{Batley:2002gn,Abouzaid:2010ny,Bai:2015nea,Abbott:2020hxn,RBC:2021acc},
 exclusive channel studies for the reduction
 of statistical error in LQCD calculations
 of the hadronic vacuum polarization contribution to
 muon $g-2$~\cite{LehnerTalkLGT16,Borsanyi:2016lpl,Bruno:2019nzm,Aoyama:2020ynm},
 and for isolating and removing excited state
 contamination in computations of nucleon form
 factors~\cite{Kronfeld:2019nfb}.

Scattering of pions is the primary testing ground
 for the LQCD multiparticle formalism.
In LQCD, particle interactions in a finite
 volume prevent the separation of lattice
 states into their asymptotic free-particle
 states when the particles are separated
 by a large spatial distance.
However, the power law corrections that are
 induced by the interactions between confined
 particles give access to the infinite volume
 scattering phase shift.
The finite volume formalism of
 L\"uscher~\cite{Luscher:1986pf,Luscher:1990ux,Luscher:1991cf}
 has been highly successful at describing
 two-particle states up to kinematic
 three- and four-particle thresholds,
 yielding predictions of the finite volume spectrum
 in good agreement with lattice results.
Since its introduction in the seminal papers by
 L\"uscher,
 the two-particle formalism has been expanded to include
 moving particles~\cite{Rummukainen:1995vs},
 unequal masses~\cite{Leskovec:2012gb},
 coupled channels~\cite{Guo:2012hv},
 nontrivial spin~\cite{Woss:2018irj,Woss:2019dnv},
 and even some three particle
 cases~\cite{Hansen:2019nir,Rusetsky:2019gyk,Mai:2021lwb}.
Additional work continues to improve the
 formalism with the goal of handling three-particle
 states in general and even a few calculations that have been
 completed using three-pion interpolating
 operators~\cite{Horz:2019rrn,Fischer:2020jzp}.
Another method to compute the scattering phase shift using
 a Bethe-Salpeter kernel also exists~\cite{Ishii:2006ec,Aoki:2009ji,HALQCD:2012aa}
 and has been used to compute $\pi\pi$ scattering phase shifts~\cite{Kawai:2017goq},
 but this method will not be explored in this analysis.

Studying pion scattering with distillation~\cite{Peardon:2009gh}
 has the additional advantage of allowing
 simple computations of
 pion transition matrix elements without
 the need for additional propagator solves.
These improvements allow for the explicit subtraction
 of nuisance terms associated with propagating
 single-pion states that contribute to thermal
 corrections to the two-pion correlation functions.
Thermal terms are often treated as negligible due
 to an exponential suppression in the temporal extent,
 an assumption that is testable with the distillation framework.
Distillation is also used to access the vacuum
 contributions to the correlation functions,
 which manifest as a large constant term
 in the correlation function.

This document is organized as follows.
In Sec.~\ref{sec:methods},
 we discuss the methods used in this analysis.
This includes a brief discussion of distillation
 and its implementation in this project.
In addition, we discuss our method for dealing
 with vacuum contributions to the isospin 0 channel
 and states that appear from the finite
 temporal boundary condition effects as well
 as a technique to improve the statistical uncertainty
 on our correlation functions.
Section~\ref{sec:ensemble} discusses the lattice setup,
 including the ensembles used and some of
 the computational details.
In Sec.~\ref{sec:analysis},
 we apply the techniques outlined in
 Sec.~\ref{sec:methods}
 to extract the spectrum of states from our data.
The observed spectra are used in Sec.~\ref{sec:results}
 with the L\"uscher formalism
 to compute and fit the lowest partial wave scattering
 phase shifts and scattering lengths for the isospin 0
 and isospin 2 channels.
In Sec.~\ref{sec:discussion}, we give results from fitting
 the phase shift curves and conclude.
In Appx.~\ref{sec:formalism},
 we go through the details of the L\"uscher formalism
 and the relation between the spectrum
 and scattering phase shift.
In particular, we write this relation
 in a basis-independent scheme to make the remnant
 rotational symmetry of the formalism manifest.

\section{Methods}
 \label{sec:methods}
 
The numerical techniques in this study seek to construct a Hermitian matrix correlator
 with propagating two-pion states.
A two-point function matrix correlator will have a general parametric form
\begin{align}
 {\cal C}_{AB}(t) &=
 \langle {\cal O}_A(t) {\cal O}^\dagger_B(0) \rangle
 \nonumber\\
 &=
 \sum_{mn} \langle m | {\cal O}_A | n \rangle
 \langle n | {\cal O}^\dagger_B | m \rangle e^{-E_n t} e^{-E_m (T-t)},
 \label{eq:generic2point}
\end{align}
 which contains the desired two-pion states as well as other unwanted contributions
 from the vacuum, one-pion states, or other higher excitations.
The large basis of operators used for the indices $A$ and $B$ is obtained
 from quark propagators obtained using the distillation methodology,
 which is outlined in Sec.~\ref{sec:distillation}.
The construction of correlation functions from distillation propagators
 and the projection of the interpolating operators to definite
 isospin quantum numbers is detailed in Sec.~\ref{sec:interpolatingoperators}.

The spectrum of states in the matrix correlator will be obtained by solving the
 Generalized EigenValue Problem (GEVP), which is outlined in Sec.~\ref{sec:gevp}.
The GEVP assumes a generic form
\begin{align}
 C_{AB}(t) &=
 \sum_{n>0} \langle 0 | {\cal O}_A | n \rangle
 \langle n | {\cal O}^\dagger_B | 0 \rangle e^{-E_n t},
 \label{eq:gevptwopoint}
\end{align}
 which does not contain any terms with states propagating through
 the periodic temporal boundary condition.
These unwanted ``thermal'' states are removed by performing
 a supplemental calculation of three-point correlation functions
 to obtain matrix elements of the form $\langle m | {\cal O}_A | n \rangle$
 and then subtracting the corresponding thermal contribution
 in the two-point function.
This procedure is described in Sec.~\ref{sec:thermal}.
Additional contributions to Eq.~(\ref{eq:gevptwopoint}) from purely vacuum terms
 in the isospin 0 channel can also be subtracted by using vacuum matrix elements
 $\langle 0 | {\cal O}_A | 0 \rangle$ from one-point correlation function data,
 described in Sec.~\ref{sec:vacuum}.
With both of these additional matrix elements computed,
 Eq.~(\ref{eq:gevptwopoint}) may be obtained from Eq.~(\ref{eq:generic2point})
 with the substractions
\begin{align}
 &C_{AB}(t) 
 \nonumber\\
 &= {\cal C}_{AB}(t)
 -
 \langle 0 | {\cal O}_A | 0 \rangle
 \langle 0 | {\cal O}^\dagger_B | 0 \rangle
 \nonumber\\
 &\phantom{=}-
 \sum_{m>0,n>0} \langle m | {\cal O}_A | n \rangle
 \langle n | {\cal O}^\dagger_B | m \rangle e^{-E_n t} e^{-E_m (T-t)},
 \nonumber\\
\end{align}
 where $m>0$ in the sum denotes all states except for the vacuum state.

After the spectrum of states is obtained through application of the GEVP,
 the scattering phase shifts and scattering lengths are computed
 by applying the L\"uscher quantization condition.
This procedure is outlined in Sec.~\ref{sec:phaseshift},
 with more detail given in Appendix~\ref{sec:formalism}.
An improvement to the momentum inputs to the L\"uscher quantization,
 which uses linear combinations of single-pion
 correlation functions to obtain the interacting-minus-noninteracting
 difference of energies in the two-pion system,
 is explained in Sec.~\ref{sec:freepicorrelation}.
The scattering lengths and phase shifts obtained from the interacting-noninteracting
 energy differences are used to produce the results in Sec.~\ref{sec:results}.

The states in this manuscript are schematically labeled as
 one-pion ($|\pi\rangle$), two-pion ($|\pi\pi\rangle$),
 or vacuum ($|0\rangle$) based on the tower of states that is expected.
Though this nomenclature is suggestive of the spectrum of states that
 should be expected in each channel, the restriction to finite volume
 complicates this picture.
The reader should use caution when identifying the discrete finite volume
 states with the corresponding asymptotic states in infinite volume.
Instead of being an exact identification of the asymptotic states,
 these labels are instead meant to represent states that transform
 under distinct representations of the symmetries of the system
 restricted to finite volume.

\subsection{Distillation}\label{sec:distillation}

Generation of the many $\pi\pi$ correlation functions used in this
 project is made simpler using the common technique in LQCD
 known as distillation~\cite{Peardon:2009gh}.
Distillation makes use of the eigenvectors of some operator,
 typically chosen to have strong overlap with the lowest-energy states
 in the spectrum.
The eigenvectors are constructed to satisfy the eigenvalue equation
\begin{equation}
 \sum_{\mathbf{y}} \Delta(\mathbf{x},\mathbf{y}) V^{n}(\mathbf{y})
 = V^{n}(\mathbf{x}) \lambda^{n} ,
\end{equation}
 where the eigenvalues are ordered such that $\lambda^m \geq \lambda^n$ for $m>n$.

The most commonly used operator is the discrete spatial Laplace operator,
 which in a three-dimensional spatial volume is given by
\begin{equation}
 \Delta_{\mathbf{x},\mathbf{y}} = -\delta_{\mathbf{x},\mathbf{y}} +\frac{1}{6a^{2}}
 \sum_{i} (
 U_{i}(\mathbf{x})
 \delta_{\mathbf{x},\mathbf{y}-a\mathbf{\hat{i}}}
 +U^{\dagger}_{i}(\mathbf{x}-a\mathbf{\hat{i}})
 \delta_{\mathbf{x},\mathbf{y}+a\mathbf{\hat{i}}} )
 \label{eq:finitelaplacian}
\end{equation}
 for discrete spacetime, where color indices and timeslices have been suppressed.
The Laplacian is intuitively a good choice because the free-field eigenmodes
 of the Laplacian are plane waves, with the lowest state corresponding
 to a particle at rest, and higher eigenmodes to particles with more momentum.

The eigenvectors of the distillation operator are provided as sources
 for solving the Dirac equation for Green's function $G$,
\begin{align}
 \sum_{y_4,\mathbf{y}}
 \slashed{D}(x_4,\mathbf{x},y_4,\mathbf{y}) G^{n}(y_4,x_4,\mathbf{y})
 = V^{n}(x_4,\mathbf{x})
\end{align}
 with both spinor and color indices suppressed.
Here, the distillation eigenvectors $V$ are now given an
 explicit Euclidean time index $x_4$.
The eigenvectors $V$ are assumed to be diagonal in the spinor
 indices so that spin representations are not mixed up by the
 application of eigenvectors.
After solving, the Green's function is contracted with a conjugate
 eigenvector to create a ``perambulator'',
\begin{align}
 {\cal G}^{mn}(y_4,x_4) = \sum_{\mathbf{y}}
 [V^m(y_4,\mathbf{y})]^\dagger G^{n}(y_4,x_4,\mathbf{y}),
 \label{eq:perambulator}
\end{align}
 where now all dependence on spatial positions has been subsumed
 by the eigenvector indices.
The perambulators still satisfy many of the same symmetry
 transformations as propagators, most importantly
 $\gamma_5$ hermiticity:
\begin{align}
 \gamma_5 [{\cal G}^{mn}(x_4,y_4)]^\dagger \gamma_5 = {\cal G}^{nm}(y_4,x_4) .
 \label{eq:gamma5hermiticity}
\end{align}
This property will be especially important for assessing how
 bias corrections may be performed in Sec.~\ref{sec:biascorrection}.

Eq.~(\ref{eq:perambulator}) has the advantage over traditional
 methods of sequential inversions in that any dependence on
 the spatial position is encoded in the eigenvector number.
This means the perambulator objects are smaller than
 traditional propagators and the perambulators can be contracted
 without the need for large amounts of computational resources.
Additionally, the contractions of more quark lines amount to
 additional matrix multiplications, meaning that the same
 set of perambulators may be used to construct arbitrary N-point
 correlation functions without the need for additional
 solutions to the Dirac matrix.
This advantage has been vital for constructing the two-pion
 correlation functions in this analysis.

\subsection{Interpolating Operator Construction}\label{sec:interpolatingoperators}

If the eigenspectrum of the distillation operator is truncated
 at some finite number of eigenvectors $N$,
 the subset of eigenvectors form a projection matrix,
\begin{equation}
 \sum_n^N V^{n}(\mathbf{x}) [V^{n}(\mathbf{y}) ]^{\dagger} = {\cal P}(\mathbf{x},\mathbf{y}) .
 \label{eq:distprojection}
\end{equation}
This projection matrix acts by smearing out the field on a lattice timeslice.
Then interpolating operators may be constructed by applying the projections
 to the quarks and Fourier phase,
\begin{align}
 {\cal O}_{i,j,\Gamma,\mathbf{P}}
 = \sum_{\mathbf{x},\mathbf{y},\mathbf{z}\in L^3}
 \bar\psi_i(\mathbf{x}) {\cal P}(\mathbf{x},\mathbf{y})
 \Gamma e^{-i\mathbf{P}\cdot \mathbf{y}}
 {\cal P}(\mathbf{y},\mathbf{z}) \psi_j(\mathbf{z}),
 \label{eq:basicinterp}
\end{align}
 where $i$ and $j$ are flavor indices,
 $\Gamma$ is a gamma spin matrix,
 and $\mathbf{P}$ is the momentum of the bilinear.
If the eigenvectors are contracted over their spacetime indices,
 elementals may be constructed of the form
\begin{align}
 \widetilde{\cal P}^{mn}(x_4,\mathbf{p})
 \equiv
 \sum_{\bm{x}}
 [V^m(x_4,\mathbf{x})]^\dagger
 e^{-i \mathbf{p}\cdot\mathbf{x}}
 V^n(x_4,\mathbf{x}),
\end{align}
 which may be used to construct, for example,
 the pion two-point function,
\begin{align}
 &\langle O_{\pi,\mathbf{p}}(t) O_{\pi,\mathbf{p}}(0) \rangle = \nonumber\\
 &\sum_{m,m',n,n'} \Big\langle \mathrm{Tr} \biggr[
 \ \mathcal G^{mn} (t,0) \gamma_5 \widetilde P^{nn'} (t, \mathbf{p})
  \nonumber\\
 &
 \phantom{\sum_{m,m',n,n'} \langle \mathrm{Tr} \Biggr[} \times
 \mathcal G^{n'm'}(0,t)  \gamma_5 \widetilde P^{m'm} (0, \mathbf{p}) \biggr]
 \Big\rangle .
\end{align}
 
The quark spinors $\psi$ in Eq.~(\ref{eq:basicinterp}) are assumed to be a 2-vector
 with isospin-symmetric up and down flavors,
\begin{align}
 \psi = \left( \begin{array}{r} u \\ d \end{array} \right) , &&
 \bar\psi^T = \left( \begin{array}{r} \bar{d} \\ -\bar{u} \end{array} \right) .
\end{align}
Isospin 0 and isospin 1 bilinears for the interpolating operators
 are constructed by contracting with Clebsch-Gordan coefficients
 to pull out the $I_z=0$ component,
\begin{align}
 {\cal I}^{(0,0)}_{i,j} &= \frac{1}{\sqrt{2}} \left(
 \begin{array}{rr} 0 &-1 \\ 1 & 0 \end{array}
 \right)
 \nonumber\\
 \implies &
 \sum_{i,j} {\cal I}^{(0,0)}_{i,j} {\cal O}_{i,j}
 \sim \frac{1}{\sqrt{2}} (\bar{d}d +\bar{u}u),
\end{align}
 and
\begin{align}
 {\cal I}^{(1,0)}_{i,j} &= \frac{1}{\sqrt{2}} \left(
 \begin{array}{rr} 0 & 1 \\ 1 & 0 \end{array}
 \right)
 \nonumber\\
 \implies &
 \sum_{i,j} {\cal I}^{(1,0)}_{i,j} {\cal O}_{i,j}
 \sim \frac{1}{\sqrt{2}} (\bar{d}d -\bar{u}u).
\end{align}
The $\pm1$ components of the isospin 1
 representation are obtained from
\begin{align}
 {\cal I}^{(1,\pm1)}_{i,j} &= \frac{1}{2}
 \left( \delta_{i,j}\pm\sigma^3_{i,j} \right),
\end{align}
 with the Pauli spin matrix $\sigma^3$.
This analysis makes use of both
 the scalar isospin 0 bilinear with zero
 center of mass momentum,
\begin{align}
 {\cal O}_{{1}}
 = \sum_{i,j} {\cal I}^{(0,0)}_{i,j}
 {\cal O}_{i,j,\mathds{1},0},
 \label{eq:scalarop}
\end{align}
 and the moving $1\pi$ operator,
\begin{align}
 {\cal O}^{(1,k)}_{\pi,\mathbf{p}_\pi}
 = \sum_{i,j} {\cal I}^{(1,k)}_{i,j}
 {\cal O}_{i,j,\gamma_5,\mathbf{p}_\pi}.
 \label{eq:pionbilinear}
\end{align}

The $\pi\pi$ interpolating operators
 are then composite operators built from
 the $1\pi$ bilinears with equal and opposite
 relative momentum $\mathbf{p}_{\text{rel}}$
 and isospin component $I_z=0$,
\begin{align}
 &{\cal O}^{(I,0)}_{\pi\pi}
 \Big(\Big|\frac{\mathbf{p}_{\text{rel}}L}{2\pi}\Big|^2\Big) \nonumber\\
 &= \frac{1}{\sqrt{N_{\{\mathbf{p}_{\text{rel}}\}}}}
 \sum_{i,j}
 \sum_{\mathbf{p} \in \{ \mathbf{p}_{\text{rel}} \} }
 {\cal I}^{(\pi\pi,I,0)}_{i,j}
 {\cal O}^{(1,i)}_{\pi, \mathbf{p}}
 {\cal O}^{(1,j)}_{\pi,-\mathbf{p}},
 \label{eq:pipiop}
\end{align}
 where $\{\mathbf{p}_{\text{rel}}\}$
 is the set of momenta that can
 be obtained by applying 
 cubic group rotations to
 $\mathbf{p}_{\text{rel}}$
 and $N_{\{\mathbf{p}_{\text{rel}}\}}$
 is the total number of momenta in
 $\{\mathbf{p}_{\text{rel}}\}$,
\begin{align}
 N_{\{\mathbf{p}_{\text{rel}}\}}
 =
 \sum_{\mathbf{p} \in \{ \mathbf{p}_{\text{rel}} \} }
 1.
 \label{eq:nprel}
\end{align}
The $\pi\pi$ interpolating operators
 in Eq.~(\ref{eq:pipiop})
 are implicitly constructed to transform
 under the trivial representation
 of the cubic rotation group,
 so will couple primarily to $\ell=0$ states.
Working in a basis where the first, second, and third rows
 correspond to the $\pi^+$, $\pi^0$, and $\pi^-$, respectively,
 the $\pi\pi$ isospin factors are
\begin{align}
 {\cal I}^{(\pi\pi,0,0)}
 &=
 \frac{1}{\sqrt{3}}
 \left( \begin{array}{rrr}
 0 & 0 & 1 \\
 0 &-1 & 0 \\
 1 & 0 & 0
 \end{array} \right),
 \nonumber\\
 {\cal I}^{(\pi\pi,2,0)}
 &=
 \frac{1}{\sqrt{6}}
 \left( \begin{array}{rrr}
 0 & 0 & 1 \\
 0 & 2 & 0 \\
 1 & 0 & 0
 \end{array} \right)
\end{align}
 for the isospin 0 and isospin 2 representations,
 respectively.

\subsection{Generalized Eigenvalue Problem}\label{sec:gevp}

The Generalized EigenValue Problem (GEVP)~\cite{Blossier:2009kd,Bulava:2014vua}
 is a technique for computing the spectrum
 of states from a matrix of correlation functions without the need to fit.
A diagonal eigenvalue matrix $\Lambda$ is obtained from solving the GEVP equation,
\begin{equation}
 C(t) V = C(t_0) V \Lambda.
 \label{eq:gevp}
\end{equation}
The matrix $C_{AB}(t)$ is a square, symmetric matrix of correlation functions,
 with source and sink operators indexed by $A$ and $B$, respectively,
 which run from 0 to $N-1$ for $N$ operators.
The GEVP makes the assumption that the correlation functions are a sum of exponentials,
\begin{equation}
 C_{AB}(t) = \sum_n \bra{0} {\cal O}_A \ket{n}
 \bra{n} {\cal O}^\dagger_B \ket{0} e^{-E_n t}.
\end{equation}
If the sum is truncated at finite $N$, where $N$ matches the rank of the matrix $C(t)$,
 then the eigenvalue matrix obtained from solving the GEVP is exactly
\begin{equation}
    \Lambda_{mn}(t_0,t) = \delta_{mn} e^{-E_n (t-t_0)} \,.
    \label{eq:gevpeigenvalueexact}
\end{equation}
If the number of states is $M>N$,
 then the eigenvalues receive exponential corrections from the higher states
 in the spectrum.
The form of Eq.~(\ref{eq:gevpeigenvalueexact})
 is recovered in the asymptotic limit of $t_0,t\to\infty$.

The usual strategy for computing the GEVP is to fix $t_0$
and compute the eigenvalues by varying $t$ for $t > t_0$.
However, the excited state contamination is primarily driven
 by the smaller of the two timeslices, or $t_0$ for this choice of ordering.
Fixing $t_0$ hides the excited state contamination,
 making for plots that are deceptively flat for increasing $t$ and small $t_0$
 despite a nonnegligible contamination.
For this reason, we choose to instead fix $\delta t = t - t_0$ and vary $t_0$.
It should be understood that this is simply a change of parameters in the context
 of the GEVP and any arguments about excited state contamination and
 asymptotic convergence of the GEVP in Ref.~\cite{Blossier:2009kd}
 will still hold.
Keeping $t_0 \geq \delta t$ ensures that the condition $t_0 \geq t/2$
 is satisfied.

One difficulty with the GEVP is the issue of eigenvalue sorting.
Nearby states with overlapping uncertainties and statistical
 fluctuations in data can make sorting of states according to
 their eigenvalues ambiguous.
To deal with this, the eigenvector sorting algorithm
 of Ref.~\cite{Fischer:2020bgv} is employed, which assigns a score 
 to every permutation of eigenvectors $\epsilon$
 according to the metric
\begin{align}
  c_\epsilon(t,t')
  = \prod_{k} \Big| {\rm det} \big[
   &V_0(t'), ..., V_{k-1}(t'), \nonumber\\
   &V_{k}(t), V_{k+1}(t'), ... V_{N-1}(t')
  \big] \Big|
  \label{eq:eigsorting}
\end{align}
 for sorting of eigenvectors on timeslice $t$ according to the reference
 vectors on timeslice $t'$.
The permutation $\epsilon$ that maximizes the score $c_\epsilon(t,t')$
 is taken as the optimal choice.
This metric favors orderings where eigenvector $V_k(t)$ 
 on the target timeslice is maximally orthogonal to the eigenvectors
 $V_{k'\neq k}(t')$ on the reference timeslice.

\subsection{Thermal Contributions to Two-Pion Correlators}\label{sec:thermal}

Correlation functions for the typical two-point $\pi\pi$ operators in general will
 contain not only the contributions from the propagating two-pion states,
 but also an unwanted contribution from a single pion that propagates
 through the temporal periodic boundary condition.
The two-point correlation function can be written as
\begin{align}
 &{\cal C}_{AB}(t)
  \nonumber\\
  &= \sum_{n} \bra{0} {\cal O}_{A} \ket{n}
  \bra{n} {\cal O}^\dagger_{B} \ket{0}
  e^{-E_{n} t}
  \nonumber\\
  &\phantom{=}+ \sum_{m'>0,m} \bra{m'} {\cal O}_{A} \ket{m}
  \bra{m} {\cal O}^\dagger_{B} \ket{m'}
  e^{-E_{m}t} e^{-E_{m'}(T-t)} ,
  \label{eq:2piexpectvalue}
\end{align}
 where the sums are taken over various choices of discrete lattice momenta.
The first term of the right-hand side is the desired correlation function,
 and the second term is a nuisance term involving $1\pi$
 transition matrix elements that must be estimated and subtracted away.

The expected number of state energies below some energy cutoff
 for the two-pion correlation functions can be estimated
 by counting distinct combinations of momenta.
Momenta that are related by discrete lattice rotations
 give degenerate spectra and therefore count toward a
 single state energy.
In the center-of-mass frame,
 at least one state contributes for each choice of
 magnitude for the back-to-back relative pion momenta
 and are schematically represented as
\begin{align}
 |n\rangle \sim \sum_{\mathbf{p}_{\text{rel}}}
 |\pi(\mathbf{p}_{\text{rel}}) \pi(-\mathbf{p}_{\text{rel}})\rangle.
\end{align}
In this schematic notation,
 the sum runs over momenta related by rotations.
This counting neglects states that may appear
 with additional particles or different particle content,
 and additional states may also appear when the interaction
 channel exhibits a resonance.
Higher-energy terms in the two-pion correlation functions
 that might result from excited states with
 other particle contents are treated as negligible due to their exponential
 mass suppression and no speculation is made about their content.

In the center of mass frame, the one-pion thermal terms
 can include any discrete momentum state
\begin{align}
 |m\rangle, |m'\rangle \sim |\pi(\mathbf{p}_\pi)\rangle
\end{align}
 with $|\mathbf{p}_\pi L/2\pi|^2 \in \mathds{Z}$,
 so long as $m=m'$ to conserve momentum.
Higher terms again might be present and are neglected.
The states $m$ and $m'$ in these thermal terms also sum over
 pions in an isospin triplet.
In general, matrix elements involving $\pi^\pm$ states do not contribute
 the same as $\pi^0$ states.
For the isospin-2 channel in particular,
 using isospin $z$-component equal 0 operators,
\begin{align}
 \langle \pi^\pm | {\cal O}_{2,0} | \pi^\pm \rangle
 = -\frac{1}{2} \langle \pi^0 | {\cal O}_{2,0} | \pi^0 \rangle.
\end{align}
For the isospin-0 channel, the prefactor is trivial,
\begin{align}
 \langle \pi^\pm | {\cal O}_{0,0} | \pi^\pm \rangle
 = \langle \pi^0 | {\cal O}_{0,0} | \pi^0 \rangle.
 \label{eq:wignereckart0}
\end{align}
For consistency, the $\pi^\pm$ contributions are converted to
 the $\pi^0$ contribution by applying Wigner-Eckart theorem.
The corresponding isospin prefactors on the thermal contributions are then
 3 and $\frac{3}{2}$ for the isospin 0 and 2 channels, respectively.
An additional prefactor counting the number of momenta,
 $N_{\{\mathbf{p}_{\rm rel}\}}$ from Eq.~(\ref{eq:nprel}),
 is also included to account for all possible momenta.

Although the thermal contributions can be sizeable for the
 spin-singlet representation at zero center-of-mass momentum,
 the thermal contributions may be removed by
 subtracting timeslices of the matrix correlation function, such as
\begin{equation}
 C'(t;dt) = C(t-dt) - C(t),
 \label{eq:timeseriessubtraction}
\end{equation}
 with fixed $dt$.
While this subtraction does remove the offending thermal contributions,
 it also increases the statistical noise of the
 desired $\pi\pi$ contribution at timeslice $t$
 and enhances the effects of excited states relative to the ground state.
This subtraction is also only exact for the $\pi\pi$ operators with
 zero total center-of-mass momentum, i.e. the rest frame.
For the moving frames, alternate subtraction schemes must be employed
 to account for the unequal masses of the two one-pion states
 and these subtractions schemes will not exactly cancel all of the
 thermal terms.

A unique feature of the distillation setup is the ability to easily compute
 the contributions from thermal terms.
These terms are particularly problematic in the S-wave representation 
 of the isospin 0 and isospin 2 channels, where it is possible to resolve the
 $1\pi$ contribution that propagates through the temporal extent of the lattice.
The transition matrix elements relating
 the $\pi\pi$ operators to the $1\pi$ states,
 $\bra{\pi} {\cal O}_{\pi\pi} \ket{\pi}$,
 can be obtained from a three-point function constructed with one-pion
 operators at the source and sink,
\begin{align}
 &{\cal C}_{3,A,\mathbf{p}_\pi}(t,\tau)
 =
 \langle {\cal O}_{\pi,\mathbf{p}_\pi}(t)
 {\cal O}_{A}(\tau)
 {\cal O}^\dagger_{\pi,\mathbf{p}_\pi}(0) \rangle
 \nonumber\\
 &= \sum_{mm'}
 \langle 0 | {\cal O}_{\pi,\mathbf{p}_\pi} |m \rangle
 \langle m | {\cal O}_{A} |m' \rangle
 \langle m'| {\cal O}^\dagger_{\pi,\mathbf{p}_\pi} |0 \rangle
 \nonumber\\
 &\phantom{= \sum_{mm'}}\times
 e^{-E_{m}(t-\tau)} e^{-E_{m'}\tau}
 +O(e^{-M_{\pi} (T-t)}),
 \label{eq:thermalcorrelator}
\end{align}
 where the $1\pi$ spectrum and matrix elements can
 be easily obtained from a one-pion two-point function.
The $1\pi$ operators are given in Eq.~(\ref{eq:pionbilinear}).
An explicit computation of these transition matrix elements
 can give insight about the size of the thermal contributions
 compared to the contributions from the desired $\pi\pi$ state propagation.

After obtaining the transition matrix elements from a dedicated study,
 the thermal terms can be effectively removed from the
 $\pi\pi$ correlation functions.
Assuming no vacuum contribution, the subtraction is defined by
\begin{align}
  &C_{AB}(t)
  \nonumber\\
  &= \langle {\cal O}_{A}(t) {\cal O}^\dagger_{B}(0) \rangle
  \nonumber\\
  &\phantom{=}- \sum_{m>0,n>0}
  \bra{n} {\cal O}_{A} \ket{m }
  \bra{m} {\cal O}^\dagger_{B} \ket{n}
  e^{-E_{m}t} e^{-E_{n}(T-t)}.
  \label{eq:thermalsubtraction}
\end{align}
Subtraction of thermal contributions gives better statistical
 precision than time series subtraction schemes without introducing
 large systematic corrections.
In calculations with moving frame data, explicit calculation of the
 transition matrix elements can give a more rigorous approach
 to removing thermal effects than an approximate subtraction
 scheme~\cite{Dudek:2012gj}.
To distinguish between these two schemes of removing
 contamination from vacuum and thermal contributions,
 we adopt the following nomenclature:
\begin{description}
 \item[``Time series subtraction'']
 Application of Eq.~(\ref{eq:timeseriessubtraction})
 to generally subtract contamination, and
 \item[``Matrix element subtraction'']
 Application of some combination of
 Eqs.~(\ref{eq:thermalsubtraction})~and~(\ref{eq:vac})--(\ref{eq:iso0subtraction})
 to subtract away specific contamination that is expected
 in the correlation functions.
\end{description}

\subsection{Vacuum Contributions to Two-Pion Correlators}\label{sec:vacuum}

In addition to the thermal corrections of the previous subsection,
 the isospin 0 channel in the center of mass frame
 also contains contributions from vacuum matrix elements,
\begin{align}
 &{\cal C}_{AB}(t)
  \nonumber\\
  &= \sum_{n>0} \bra{0} {\cal O}_{A} \ket{n}
  \bra{n} {\cal O}^\dagger_{B} \ket{0}
  e^{-E_{n} t}
  \nonumber\\
  &\phantom{=}+ \sum_{m'>0,m} \bra{m'} {\cal O}_{A} \ket{m}
  \bra{m} {\cal O}^\dagger_{B} \ket{m'}
  e^{-E_{m}t} e^{-E_{m'}(T-t)}
  \nonumber\\
  &\phantom{=}+ \bra{0} {\cal O}_{A} \ket{0} \bra{0} {\cal O}^\dagger_{B} \ket{0}.
  \label{eq:2piexpectvalueI0}
\end{align}
The vacuum matrix element can be obtained
 from a one-point correlation function,
\begin{align}
 {\cal C}_{1,A} \equiv \langle {\cal O}_{A} \rangle =
 \langle 0 | {\cal O}_{A} | 0 \rangle
 + \sum_{m>0} \langle m | {\cal O}_{A} | m \rangle e^{-E_{m} T},
 \label{eq:1pt}
\end{align}
 where the second term on the right hand side is a thermal
 contribution to the one-point function from thermal states
 propagating through the periodic boundary conditions.

Though in principle any momentum can appear for a
 $1\pi$ state thermal term,
 in practice the exponential suppression is sufficient
 to remove all but the 0-momentum contribution.
Like the thermal terms,
 summing over $m$ also produces an isospin factor to
 account for a triplet of pions propagating
 through the periodic boundary condition.
 From Eq.~(\ref{eq:wignereckart0}), this scales
 the transition matrix element for $\pi^0$ by 3.

The vacuum matrix element contributions tend to be statistically noisy,
 so it is beneficial to take advantage of statistical cancellation
 by correlating the vacuum contributions with
 the $\pi\pi$ two-point functions timeslice-by-timeslice.
The vacuum matrix element first has the thermal term removed
 (retaining the $t$-dependence of the vacuum term but
 taking the time-averaged thermal contribution),
\begin{align}
 &C_{1,A}(t)\equiv
 {\cal C}_{1,A}(t)
 -\sum_{m>0} \langle m | {\cal O}_{A} | m \rangle e^{-E_{m} T}.
 \label{eq:vac}
\end{align}
A subtracted two-point correlation function can be formed
 to remove the vacuum matrix element, correlating the times
 of the two-point functions with those of the one-point function,
\begin{align}
  &C_{AB}(t,t_0)\equiv \nonumber\\
  & \langle {\cal O}_{A}(t+t_0) {\cal O}^\dagger_{B}(t_0) \rangle
  -C_{1,A}(t+t_0) C_{1,B}(t_0).
  \label{eq:vsub}
\end{align}
After this correction,
 Eq.~(\ref{eq:vsub}) is averaged over $t_0$ and
 the remaining thermal term for the two-point function is subtracted away,
\begin{align}
  &C_{AB}(t)
  \equiv  \frac{1}{T} \sum_{t_0}
  C_{AB}(t,t_0)
  \nonumber\\
  &\phantom{=}- \sum_{m>0,n>0}
  \bra{n} {\cal O}_{A} \ket{m }
  \bra{m} {\cal O}^\dagger_{B} \ket{n}
  e^{-E_{m}t} e^{-E_{n}(T-t)}.
  \label{eq:iso0subtraction}
\end{align}
Combining all of these steps yields a correlation function
 that contains only the two-pion contributions of interest.

\subsection{Scattering Phase Shift Formalism}\label{sec:phaseshift}
The $\pi\pi$ scattering phase shift is computed from the energy spectrum in
 the finite, periodic volume assuming the center-of-mass energy is small enough
 that the pions elastically scatter with each other.
Large shifts in energy are possible due to finite periodic volume that the pions are confined to.
These energy shifts are the result of pions being unable to exist in
 wavefunctions that are asymptotically separated from each other,
 resulting in reinteractions as the particles traverse the periodic boundary conditions.
This energy shift can be expressed as a power-law correction to the two-particle spectrum states,
 unlike the usual exponential corrections that are seen for one-particle states.
In this section, only the most important formulae for computing the scattering phase shifts
 will be given.
A more detailed derivation has been performed in Appendix~\ref{sec:formalism}.

The derivation of the scattering phase shift relation comes
 from solving the determinant equation
\begin{equation}
  {\rm det}\left[e^{2i\delta(q)} -U (q)\right] = 0 ,
\end{equation}
 where the matrix $U(q)$ encodes the restriction of states with definite momentum
 and angular momentum to a finite box with cubic geometry.
The scattering phase shift $\delta$ depends on the spin and isospin channels.
For the purposes of this paper, the phase shift is written
 with subscripts as $\delta_{I\ell}(q)$, where $I$ is the
 isospin and $\ell$ is the angular momentum.
The dimensionless momentum $\mathbf{q}$ is given by
\begin{align}
 \mathbf{q} = {\mathbf{p}L}/{2\pi}.
\end{align}
The center-of-mass relative momentum of the two-pion states, $\mathbf{p}$,
 is deduced from the two-pion state energy,
\begin{equation}
  \mathbf{p}^2 = \frac{1}{4} \left( E_{\pi\pi}^2
  -(2 M_\pi)^2 -\mathbf{P}^2 \right) ,
  \label{eq:2pimomentum}
\end{equation}
 where the energy $E_{\pi\pi}$ is fit from lattice data and
 $\mathbf{P}$ is the momentum of the center of mass.
The matrix $U(q)$ is related to the scattering matrix ${\cal M}$ by
\begin{equation}
    U = ({\cal M} +i\mathds{1})({\cal M} -i\mathds{1})^{-1} .
\end{equation}
${\cal M}$ is determined on the lattice from a series of subduction coefficients $S$,
 Wigner ${\cal D}$-matrices, and a cubic rotation scattering matrix $M$,
 as a function of $q \propto p$,
\begin{align}
    {\cal M}^{\mathbf{P},\Lambda}_{\ell,n;\ell',n'}(q)
    =& \frac{1}{{\rm dim}\Lambda_{\mathbf{P}}}
    \sum_{\mu}
    \sum_{\lambda,\lambda'}
    \sum_{mm'} \nonumber\\
    &\times
    S^{\mathbf{P},\ell,\lambda\;\ast}_{\Lambda^{(n )},\mu}
    S^{\mathbf{P},\ell',\lambda'}_{\Lambda^{(n')},\mu} \nonumber\\
    &\times
    {\cal D}^{\ell\;\ast}_{m \lambda }(\hat{R}_0)
    {\cal D}^{\ell'     }_{m'\lambda'}(\hat{R}_0) \nonumber\\
    &\times
    M^{\mathbf{P}L/2\pi}_{\ell,m;\ell',m'} (q) .
    \label{eq:calMfirst}
\end{align}
Here, ${\rm dim}\Lambda_{\mathbf{P}}$ is the number of elements of
 the cubic rotation group irrep $\Lambda$
 that have total center-of-mass momentum $\mathbf{P}$.
Index $\mu$ is the ``row'' of the irrep $\Lambda$,
 and is an element of the irreducible representation vector space.
The indices $\ell^{(\prime)}$ are total angular momentum quantum numbers,
 $\lambda^{(\prime)}$ are helicity magnitude quantum numbers,
 and $m^{(\prime)}$ are $z$-component angular momentum quantum numbers.
The indices $n^{(\prime)}$ are used to denote distinct replicas
 of the irrep $\Lambda$ within the decomposition of $\ell^{(\prime)}$,
 since more than one of the same irrep may be obtained from a single angular
 momentum decomposition.

The matrix $M$ can then be reduced to a series of Clebsch-Gordan coefficients $C$
 of \sutwo~spin and a known geometric function ${\cal Z}$,
\begin{align}
  &M^{\mathbf{d}}_{\ell m, \ell^\prime m^\prime} (q) = -i\gamma^{-1}
  \frac{(-1)^\ell}{\pi^{3/2}} \nonumber\\
  &\phantom{M}\times
  \sum_{j=|\ell-\ell^\prime|}^{\ell+\ell^\prime}
  \sum_{s=-j}^{j} \left( -iq \right)^{-(j+1)}
  {\cal Z}_{js}^{\mathbf{d}}\left(1;q\right)
  C_{\ell m,js, \ell^\prime m^\prime} \,.
  \label{eq:Msecond}
\end{align}
The factor $\gamma$ is the Lorentz contraction factor
 computed with the momentum $\mathbf{P}$.

Although Eq.~(\ref{eq:calMfirst})
 is written in a manifestly basis-independent form
 due to the sum over irrep row $\mu$,
 Schur orthogonality relations may be used to prove that the computation
 with a single irrep row is also basis independent.

\subsection{Correlated Subtraction of One-Pion States}\label{sec:freepicorrelation}

To improve upon the precision of the extracted spectrum,
 it is beneficial to exploit the correlation between statistical fluctuations
 of the one-pion and two-pion correlation functions.
The phase shift may be computed using the energy difference
 between the usual interacting two-pion system
 and a ``noninteracting'' two-pion system,
 where noninteracting here means that the two pions are not allowed to
 exchange gluons or quarks.
This is an improvement over strategies that implement the ratio
 of correlation functions, such as in
 Refs.~\cite{Sharpe:1992pp,Gupta:1993rn,Kuramashi:1993ka}.

To make this procedure more concrete, consider the one-pion two-point
 correlation function with definite momentum $\mathbf{P}_0$,
\begin{align}
 C_{\pi,\mathbf{P}_0}(t)
 = \langle {\cal O}_{\pi,\mathbf{P}_0}(t)
 {\cal O}^\dagger_{\pi,\mathbf{P}_0}(0) \rangle.
\end{align}
This correlation function will have as its ground state the energy
 of a pion moving with momentum $\mathbf{P}_0$, $E_{\pi,\mathbf{P}_0}$.
This is combined in a product with another correlator
 with momentum $\mathbf{P}_1$ such that the two correlators
 form a noninteracting two-pion correlator with total momentum
 $\mathbf{P}=\mathbf{P}_0+\mathbf{P}_1$,
\begin{align}
 C^{\text{NI}}_{\pi\pi,\mathbf{P}_0,\mathbf{P}_1}(t)
 = C_{\pi,\mathbf{P}_0}(t) C_{\pi,\mathbf{P}_1}(t).
 \label{eq:noninteractingcorrelator}
\end{align}
Since no particles are exchanged, the noninteracting two-pion system
 only yields nonzero correlation functions when momentum is conserved
 for each pion independently.
Therefore, this correlator will have as its ground state the sum of the
 two one-pion energies,
\begin{align}
 E^{\text{NI}}_{\pi\pi,\mathbf{P}_0,\mathbf{P}_1} =
 E_{\pi,\mathbf{P}_0} + E_{\pi,\mathbf{P}_1}.
\end{align}
Eq.~(\ref{eq:noninteractingcorrelator}) is then averaged over all
 momentum combinations that are connected by rotations
 to optimize the statistical precision while retaining the
 statistical fluctuations of the interacting two-pion correlators.

In contrast to the noninteracting two-pion correlator,
 the usual interacting two-pion correlation functions will
 contain contributions where the pions exchange momentum.
This means that the interacting two-pion correlation functions will
 have a tower of two-pion states.
Provided that there is no resonance in the two-pion channel
 and that the energy is low enough that neither single pion
 can be promoted to an excited state,
 the states of the spectrum are in one-to-one correspondence with
 the noninteracting energy levels.
The interacting energy levels may therefore be identified
 by the momenta of the corresponding noninteracting momenta
 and so may be written as $E_{\pi\pi,\mathbf{P}_0,\mathbf{P}_1}$.
Then the interacting-noninteracting energy difference is
\begin{equation}
 \Delta E_{\pi\pi,\mathbf{P}_0,\mathbf{P}_1} =
 E_{\pi\pi,\mathbf{P}_0,\mathbf{P}_1} -
 E^{\text{NI}}_{\pi\pi,\mathbf{P}_0,\mathbf{P}_1}.
 \label{eq:niediff}
\end{equation}

For the purposes of evaluating the L\"uscher quantization condition,
 the center-of-mass momentum of the two pion system must be
 obtained from the relation
\begin{equation}
 \mathbf{p}^{2} =
 \frac{1}{4} \left( (E_{\pi\pi})^2 - (2M_\pi)^2 - \mathbf{P}^2 \right).
  \label{eq:ppicom}
\end{equation}
To leverage the precise energy difference obtained from Eq.~(\ref{eq:niediff}),
 the two-pion energy in Eq.~(\ref{eq:ppicom}) is replaced with
\begin{align}
 E_{\pi\pi} \to
 E_{\pi,\mathbf{P}_0} +E_{\pi,\mathbf{P}_1}
 +\Delta E_{\pi\pi,\mathbf{P}_0,\mathbf{P}_1}
\end{align}
 to obtain
\begin{widetext}
\begin{equation}
  \mathbf{p}^{2} =
  \frac{1}{4} \left(
  (\Delta E_{\pi\pi,\mathbf{P}_0,\mathbf{P}_1})^2
  + 2(\Delta E_{\pi\pi,\mathbf{P}_0,\mathbf{P}_1})
  ( E_{\pi,\mathbf{P}_0} +E_{\pi,\mathbf{P}_1})
  + ( E_{\pi,\mathbf{P}_0} +E_{\pi,\mathbf{P}_1})^2
  - (2M_\pi)^2 - \mathbf{P}^2 \right).
 \label{eq:ppicomdifference}
\end{equation}
\end{widetext}
The one-pion energies inserted into Eq.~(\ref{eq:ppicomdifference})
 may additionally be replaced by their continuum counterpart
 obtained by applying the dispersion relation to the pion mass,
\begin{equation}
 E_{\pi,\mathbf{P}_i} \to \sqrt{ M_\pi^2 + \mathbf{P}_i^2 }.
\end{equation}
This has the added benefit of removing some of the discretization effects
 that appear in the energy of the moving one-pion state,
 making the energy difference
 $\Delta E_{\pi\pi,\mathbf{P}_0,\mathbf{P}_1}$
 the dominant source of discretization errors.

\section{Computation Details}
\label{sec:ensemble}

\subsection{Ensemble Details}

\begin{table}[ht]
\centering
\begin{tabular}{c|ccccccc}
 Ensemble & $L^3\times T$ & $a^{-1}~[{\rm GeV}]$ & \# Conf. & \# Evecs \\
 \hline
 24ID & $24^3\times64$  & 1.015(15) & 33 & 120 \\
 48I  & $48^3\times96$  & 1.730(4)  & 27 & 60  \\
\end{tabular}
\caption{
 Details of the ensembles used in this study.
 Both ensembles used have physical $M_\pi$~\cite{Tu:2020vpn}.
\label{tab:ensembledetails}
}
\end{table}

This project uses the physical pion mass 2+1 flavor
 M\"obius Domain Wall Fermion gauge configurations
 generated by the RBC-UKQCD collaboration,
 labeled as 24ID and 48I.
The 48I ensemble uses an Iwasaki gauge action~\cite{RBC:2014ntl},
 while the 24ID uses the Iwasaki action combined with a
 Dislocation-Suppressing Determinant Ratio (Iwasaki+DSDR)
 method~\cite{Vranas:1999rz,Vranas:2006zk,Fukaya:2006vs,Renfrew:2008zfx}.
The lattice dimensions and lattice spacings are provided in
 Table~\ref{tab:ensembledetails}.
Both gauge ensembles are generated with physical quark masses,
 and all Dirac matrix solves are computed with the valence mass equal
 to the sea light-quark mass.

\subsection{Dirac Equation Solutions}
 
The distillation perambulators are computed using eigenvectors
 of the 3-dimensional Laplacian operator in Eq.~(\ref{eq:finitelaplacian}).
The links $U_i$ in the Laplacian have been Gaussian smeared
 in both space and time with $\rho=0.1$ and $N=30$.
Smearing in the time direction is kept ultra-local by
 zeroing out all but timeslices $t-1$, $t$, and $t+1$, smearing
 in all 4 dimensions, and then retaining only timeslice $t$.
This procedure is repeated for every timeslice $t$ over the entire configuration.%
\footnote{The source code for generating this smearing can be found online:
\url{https://github.com/lehner/gpt/blob/master/applications/distillation/booster-96I-basis.py}
}
The effect of smearing in the time direction smooths out fluctuations
 in the correlation functions but otherwise does not affect the final
 precision of measurements.

All measurements obtained on a single configuration are averaged together
 and treated as a single data sample to avoid correlation between measurements
 on the same configuration.
The sample configurations are separated by enough Markov Chain
 Monte Carlo trajectories so that no autocorrelation
 is observed in the correlator data.
For this reason, each configuration is treated as independent
 and no blocking over configurations is performed.
Statistical uncertainties on correlation functions and derived quantities
 are computed using jackknife resampling.

Correlation functions used in this analysis are computed with interpolating
 operators that couple to $\pi\pi$ states in the isospin 0 and isospin 2 channels.
Distillation operators are constructed with unit weights applied equally
 to all eigenmodes in the distillation basis.
Other interpolating operators may be achieved by relaxing this constraint,
 but no such study was performed here.
All of the $\pi\pi$ interpolating operators used in this analysis
 are constructed from two quark bilinears with equal and opposite Fourier phases,
 corresponding to  0 total center-of-mass momentum.
In the isospin 0 channel, an additional distillation-smeared scalar current was
 also included in the basis.
No links were used to point split the operators,
 which could be used to generate operators in nontrivial spin representations.

As a rule of thumb, it is possible to guess how many $\pi\pi$ states
 are accessible within statistics based on the number of distillation eigenvectors
 included in the interpolating operators.
In a free theory, the eigenvectors of the Laplacian operator used for the distillation basis
 are plane waves, with eigenvalues corresponding to the momentum squared.
With $N_v$ eigenvectors, it is then possible to access momenta up to $P_{\text{max}}$
 given by
\begin{align}
 |P_{\text{max}}|^2 \approx (N_v/3)^{\frac{1}{3}},
\end{align}
 where the factor of 3 accounts for color dilution and the power of $1/3$
 comes from filling all eigenvector momenta in three spatial directions.
From this counting, momenta up to $|P_{\text{max}}|^2 \approx 3.4$
 for the 24ID ensemble and $|P_{\text{max}}|^2 \approx 2.7$ on the 48I ensemble
 are expected to be well-constrained by the distillation data.
Empirically, momenta up to $|P_{\text{max}}|^2 = 4$ were found to be precise
 enough to use in this analysis.

\subsection{AMA Bias Correction of Correlation Functions}
\label{sec:biascorrection}

The All Mode Averaging (AMA)~\cite{Shintani:2014vja}
 bias correction procedure is performed on
 one-, two-, and three-point correlation functions.
Each of these correlation function types have a different
 set of available timeslices on which to compute the bias
 correction, so some description of the procedure is in order.
On both ensembles, ``exact'' full-precision solves are computed
 for all eigenvectors on timeslices in the set
 ${\cal T}_{\text{exact}} = \{t/a \in \{0, ..., 15\}\}$.
The low-precision ``sloppy'' solves are computed on every timeslice
 of each configuration, i.e.
 ${\cal T}_{\text{sloppy,24ID}} = \{t/a \in \{0, ..., 63\}\}$
 and
 ${\cal T}_{\text{sloppy,48I}} = \{t/a \in \{0, ..., 95\}\}$.

For the $1\pi$ two-point and three-point functions,
 the sink operator consists of a single quark bilinear operator.
The valence quark lines propagating outward from the sink operator
 does not connect the sink timeslice back to itself.
As a consequence, the antiquark line may always be obtained
 using $\gamma_5$ hermiticity in Eq.~(\ref{eq:gamma5hermiticity}),
 thereby avoiding the need for inversions on the sink timeslice.
So long as the source operator (and insertion operator for the
 three-point function) are contained within ${\cal T}_\text{exact}$,
 a viable full-precision solve may be obtained for the
 bias correction of the correlation function.
Therefore, all sink timeslices for the $1\pi$ two-point functions
 and all sink timeslices for $1\pi$ three-point functions with
 insertion time $\tau/a\in{\cal T}_{\text{exact}}$,
 which is satisfied by all data in this manuscript,
 may be bias corrected with the AMA procedure.

In $1\pi$ two-point correlation functions,
 the bias correction is computed and averaged over all source
 times before being added to the sloppy solves.
For ``sloppy'' solves $S$ and ``exact'' solves $X$,
 the bias correction procedure is written as
\begin{align}
 B(t) &= \big\langle X_{t_0} (t) -S_{t_0} (t)
 \big\rangle_{t_0\in {\cal T}_{\text{exact}}},
 \nonumber\\
 A(t) &= B(t) + \big\langle S_{t_0} (t)
 \big\rangle_{t_0\in {\cal T}_{\text{sloppy}}},
 \label{eq:biascorrectionsimple1pi2pt}
\end{align}
 where $A$ and $B$ denote the full bias-corrected data
 and the bias correction term, respectively.
In this equation, $t_0$ denotes the source time
 for equivalent correlators related by time translation.
The $1\pi$ three-point correlation functions are similarly computed,
 being sure to keep both the source time and insertion time within
 the allowed ranges:
\begin{align}
 B(t,\tau) &= \big\langle X_{t_0} (t,\tau) -S_{t_0} (t,\tau)
 \big\rangle_{\{t_0,t_0+\tau\}\in {\cal T}_{\text{exact}}},
 \nonumber\\
 A(t,\tau) &= B(t,\tau) + \big\langle S_{t_0} (t,\tau)
 \big\rangle_{\{t_0,t_0+\tau\}\in {\cal T}_{\text{sloppy}}}.
 \label{eq:biascorrectionsimple1pi3pt}
\end{align}

In contrast with the $1\pi$ correlation functions,
 the $\pi\pi$ interpolating operators used in this analysis
 require solutions to the Dirac equation
 originating from both the source and sink timeslices.
The bias correction may therefore only be applied to timeslices where it is possible
 to construct both of the $\pi\pi$ interpolating operators
 on timeslices within the set ${\cal T}_{\text{exact}}$.
When these correlators are substituted into the GEVP equation,
 the range of available bias correction data is reduced further.
For larger time separations, only the average of the sloppy measurements is available,
 which could introduce a bias in the result.
The data out beyond the region where the AMA bias correction is available
 is used only for checking consistency and to demonstrate general trends
 of the data beyond the available time range.

The bias-correction procedure for the $\pi\pi$ two-point correlation functions
 may then be written
\begin{align}
 B(t) &= \big\langle X_{t_0} (t) -S_{t_0} (t)
 \big\rangle_{\{t_0,t_0+t\}\in {\cal T}_{\text{exact}}},
 \nonumber\\
 A(t) &= B(t) + \big\langle S_{t_0} (t)
 \big\rangle_{\{t_0,t_0+t\}\in {\cal T}_{\text{sloppy}}}.
 \label{eq:biascorrectionsimple2pi2pt}
\end{align}
This equation is applicable for the isospin 2 channel $\pi\pi$
 correlation functions, even with thermal corrections applied
 as in Eq.~(\ref{eq:thermalsubtraction}).

The only remaining bias correction to discuss is the isospin 0 channel
 with the additional vacuum subtraction, as detailed in
 Eqs.~(\ref{eq:vac})--(\ref{eq:iso0subtraction}).
This is made more complicated by the correlation of source times
 between the one-point function and the two-point function.
The correct way to perform this subtraction is to apply
 Eq.~(\ref{eq:biascorrectionsimple2pi2pt})
 to the vacuum-subtracted correlation function in
 Eq.~(\ref{eq:vsub}).
This ensures that both timeslices in the product of one-point functions
 are contained within ${\cal T}_{\text{exact}}$
 when the bias correction is performed.
After the correlation function is bias-corrected,
 the remaining thermal subtraction in Eq.~(\ref{eq:iso0subtraction})
 may be performed on the corrected correlators.

\section{Spectrum Analysis}
 \label{sec:analysis}

\subsection{Corrections to Two-Point Correlation Functions}
\label{sec:subtractions}

As discussed in Sec.~\ref{sec:thermal},
 the $\pi\pi$ correlation functions have significant contributions from
 thermal contaminations due to $1\pi$ states that propagate through the
 periodic temporal boundary.
These manifest as exponential correlator contributions with small energies
 that are exponentially suppressed by the lattice temporal extent.
If not properly removed, these terms could cause the spectrum to be underestimated
 or fail to reach a plateau.

The temporal contributions in this analysis are removed by computing them
 in a dedicated calculation and subtracting them from the $\pi\pi$ correlators.
This is accomplished in two steps.
The first step is to compute the spectrum and matrix elements for
 one-pion (denoted $1\pi$) two-point
 correlation functions for each lattice momentum.
The second step is to insert a $\pi\pi$ operator between two $1\pi$ interpolating operators
 to compute a three-point function.
Using the spectrum and matrix elements from the first step,
 the three-point function is used to compute a transition matrix element
 that connects a $1\pi$ state back to itself.

\subsubsection{One-Pion Fits}
\label{sec:onepion}

For the first step in the process, the $1\pi$ two-point correlation functions
 are computed to extract the spectrum and matrix elements for the distillation-smeared
 pseudoscalar source and sink operators.
These correspond to the expectation values
\begin{align}
 &\langle
 {\cal O}_{\pi,\mathbf{p}_\pi}(t)
 {\cal O}^\dagger_{\pi,\mathbf{p}_\pi}(0)
 \rangle
 \nonumber\\
 &= \sum_m^\infty
 \langle 0 | {\cal O}_{\pi,\mathbf{p}_\pi} | \pi^0 \rangle_m
 \langle \pi^0 |_m {\cal O}^\dagger_{\pi,\mathbf{p}_\pi} | 0 \rangle
 e^{-E_{\pi,m} t}
 \nonumber\\
 &+ \sum_m^\infty
 \langle \pi^0 |_m {\cal O}_{\pi,\mathbf{p}_\pi} | 0 \rangle
 \langle 0 | {\cal O}^\dagger_{\pi,\mathbf{p}_\pi} | \pi^0 \rangle_m
 e^{-E_{\pi,m} (T-t)}.
 \label{eq:1pifit}
\end{align}
The pion states are given a superscript to indicate that they are
 computed for the neutral pion.
The first and second terms of the RHS are the forward- and backward-propagating contributions,
 respectively, with $T$ being the total time extent of the lattice.

The $1\pi$ two-point correlation functions are fit to the sum of exponentials
 in Eq.~(\ref{eq:1pifit}) using a correlated $\chi^2$ test statistic.
The fit ansatz assumes that the sum is truncated to at most two states $|\pi^0\rangle_m$,
 corresponding to the pion ground state and a single excited state,
 both with momentum
 $\mathbf{p}_{\text{in}} = \mathbf{p}_{\text{out}} \equiv \mathbf{p}_{\pi}$.
The fits included all data in a range between \tmin~and \tmax,
 with a separate choice of \tmin~and \tmax~for both 1-state and 2-state fits.
The difference between the 1-state and 2-state fits was used
 as a systematic uncertainty for the contamination due to excited states.

The results of the fits are given in
 Tables~\ref{tab:1pifit24ID}~and~\ref{tab:1pifit48I}
 for the 24ID and 48I ensembles, respectively.
All the fits on the 24ID ensemble were performed with
 $(\tmin,\tmax)=(3,14)$ for the 2-state fits and $(\tmin,\tmax)=(6,14)$ for the 1-state fits,
 corresponding to 7 and 6 degrees of freedom, respectively.
Similarly, the 48I ensemble were performed with
 $(\tmin,\tmax)=(5,16)$ for the 2-state fits and $(\tmin,\tmax)=(10,16)$ for the 1-state fits,
 corresponding to 7 and 4 degrees of freedom, respectively.
These fit ranges are small enough that the covariance matrix is well-conditioned.

\begin{table}
\begin{tabular}{c|rrc}
 $\Big(\frac{\mathbf{p}L}{2\pi}\Big)^{2}$ & $aE_{\pi}$(stat)(syst) & $\langle 0 | {\cal O}_{\pi} | \pi^0 \rangle$(stat)(syst) & $p$-val \\
 \hline
0 & $0.14015(32)(00)$ & $0.24643(61)(03)$ & 0.950 \\
1 & $0.29590(48)(10)$ & $0.12878(31)(03)$ & 0.656 \\
2 & $0.39491(94)(30)$ & $0.09100(26)(13)$ & 0.019 \\
3 & $0.4748(20)(03)$ & $0.06792(34)(09)$ & 0.573 \\
4 & $0.5330(42)(14)$ & $0.05104(65)(19)$ & 0.392 \\
\end{tabular}
\caption{
 Pion masses and matrix elements from $1\pi$ two-point function fits
  on the 24ID ensemble.
 The first column denotes the squared magnitude of the lattice momentum, in units of $(2\pi/L)^2$.
 For each momentum, the mass and matrix elements of the ground state,
  in lattice units, and the $p$-value for the 2-state fit are given.
 Both statistical and systematic uncertainties are listed,
  with the only systematic arising from the difference between
  the 1-state and 2-state fits.
 The 2-state fits all have 7 degrees of freedom.
 \label{tab:1pifit24ID}
}
\end{table}

\begin{table}
\begin{tabular}{c|rrc}
 $\Big(\frac{\mathbf{p}L}{2\pi}\Big)^{2}$ & $aE_{\pi}$(stat)(syst) & $\langle 0 | {\cal O}_{\pi} | \pi^0 \rangle$(stat)(syst) & $p$-val \\
 \hline
0 & $0.08027(20)(06)$ & $0.04841(11)(02)$ & 0.388 \\
1 & $0.15358(37)(19)$ & $0.021413(48)(71)$ & 0.540 \\
2 & $0.20320(38)(29)$ & $0.013060(35)(30)$ & 0.383 \\
3 & $0.2408(09)(08)$ & $0.008396(36)(44)$ & 0.519 \\
4 & $0.2743(19)(29)$ & $0.00558(05)(09)$ & 0.762 \\
\end{tabular}
\caption{
 The same as Table~\ref{tab:1pifit24ID}, but for the 48I ensemble.
 Like the 24ID fits, the 2-state fits all have 7 degrees of freedom.
 \label{tab:1pifit48I}
}
\end{table}

\subsubsection{Pion Transition Matrix Elements}
\label{sec:piontransition}

The matrix elements that appear as contamination of the two-point function
 in Eq.~(\ref{eq:2piexpectvalue}) are squares of the $1\pi$-to-$1\pi$
 matrix elements times an exponential suppression factor.
These contributions are observed to be at most at the 1\% level,
 so the precision requirements on these matrix elements is not as great
 as on the $\pi\pi$ two-point functions themselves.
For this reason, a simplified analysis method is sufficient for extracting
 the $1\pi$-to-$1\pi$ transition matrix elements.

The transition matrix elements needed
 to complete the subtraction are obtained from a ratio,
\begin{widetext}
\begin{align}
 \bra{\pi^0} {\cal O}_{X} \ket{\pi^0}
 \overset{t,\tau\to\infty}{\approx}
 \sqrt{ 
 \frac{
 \langle {\cal O}_{\pi,\mathbf{p}_\pi}(t) {\cal O}_{X}(\tau)
 {\cal O}^{\dagger}_{\pi,\mathbf{p}_\pi}(0) \rangle
 \langle {\cal O}_{\pi,\mathbf{p}_\pi}(t) {\cal O}_{X}(t-\tau)
 {\cal O}^{\dagger}_{\pi,\mathbf{p}_\pi}(0) \rangle
 }{ \langle {\cal O}_{\pi,\mathbf{p}_\pi}(t)
  {\cal O}^{\dagger}_{\pi,\mathbf{p}_\pi}(0) \rangle
 \langle {\cal O}_{\pi,\mathbf{p}_\pi}(t)
  {\cal O}^{\dagger}_{\pi,\mathbf{p}_\pi}(0) \rangle }
 } 
 \equiv R(t,\tau)
 .
 \label{eq:c3ptratio}
\end{align}
\end{widetext}
Here, ${\cal O}_{X}$ is a stand-in for either
 the distillation-smeared scalar
 operator ${\cal O}_1$ in Eq.~(\ref{eq:scalarop})
 or the $\pi\pi$ operators
 ${\cal O}_{\pi\pi} \Big(\big|\frac{\mathbf{p}_{\text{rel}}L}{2\pi}\Big|^2\Big)$
 with back-to-back momenta $\mathbf{p}_{\text{rel}}$
 in Eq.~(\ref{eq:pipiop}).
The scalar operator is only permitted
 in the isospin 0 channel.
The pion operators ${\cal O}_{\pi,\mathbf{p}_\pi}$ are constructed
 with momentum $\mathbf{p}_{\pi}$,
 to give the ingoing and outgoing $1\pi$ states momentum.
The times $t$ and $\tau$ correspond to the source-sink and source-insertion times,
 respectively.
Data for the three-point functions are generated for source-insertion times
 $\tau/a\in\{3,6,9,12\}$, and source-sink separation times up to $t/a=24$.
The correlators are computed for all of the $\pi\pi$ and scalar operators
 that are included in the analysis of $\pi\pi$ two-point function analysis
 and for ingoing/outgoing momenta $|\mathbf{p}_{\pi}L/2\pi|^2 \leq 4$.

Symmetrizing over the three-point function times in the numerator
 of Eq.~(\ref{eq:c3ptratio}) helps by canceling the exponential
 falloff of the numerator and denominator.
This ratio assumes that the correlation functions in the numerator
 and the denominator have only one state contributing to the correlation function.
While this will not be strictly true, the correlation function does indeed
 appear to be dominated by the ground state.
The excited state pion has a mass of 
 $\sim1300~{\rm GeV}$~\cite{Tanabashi:2018oca}, indicating that it is expected
 to be highly suppressed compared to the single, unexcited pion state.

\begin{figure}[t!]
 \includegraphics[width=\figsize\textwidth]{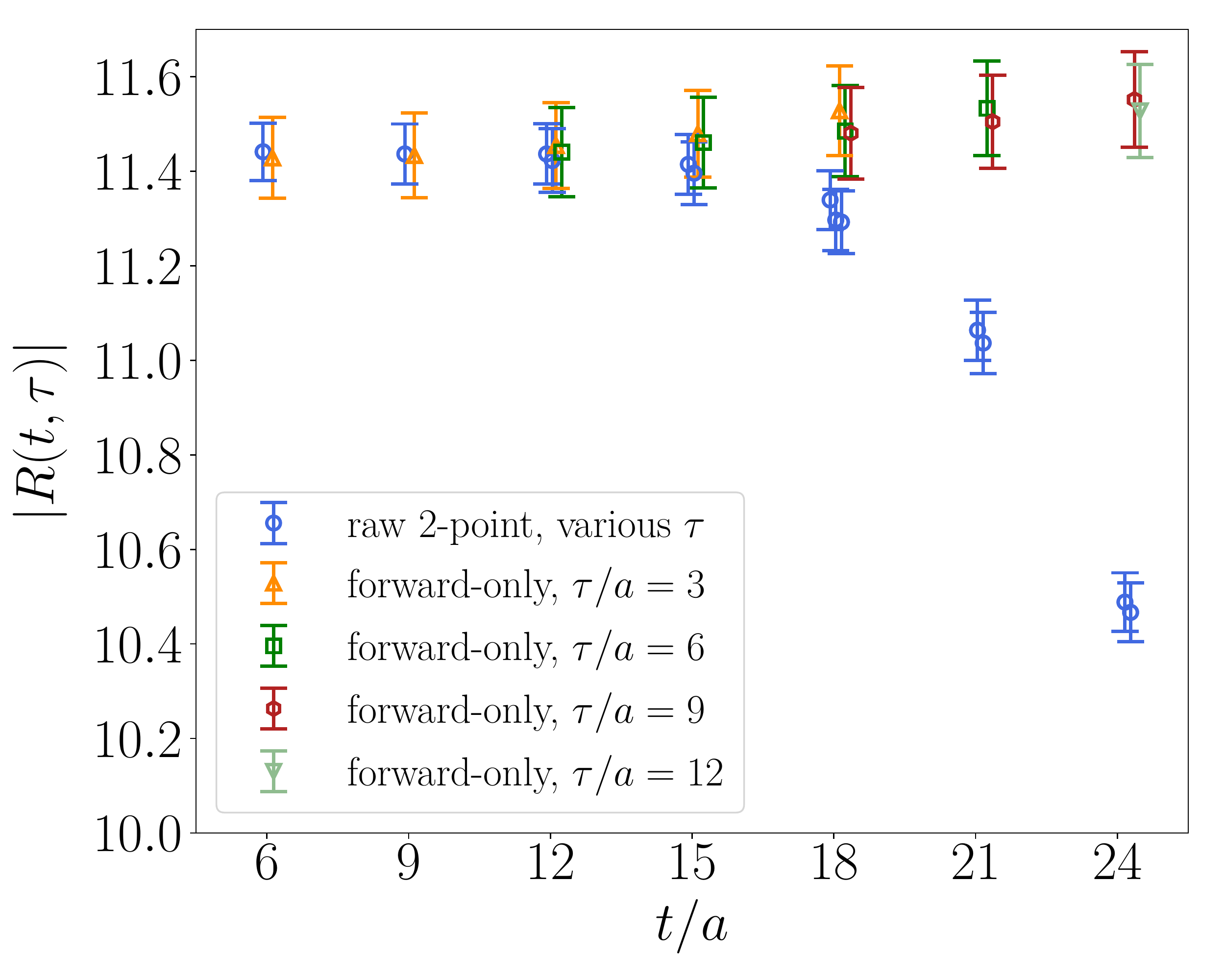}
 \caption{
 Transition matrix element for the three-point correlation function
  in the isospin 2 channel on the 24ID ensemble,
  where the effect of changing the denominator is most prominent.
 The ratio $|R(t,\tau)|$ for this plot asymptotes to
  $\langle \pi^0 |
  {\cal O}_{\pi\pi} (0)
  | \pi^0 \rangle$
  with ingoing and outgoing pion momentum
  $\mathbf{p}_{\pi} = 0$.
 The data labeled ``raw 2-point'' use the raw correlation function
  in the denominator of $R(t,\tau)$ given in Eq.~(\ref{eq:c3ptratio}).
 The ``forward-only'' data include only the forward-propagating
  exponential in the denominator, as described by
  Eq.~(\ref{eq:c3ptreplacementden}).
 For the ``forward-only'' ratio data, values with distinct $\tau$
  are plotted with separate symbols and colors.
 \label{fig:atwsub0}}
\end{figure}

For many of the choices of operator and momenta,
 this ratio exhibits a distinct dependence on the source-sink time $t$
 as $t$ becomes large, as seen for the blue
 circles plotted in Fig.~\ref{fig:atwsub0}.
This is due to the backward-propagating pion term in the
 correlation function in Eq.~(\ref{eq:1pifit}),
 which contributes nonnegligibly to the denominator of Eq.~(\ref{eq:c3ptratio}).
To correct for this, the $1\pi$ two-point correlation functions
 in the denominator of Eq.~(\ref{eq:c3ptratio})
 are replaced by only the forward-propagating exponential from the
 $1\pi$ ground state obtained from fits in the previous section,
\begin{align}
 &\langle {\cal O}_{\pi}(t) {\cal O}^\dagger_{\pi}(0) \rangle
 \to |\bra{0} {\cal O}_{\pi} \ket{\pi^0}|^2 e^{-E_\pi t} \,.
 \label{eq:c3ptreplacementden}
\end{align}

The ratio using only the forward-propagating exponential is plotted
 in Fig.~\ref{fig:atwsub0} using various symbols and colors,
 versus the ratio with the unmodified two-point correlator
 shown as blue circles.
After correcting for the backward-propagating term
 in the denominator, all of the ratio data for various
 $t/a$ and $\tau/a$ are consistent with each other within uncertainties.
This agreement is seen for all choices of operators and momentum
 in the isospin 2 channel, with the exception of only the smallest
 $\tau/a$ on the 48I ensemble where excited state contamination is visible.

In addition to the backward propagating pion in the denominator
 of the ratio of Eq.~(\ref{eq:c3ptratio}),
 the numerator of the isospin 0 three-point functions also exhibit
 a significant contribution from a backward-propagating pion.
This contamination is due to a vacuum matrix element
 that is not present in the isospin 2 channel,
\begin{align}
 &\langle {\cal O}_{\pi}(t) {\cal O}_{X}(\tau)
 {\cal O}^{\dagger}_{\pi}(0) \rangle
 \nonumber\\
 & \approx
 \langle 0 | {\cal O}_{\pi} | \pi^0 \rangle
 \langle \pi^0 | {\cal O}_{X} | \pi^0 \rangle
 \langle \pi^0 | {\cal O}^{\dagger}_{\pi} | 0 \rangle
 e^{-E_\pi t}
 \nonumber\\
 &+
 \langle \pi^0 | {\cal O}_{\pi} | 0 \rangle
 \langle 0 | {\cal O}_{X} | 0 \rangle
 \langle 0 | {\cal O}^{\dagger}_{\pi} | \pi^0 \rangle
 e^{-E_\pi (T-t)}.
\end{align}
For all but the lightest pion state,
 these contributions are safely negligible.
However, the pion at rest propagating through the temporal boundary 
 condition has a small enough mass to contribute nonnegligibly.
These contributions are estimated using the vacuum matrix elements
 approximated by the raw one-point correlation function,
\begin{align}
 \langle 0 | {\cal O}_{X} | 0 \rangle
 \approx \langle {\cal O}_{X} \rangle,
\end{align}
 and the overlaps and energies computed from the
 $1\pi$ two-point correlation functions.
The three-point correlation functions in the numerator
 are then replaced by the subtracted correlator,
\begin{align}
 &\langle {\cal O}_{\pi}(t) {\cal O}_{X}(\tau)
 {\cal O}^{\dagger}_{\pi}(0) \rangle \to
 \nonumber\\
 &\quad\langle {\cal O}_{\pi}(t) {\cal O}_{X}(\tau)
 {\cal O}^{\dagger}_{\pi}(0) \rangle
 -\langle {\cal O}_{X} \rangle
 |\bra{0} {\cal O}_{\pi} \ket{\pi^0}|^2 e^{-E_\pi (T-t)}.
 \label{eq:c3ptreplacementnum}
\end{align}

\begin{figure}[t!]
 \includegraphics[width=\figsize\textwidth]{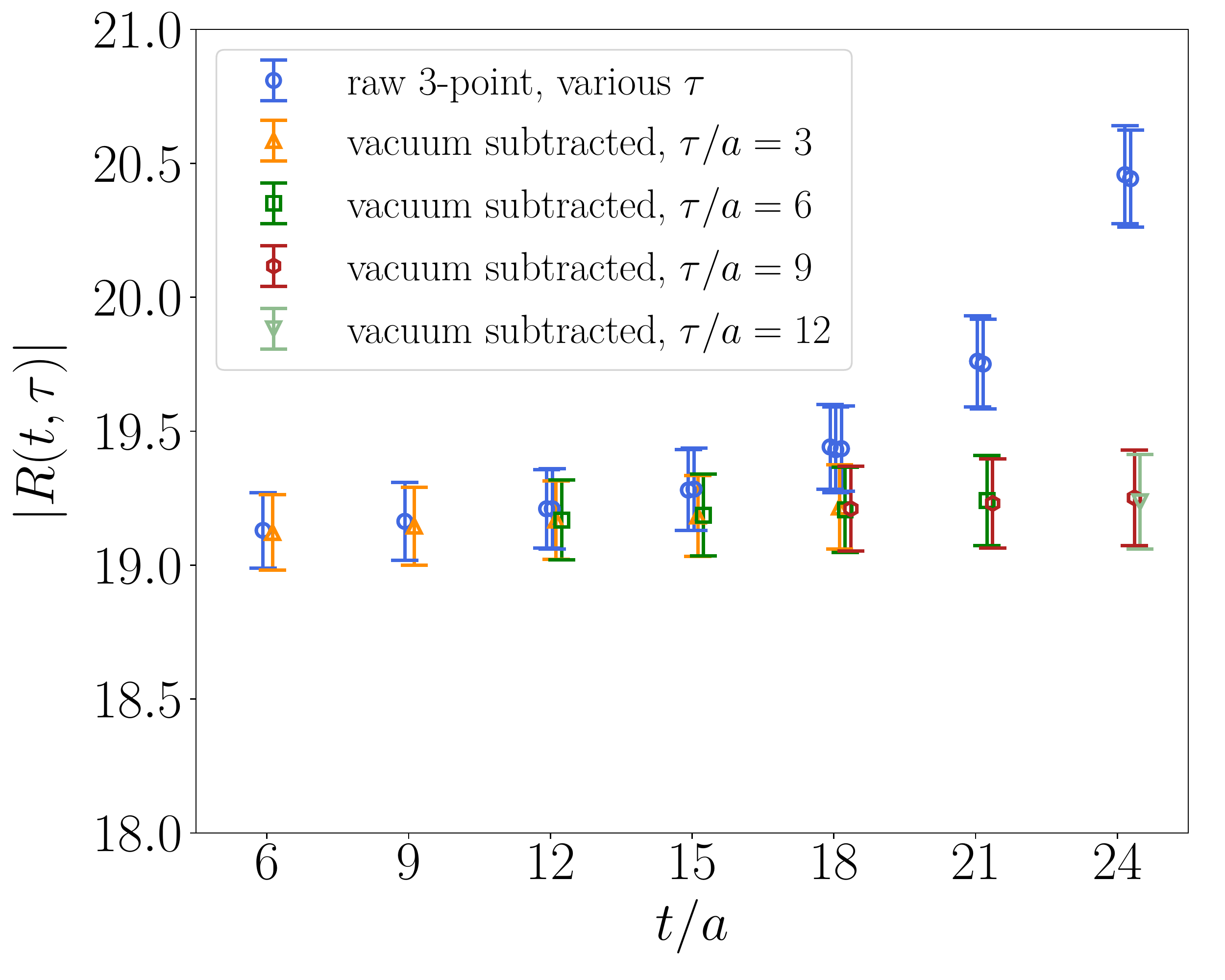}
 \caption{
 Transition matrix element for the three-point correlation function
  in the isospin 0 channel on the 24ID ensemble,
  where the effect of changing the numerator is most prominent.
 The ratio $|R(t,\tau)|$ for this plot asymptotes to
  $\langle \pi^0 |
  {\cal O}_{\pi\pi} (0)
  | \pi^0 \rangle$
  with ingoing and outgoing pion momentum
  $\mathbf{p}_{\pi} = 0$.
 The data labeled ``raw 3-point'' use the raw correlation function
  in the numerator of $|R(t,\tau)|$ given in Eq.~(\ref{eq:c3ptratio}).
 The ``vacuum subtracted'' data include the subtraction
  of the term proportional to the vacuum matrix element
  in the numerator, as described by Eq.~(\ref{eq:c3ptreplacementnum}).
 All data plotted here have the backward-propagating
  term removed from the denominator,
  as described in Eq.~(\ref{eq:c3ptreplacementden}).
 For the ``vacuum subtracted'' ratio data, values with distinct $\tau$
  are plotted with separate symbols and colors.
 \label{fig:atwsub1}
 }
\end{figure}

The ratio data using the raw three-point correlation functions
 and the ratios after the subtraction of Eq.~(\ref{eq:c3ptreplacementnum})
 are plotted in Fig.~\ref{fig:atwsub1}.
In this figure, the ratio data without the subtraction exhibit a definitive
 upward trend with increasing $t/a$ due to the backward-propagating one-pion state
 with the vacuum contribution.
This trend vanishes after the subtraction and all data are very consistent
 over the entire set of times plotted.
After performing the subtractions of the backward-propagating
 one-pion state in both the numerator and denominator,
 the isospin 0 ratio data for all operators and momenta
 are in good agreement within their quoted uncertainties
 except for the smallest choice of $\tau/a$ on the 48I ensemble.

The ratio values $R(18,9)$ for all choices of momentum and isospin
 are taken as the nominal value for the transition matrix elements.
To account for the remaining time dependence due to
 neglected contaminations from excited states,
 a systematic uncertainty is included corresponding to the
 difference $\delta R = R(18,9)-R(12,6)$.

\begin{figure*}[t!]
 \centering
 \begin{tabular}{cc}
 \includegraphics[width=\figsize\textwidth]{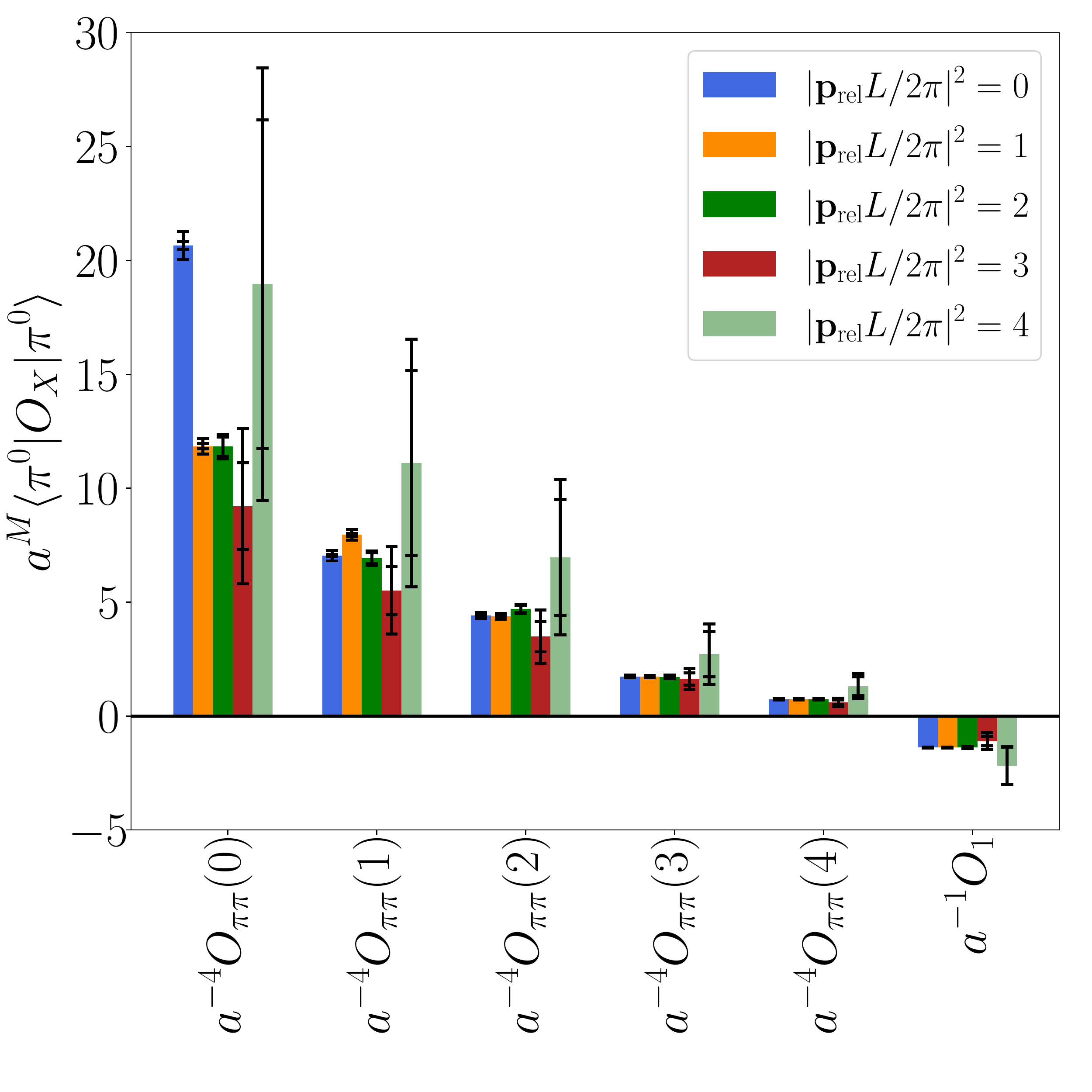} &
 \includegraphics[width=\figsize\textwidth]{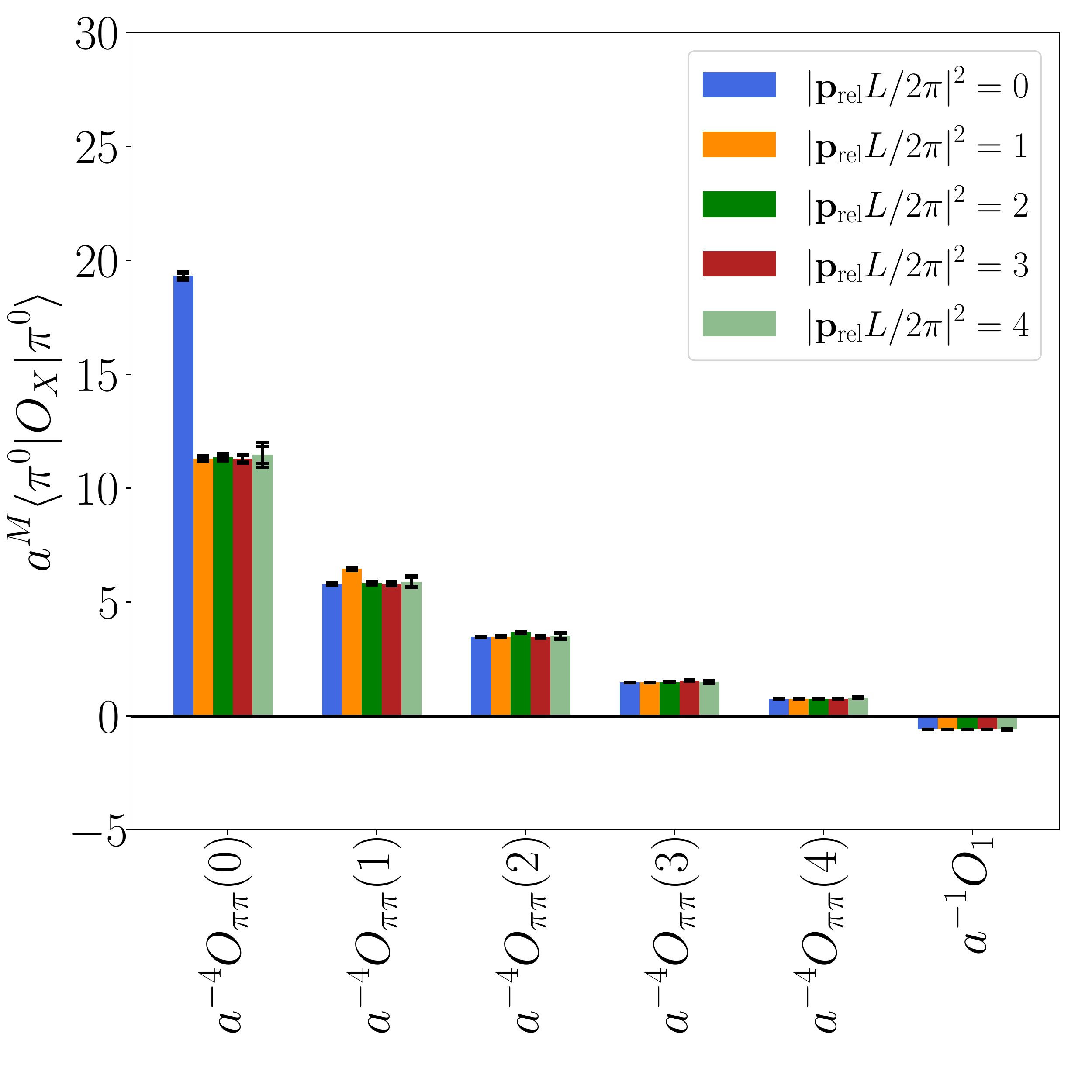} \\
 \includegraphics[width=\figsize\textwidth]{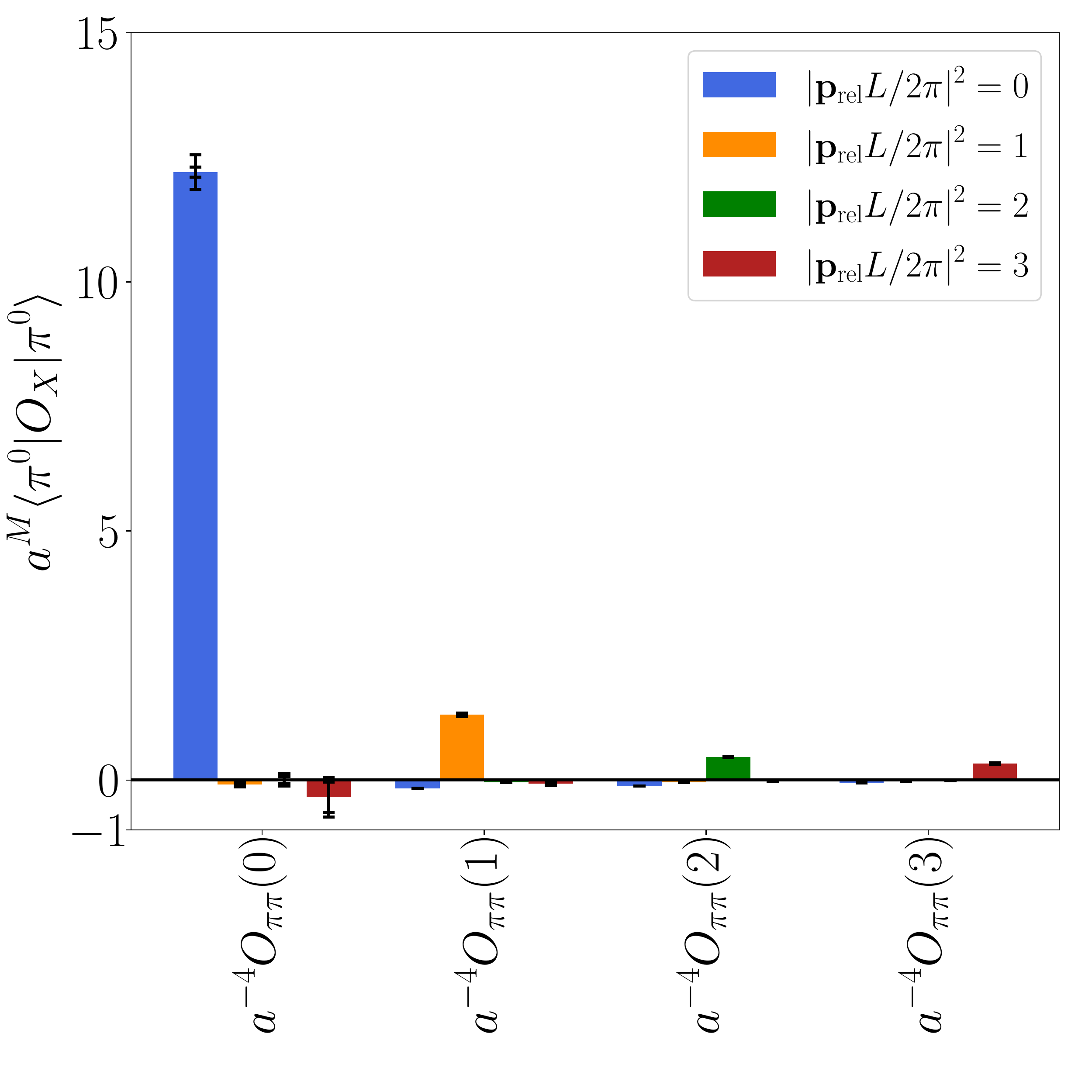} &
 \includegraphics[width=\figsize\textwidth]{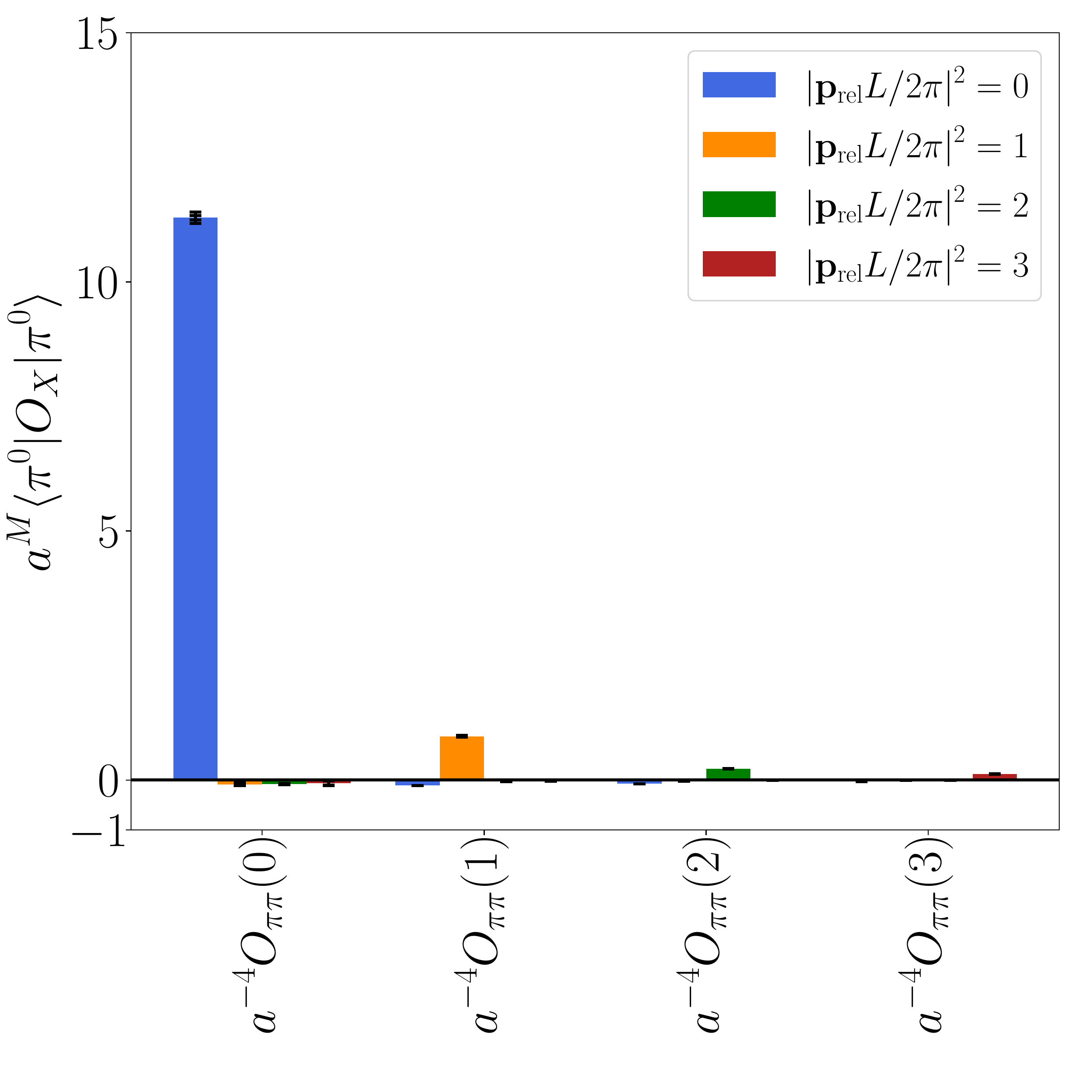}
 \end{tabular}
 \caption{
 Summary plots of the size of the $1\pi$ transition matrix elements,
  $a^{-4} \langle \pi^0 |
   {\cal O}_{\pi\pi}
  | \pi^0 \rangle$
  and $a^{-1} \langle \pi^0 | {\cal O}_{1} | \pi^0 \rangle$,
  over all choices of external pion momentum
  $\mathbf{p}_{\text{rel}}$ (bar color)
  and over the operator construction (horizontal axis displacement).
 The $\pi\pi$ operators are labeled by the values of
  $|\mathbf{p}_{\text{rel}}L/2\pi|^2$ 
  for the bilinears that make up the operator (in parentheses).
 The top (bottom) row contains the $I=0$ ($I=2$) results.
 For the isospin 0 channel,
  the scalar operator from Eq.~(\ref{eq:scalarop}) is also included.
 The left column contains results for the 24ID ensemble, 
  and the right column for 48I ensemble.
 Both statistical and total uncertainties are plotted.
 \label{fig:atwsummary}}
\end{figure*}

A summary of the transition matrix elements obtained from the ratio in Eq.~(\ref{eq:c3ptratio})
 with the replacements of the denominator from Eq.~(\ref{eq:c3ptreplacementden})
 and, for the isospin 0 channel, the numerator from Eq.~(\ref{eq:c3ptreplacementnum})
 are depicted in Fig.~\ref{fig:atwsummary} and their values listed in
 Tables~\ref{tab:24idI0matrixelement}--\ref{tab:48iI2matrixelement}.
The isospin 0 transition matrix elements, depicted in the top row of Fig.~\ref{fig:atwsummary},
 are more agnostic in regards to how strongly the operators couple to any given state
 compared to the isospin 2 channel.
The isospin 2 matrix elements, in the bottom row,
 exhibit significant suppression of operator-state combinations where
 $\mathbf{p}_{\pi} \neq \mathbf{p}_{\text{rel}}$.
In all cases, the thermal terms associated with these matrix elements
 will also be suppressed by exponentials of size
 $e^{-E_{\mathbf{p}_{\pi}}T}$.
The only terms that produce a nonnegligible contribution to the correlation
 function are those with $\mathbf{p}_{\text{rel}}=(0,0,0)$,
 which are the left-most (blue) bar for each operator choice.

\begin{table*}[!htb]
\begin{tabular}{c|rrrrrrrrrrr|rr}
 && \multicolumn{11}{c}{ $\langle \pi^0 | {\cal O}_{X} | \pi^0 \rangle$(stat)(syst) } \\
 \hline
 $\mathbf{p}_{\pi}$ & $\mathbf{p}_{\text{rel}}$
& \multicolumn{2}{c}{(0,0,0)}
& \multicolumn{2}{c}{(1,0,0)}
& \multicolumn{2}{c}{(1,1,0)}
& \multicolumn{2}{c}{(1,1,1)}
& \multicolumn{2}{c}{(2,0,0)}
& \multicolumn{2}{c}{scalar}
 \\
 \hline
(0,0,0) && $19.43(16)(22)$ && $6.62(05)(10)$ && $4.146(30)(63)$ && $1.636(13)(24)$ && $0.694(07)(10)$ && $-1.361(09)(14)$ & \\
(1,0,0) && $11.14(11)(05)$ && $7.484(61)(13)$ && $4.113(38)(06)$ && $1.625(15)(02)$ && $0.6915(72)(10)$ && $-1.359(12)(03)$ & \\
(1,1,0) && $11.13(40)(10)$ && $6.52(22)(10)$ && $4.42(14)(05)$ && $1.619(56)(22)$ && $0.690(24)(09)$ && $-1.354(47)(27)$ & \\
(1,1,1) && $8.7(1.8)(2.7)$ && $5.2(1.0)(1.5)$ && $3.3(0.6)(0.9)$ && $1.53(25)(35)$ && $0.56(10)(14)$ && $-1.09(21)(30)$ & \\
(2,0,0) && $17.8(6.8)(5.8)$ && $10.5(3.8)(3.4)$ && $6.6(2.4)(2.1)$ && $2.6(0.9)(0.8)$ && $1.23(39)(34)$ && $-2.14(80)(17)$ & \\
\end{tabular}
\caption{
 Table of $1\pi$ transition matrix elements for the 24ID ensemble
 in the isospin 0 channel.
 The momenta
 $\mathbf{p}_\pi$ and $\mathbf{p}_{\text{rel}}$
 are defined in
 Eqs.~(\ref{eq:thermalcorrelator})~and~(\ref{eq:pipiop}),
 respectively.
 The matrix elements reported here are obtained from the ratio
 in Eq.~(\ref{eq:c3ptratio}).
 The systematic uncertainty comes from the fits to $1\pi$
 two-point correlation functions and from the difference
 of ratio times.
 \label{tab:24idI0matrixelement}
}
\end{table*}

\begin{table*}[!hbt]
\begin{tabular}{c|rrrrrrrrrrr|rr}
 && \multicolumn{11}{c}{ $\langle \pi^0 | {\cal O}_{X} | \pi^0 \rangle$(stat)(syst) } \\
 \hline
 $\mathbf{p}_{\pi}$ & $\mathbf{p}_{\text{rel}}$
& \multicolumn{2}{c}{(0,0,0)}
& \multicolumn{2}{c}{(1,0,0)}
& \multicolumn{2}{c}{(1,1,0)}
& \multicolumn{2}{c}{(1,1,1)}
& \multicolumn{2}{c}{(2,0,0)}
& \multicolumn{2}{c}{scalar}
 \\
 \hline
(0,0,0) && $2.159(08)(07)$ && $0.6467(27)(34)$ && $0.3876(18)(17)$ && $0.1649(09)(07)$ && $0.08408(49)(33)$ && $-0.3379(09)(08)$ & \\
(1,0,0) && $1.2622(63)(44)$ && $0.7212(40)(27)$ && $0.3884(24)(13)$ && $0.1650(11)(06)$ && $0.08396(63)(29)$ && $-0.3398(15)(11)$ & \\
(1,1,0) && $1.268(08)(12)$ && $0.6519(44)(55)$ && $0.4093(30)(27)$ && $0.1658(13)(13)$ && $0.08436(72)(64)$ && $-0.3418(18)(21)$ & \\
(1,1,1) && $1.261(16)(07)$ && $0.6476(78)(34)$ && $0.3878(47)(19)$ && $0.1743(21)(08)$ && $0.0838(11)(04)$ && $-0.3394(39)(18)$ & \\
(2,0,0) && $1.280(42)(41)$ && $0.657(21)(20)$ && $0.394(12)(12)$ && $0.1673(53)(50)$ && $0.0898(28)(23)$ && $-0.344(11)(12)$ & \\
\end{tabular}
\caption{
 Same as Table~\ref{tab:24idI0matrixelement},
 but for the 48I ensemble.
 \label{tab:48iI0matrixelement}
}
\end{table*}

\begin{table*}[!hbt]
\begin{tabular}{c|rrrrrrrrrrr}
 && \multicolumn{9}{c}{ $\langle \pi^0 | {\cal O}_{\pi\pi} | \pi^0 \rangle$(stat)(syst) } \\
 \hline
 $\mathbf{p}_{\pi}$ & $\mathbf{p}_{\text{rel}}$
& \multicolumn{2}{c}{(0,0,0)}
& \multicolumn{2}{c}{(1,0,0)}
& \multicolumn{2}{c}{(1,1,0)}
& \multicolumn{2}{c}{(1,1,1)}
& \multicolumn{2}{c}{(2,0,0)}
 \\
 \hline
(0,0,0) && $11.48(10)(04)$ && $-0.1563(17)(03)$ && $-0.1125(11)(04)$ && $-0.05489(58)(21)$ && $-0.03071(35)(11)$ & \\
(1,0,0) && $-0.083(17)(43)$ && $1.2352(61)(24)$ && $-0.04440(53)(16)$ && $-0.01986(26)(03)$ && $-0.01049(17)(11)$ & \\
(1,1,0) && $0.00(07)(09)$ && $-0.0445(44)(18)$ && $0.4330(40)(42)$ && $-0.0112(09)(14)$ && $-0.00645(49)(26)$ & \\
(1,1,1) && $-0.33(29)(23)$ && $-0.067(17)(32)$ && $-0.012(08)(10)$ && $0.3100(46)(70)$ && $-0.0025(22)(14)$ & \\
\end{tabular}
\caption{
 Same as Table~\ref{tab:24idI0matrixelement},
 but for the isospin 2 channel.
 \label{tab:24idI2matrixelement}
}
\end{table*}

\begin{table*}[!hbt]
\begin{tabular}{c|rrrrrrrrrrr}
 && \multicolumn{9}{c}{ $\langle \pi^0 | {\cal O}_{\pi\pi} | \pi^0 \rangle$(stat)(syst) } \\
 \hline
 $\mathbf{p}_{\pi}$ & $\mathbf{p}_{\text{rel}}$
& \multicolumn{2}{c}{(0,0,0)}
& \multicolumn{2}{c}{(1,0,0)}
& \multicolumn{2}{c}{(1,1,0)}
& \multicolumn{2}{c}{(1,1,1)}
& \multicolumn{2}{c}{(2,0,0)}
 \\
 \hline
(0,0,0) && $1.2603(51)(14)$ && $-0.01221(09)(11)$ && $-0.008082(52)(03)$ && $-0.003408(38)(02)$ && $-0.001609(21)(02)$ & \\
(1,0,0) && $-0.0096(05)(28)$ && $0.0983(06)(14)$ && $-0.002313(26)(25)$ && $-0.000730(09)(08)$ && $-0.0002245(66)(43)$ & \\
(1,1,0) && $-0.0092(08)(12)$ && $-0.00283(09)(23)$ && $0.02545(17)(55)$ && $-0.000339(12)(03)$ && $-0.0000724(53)(54)$ & \\
(1,1,1) && $-0.0065(51)(24)$ && $-0.00220(22)(04)$ && $-0.000923(67)(20)$ && $0.01326(12)(54)$ && $-0.000068(16)(19)$ & \\
\end{tabular}
\caption{
 Same as Table~\ref{tab:24idI2matrixelement},
 but for the 48I ensemble.
 \label{tab:48iI2matrixelement}
}
\end{table*}

\subsubsection{Vacuum Matrix Elements}

For the isospin 0 channel,
 the vacuum matrix element terms that appear
 in Eq.~(\ref{eq:2piexpectvalueI0}) must be accounted for
 and subtracted.
Since these terms couple only to the vacuum,
 they have no time dependence and appear
 as a constant term.
These terms can be exactly cancelled from combinations
 like a simple nearest-neighbor subtraction,
 Eq.~(\ref{eq:timeseriessubtraction}),
 but at the cost of additional statistical noise.
Instead, this analysis uses an explicit computation of the
 $1\pi$ one-, two-, and three-point correlation
 functions to reconstruct and subtract the vacuum matrix
 elements, as outlined in Eqs.~(\ref{eq:vac})--(\ref{eq:iso0subtraction}).

Care must be taken to account for possible thermal terms
 in the vacuum contribution.
The one-point correlation function in Eq.~(\ref{eq:1pt})
 in a finite temporal volume may be expanded in terms of
 a complete set of eigenstates, which yields
 a small, exponentially suppressed thermal
 contribution to the vacuum matrix element of interest.
Assuming the one-point function amplitude is dominated by
 the vacuum matrix element,
 the size of the thermal contribution may be estimated
 from the transition matrix elements
 computed in the ratio of Eq.~(\ref{eq:c3ptratio})
 by comparing the uncertainty to the total ratio
\begin{align}
 \delta \langle {\cal O}_{X} \rangle
 \sim
 3\sum_{p}
 \langle \pi^0 | {\cal O}_{X} | \pi^0 \rangle e^{-E_\pi T}.
 \label{eq:vacuumthermal}
\end{align}
In this equation,
 the factor of 3 is an isospin factor that
 is described in the discussion around Eq.~(\ref{eq:wignereckart0}).

The one-point correlation function data and the corresponding
 thermal contamination for the 0-momentum ingoing and outgoing $1\pi$ states
 are listed in Table~\ref{tab:onepointvalues}.
For most of the thermal terms, the exponent mass $E_n$
 is large enough that the exponential completely
 suppresses the contribution.
For the 0-momentum pion, however, the term is still
 comparable to the uncertainty on the one-point
 correlation function.
While the thermal correction to the vacuum matrix element is completely
 negligible for the 24ID ensemble,
 the thermal contamination can be as large as the statistical uncertainty
 for the 48I ensemble.
After performing correlated subtraction of the vacuum terms
 from the $\pi\pi$ two-point functions, the statistical
 uncertainty is reduced and the effect
 of the additional thermal contribution becomes
 significantly more problematic.
For the correlation function with
 $\mathbf{p}_{\pi} = \mathbf{p}_{\text{rel}} = 0$,
 subtraction of this additional thermal term results in 
 a correction of about 3.2\% on the correlation function at $t/a=10$,
 while the uncertainty is at the level of 0.8\%.

\begin{table*}
\begin{tabular}{c|rrrrrrrrrrrrr}
 &&& \multicolumn{9}{c}{ $\langle {\cal O}_{X} \rangle$ (stat) } \\
 \hline
 Ens & $\mathbf{p}_{\text{rel}}$
& \multicolumn{2}{c}{(0,0,0)}
& \multicolumn{2}{c}{(1,0,0)}
& \multicolumn{2}{c}{(1,1,0)}
& \multicolumn{2}{c}{(1,1,1)}
& \multicolumn{2}{c}{(2,0,0)}
& \multicolumn{2}{c}{scalar}
 \\
 \hline
24ID && $11.358(58)$     && $6.644(29)$     && $4.162(20)$     && $1.6475(94)$     && $0.7033(52)$     && $-1.3806(49)$     & \\
48I && $1.2775(48)$     && $0.6527(28)$     && $0.3910(19)$     && $0.16625(92)$     && $0.08464(52)$     && $-0.34212(89)$     & \\
\hline \\[-1em]
\hline
 &&& \multicolumn{9}{c}{ $10^3\times 3\langle \pi^0|{\cal O}_{X}|\pi^0 \rangle e^{-E_\pi T}$ (stat)(syst) } \\
 \hline
 Ens & $\mathbf{p}_{\text{rel}}$
& \multicolumn{2}{c}{(0,0,0)}
& \multicolumn{2}{c}{(1,0,0)}
& \multicolumn{2}{c}{(1,1,0)}
& \multicolumn{2}{c}{(1,1,1)}
& \multicolumn{2}{c}{(2,0,0)}
& \multicolumn{2}{c}{scalar}
 \\
 \hline
24ID && $7.42(16)(09)$ && $2.527(54)(40)$ && $1.583(34)(24)$ && $0.624(14)(09)$ && $0.2648(60)(40)$ && $-0.519(11)(05)$ & \\
48I && $2.915(57)(18)$ && $0.873(16)(06)$ && $0.523(10)(04)$ && $0.2227(42)(15)$ && $0.1135(21)(07)$ && $-0.4561(87)(26)$ & \\
\end{tabular}
\caption{
 Values of one-point correlation functions (upper)
  averaged over all timeslices and their
  corresponding thermal corrections,
  scaled by a factor of $10^3$
  (lower) on both ensembles.
 The factor of 3 on the matrix element is a isospin counting factor
  that is described in the text.
 These values are strongly correlated timeslice-by-timeslice
  with the $\pi\pi$ two-point correlation functions,
  so the statistical precision of the subtracted two-point correlation
  function can be smaller than the statistical uncertainty here.
 \label{tab:onepointvalues}
}
\end{table*}

The contribution of the thermal correction to the vacuum matrix element
 is removed using the one-pion transition matrix elements from Sec.~\ref{sec:thermal},
 estimated with the ratio in Eq.~(\ref{eq:c3ptratio}).
Both of the corrections in
 Eqs.~(\ref{eq:c3ptreplacementden})~and~(\ref{eq:c3ptreplacementnum})
 are applied to the transition matrix elements.
The corrected vacuum matrix element is then
 given by Eq.~(\ref{eq:vac}), where the $\mathbf{p}_\pi = 0$
 pion thermal contribution is subtracted away.

Though the subtraction schemes for the vacuum matrix elements
 and the one-pion transition matrix elements both depend on each other,
 any thermal correction to the vacuum matrix element
 in Eq.~(\ref{eq:c3ptreplacementnum}) is neglected.
This correction on the correction would be proportional to
 $e^{-E_\pi (2T-t)}$, which would make it doubly-suppressed
 in the temporal extent of the lattice.
An improperly estimated vacuum matrix element would express
 itself as a residual time dependence of the ratio
 that grows with $t/a$, which could be seen in,
 for example, Fig.~\ref{fig:atwsub1}.
No evidence of these neglected contributions is visible.

\subsection{Two-pion Spectrum Results}

The spectrum of states for the $\pi\pi$ two-point correlation functions
 is determined from the GEVP using a basis of operators constructed using the
 distillation smearing.
For the isospin 0 and isospin 2 channels in the rest frame,
 $\pi\pi$ operators may be constructed with back-to-back momenta of magnitude
 $|\mathbf{p}_{\text{rel}}{L}/{2\pi}|^2\in \{0,1,2,3,4\}$.
In addition, the isospin 0 channel includes a distillation-smeared scalar current bilinear operator.

Before the spectrum is computed, subtraction schemes are applied to the correlation
 function to remove nuisance terms.
These subtractions are described in detail in Sec.~\ref{sec:methods}.
There are the two terms that are the target of the subtractions;
 first are the thermal terms discussed in Sec.~\ref{sec:thermal},
 and the second are the vacuum terms discussed in Sec.~\ref{sec:vacuum}.
Details about the computation of the matrix elements
 needed to construct the subtraction terms
 are given in Sec.~\ref{sec:subtractions}.

\begin{figure*}[htb!]
 \centering
 \begin{tabular}{cc}
 \includegraphics[width=\figsize\textwidth]{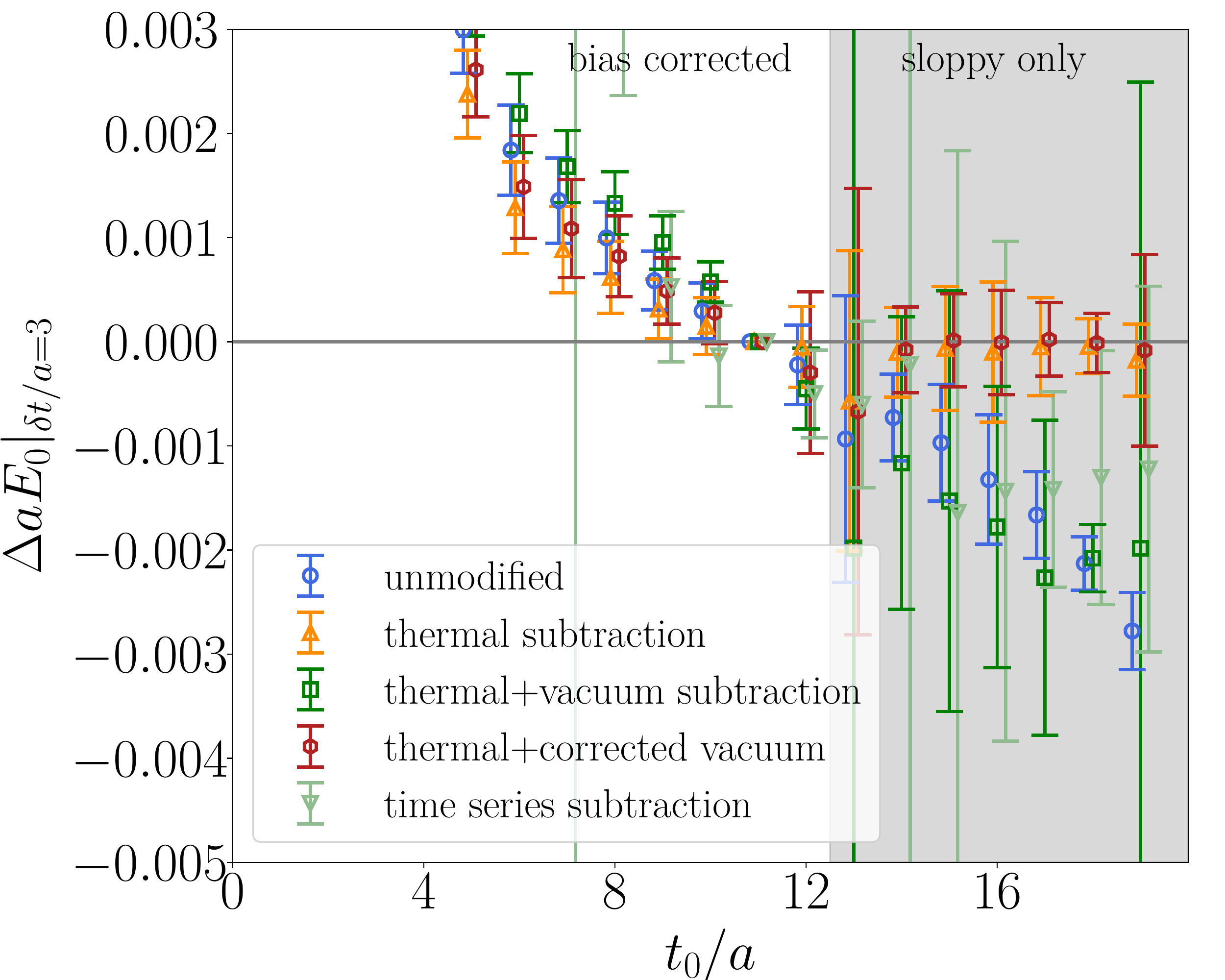} &
 \includegraphics[width=\figsize\textwidth]{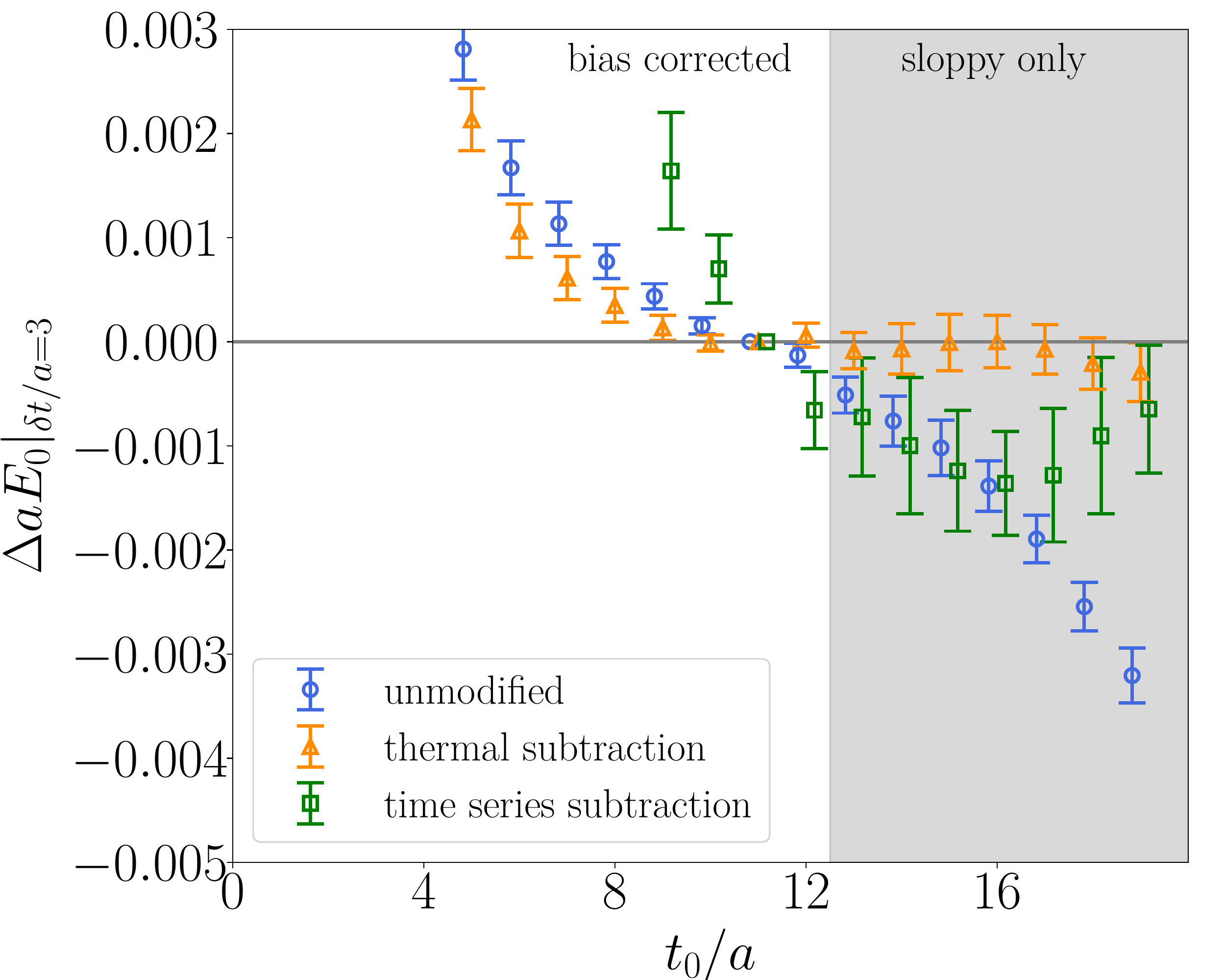} \\
 \end{tabular}
 \caption{
  Comparison of different subtraction schemes.
  Both plots are on the 48I ensemble;
   the left plot shows the first $\pi\pi$ state obtained from the GEVP in the isospin 0 channel,
   and the right plot shows the ground state for the I=2.
  The horizontal axis is the correlator time $t_0$,
   and both plots are shown with fixed $\delta t=3$.
  The vertical axis is the energy difference in lattice units
   implied from the GEVP eigenvalue as described
   in the text around Eq.~(\ref{eq:effectiveenergydiff}).
  The bias-corrected data are only available up to $t_0/a=12$
   on these plots, but sloppy only data are plotted in the gray
   shaded region to demonstrate convergence to a plateau.
  In the isospin 0 channel, the vacuum state is seen in the unmodified
   and thermal subtraction-only schemes, so the ``first $\pi\pi$'' state corresponds
   to the first excited state energy instead of the ground state.
  \label{fig:subtractionscheme}
 }
\end{figure*}

The effects of various subtraction schemes
 are plotted in Fig.~\ref{fig:subtractionscheme}.
To construct this plot,
 the GEVP eigenvalues are converted to energies by inverting
 the formula in Eq.~(\ref{eq:gevpeigenvalueexact}) assuming $m=n$,
\begin{equation}
 aE_n(t_0,\delta t) = -\frac{1}{\delta t/a} \log \Lambda_{nn} (t_0,t_0+\delta t).
 \label{eq:effectiveenergy}
\end{equation}
The effective energies from Eq.~(\ref{eq:effectiveenergy}) are computed
 and the difference with a fixed timeslice,
\begin{align}
 \Delta a E_n (t_0, \delta t) = a E_n (t_0, \delta t) - a E_n (\bar{t}_0, \delta t)
 \label{eq:effectiveenergydiff}
\end{align}
 with $\bar{t}_0$ and $\delta t$ held fixed.
In Fig.~\ref{fig:subtractionscheme},
 $\bar{t}_0/a=11$ for both choices of isospin.
This energy difference is applied to reduce the difference
 between the sloppy-only and bias-corrected results
 so stability over an extended time range can be demonstrated.
Because of this subtraction, the plateaux for various subtraction schemes
 are not required to be in agreement with each other.
The isospin 0 channel is plotted on the left side of Fig.~\ref{fig:subtractionscheme},
 and the isospin 2 channel is plotted on the right side.

In Fig.~\ref{fig:subtractionscheme},
 the ``unmodified'' correlator is the spectrum obtained from the
 raw data for the $\pi\pi$ two-point correlation functions.
``Thermal subtraction'' corresponds to the subtraction of $1\pi$ state
  contributions that propagate through the temporal time extent,
  written in Eq.~(\ref{eq:thermalsubtraction}).
``Thermal+vacuum subtraction'' additionally removes the terms from
  the isospin 0 channel that have no propagating states, written in
  Eqs.~(\ref{eq:vsub})~and~($\ref{eq:iso0subtraction})$
  and neglecting the thermal correction to the vacuum term
  by assuming $C_{\text{vac}}=\langle {\cal O}_{X} \rangle$.
``Thermal+corrected vacuum'' applies all of the corrections
  in Eqs.~(\ref{eq:vac})--(\ref{eq:iso0subtraction}) as written.

The ``time series subtraction'' method is the commonly employed
 strategy for removing contaminants.
This strategy applies the modification
 from Eq.~(\ref{eq:timeseriessubtraction})
 to the matrix of correlation functions before they are
 substituted into the GEVP.
This strategy exactly removes terms that are constant with time,
 which for correlation functions with 0 total center-of-mass momentum
 includes all of the thermal and vacuum corrections that appear in both
 the isospin 0 and 2 channels.
The difference is a discrete derivative with respect to Euclidean time,
 which pulls out an energy-dependent prefactor for the contributions of all of the states.
As a result, the effects of excited state contamination at larger Euclidean times
 are expected to be enhanced from the time series subtraction scheme.

In the isospin 2 channel,
 the ``unmodified'' data is expected to have a contamination
 in the ground state from the thermal terms.
This correction is a small constant term in the
 correlation function that prevents the ground state effective
 energy from plateauing, which is clearly seen at late times.
The ``thermal subtraction'' removes the thermal terms by
 applying the subtraction of Eq.~(\ref{eq:thermalsubtraction}),
 which removes the visible contamination.
The ``time series subtraction,'' Eq.~(\ref{eq:timeseriessubtraction}) with $dt=3$,
 is also provided as a reference point.
The thermal subtraction is used as the nominal choice for the isospin 2 channel,
 and will be referred to as the ``matrix element subtraction'' method for isospin 2
 in the remainder of the manuscript.

In the isospin 0 channel,
 the same patterns are seen as in the isospin 2 channel.
The ``unmodified'' data still exhibit a contamination from thermal effects,
 which produces effective energies that never reach a plateau at large times.
This contaminant is removed by applying the ``thermal subtraction'' of
 Eq.~(\ref{eq:thermalsubtraction}).
The thermal subtraction still retains an effective energy consistent with 0,
 which absorbs all of the contamination from the vacuum term,
 so no slow falloff of the effective energy is observed.
The ``thermal+vacuum subtraction'' removes most of the vacuum contribution,
 but reintroduces thermal effects that again prevent a plateau at large time,
 similar to the ``unmodified'' data.
Correction of the vacuum term, shown as ``thermal+corrected vacuum,''
 removes the new thermal effects and restores the plateau behavior
 that is seen in the ``thermal subtraction'' but without the additional vacuum state
 in the GEVP.
The ``time series subtraction'' method is again shown for reference with the other subtraction schemes,
 with $dt=3$.
The ``thermal+corrected vacuum'' is used as the nominal choice,
 for the isospin 0 channel in the remainder of the manuscript,
 referred to as the ``matrix element subtraction'' for isospin 0 from this point on.

\begin{figure*}[htb!]
 \begin{tabular}{cc}
 \includegraphics[width=\figsize\textwidth]{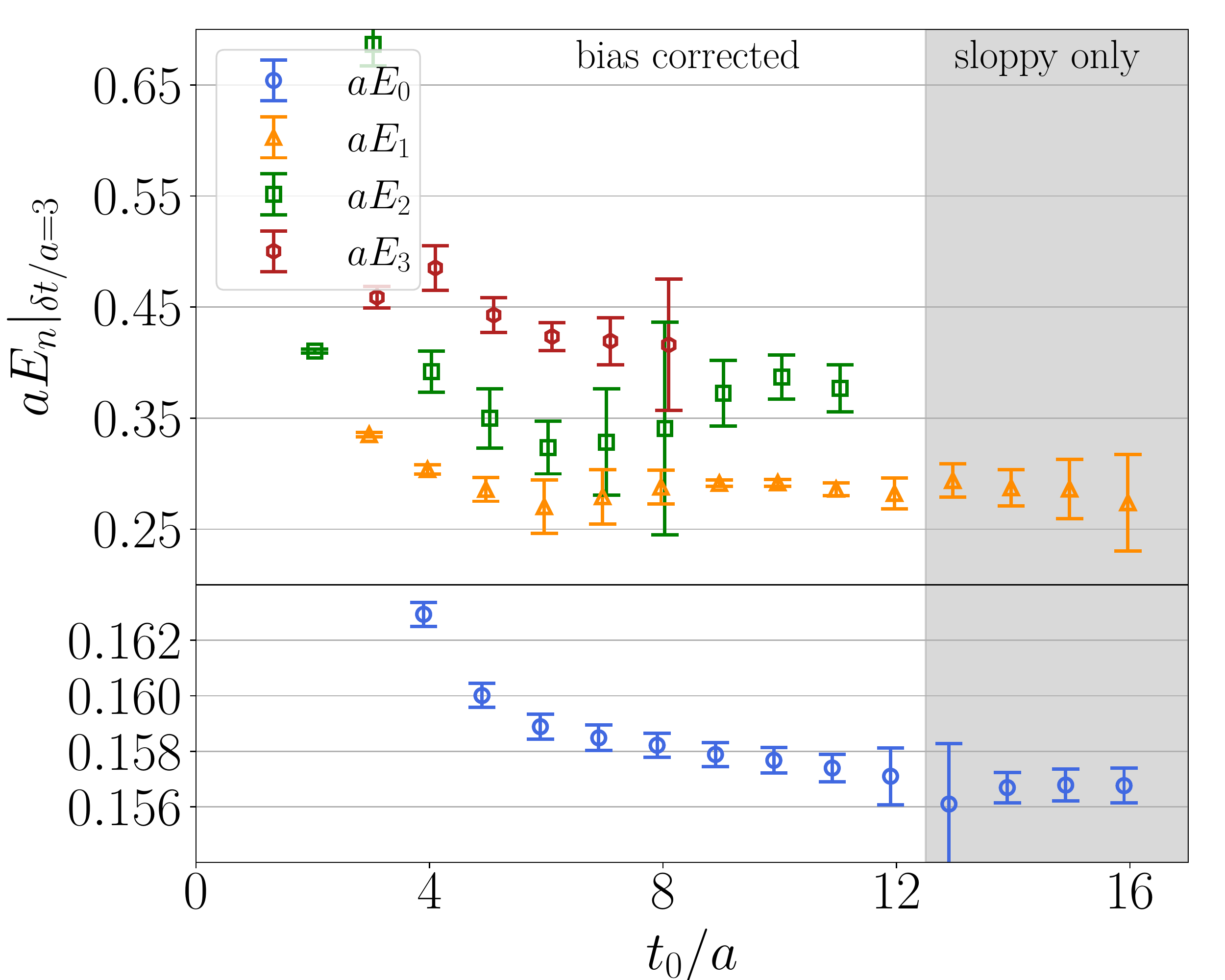} &
 \includegraphics[width=\figsize\textwidth]{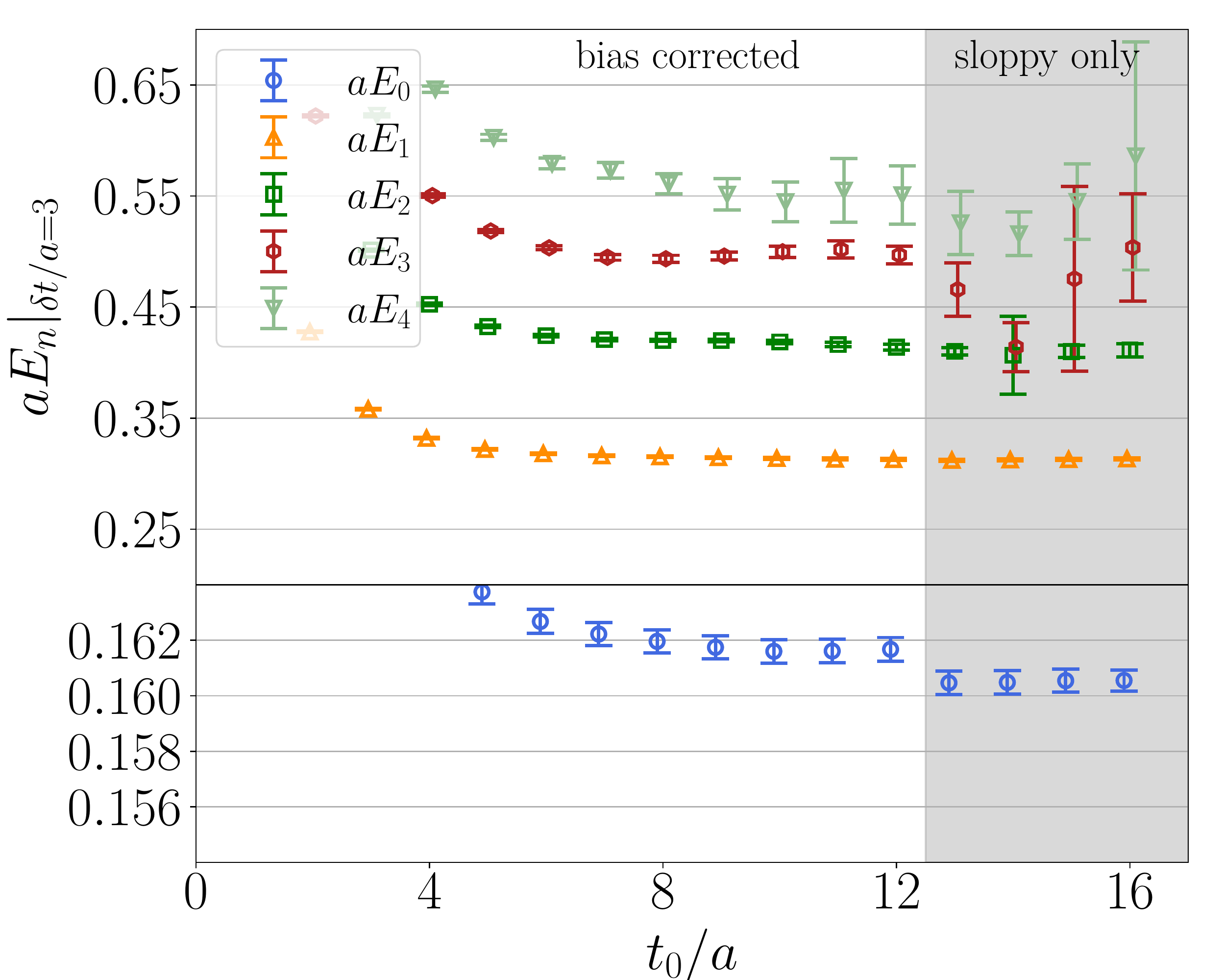} \\
 \includegraphics[width=\figsize\textwidth]{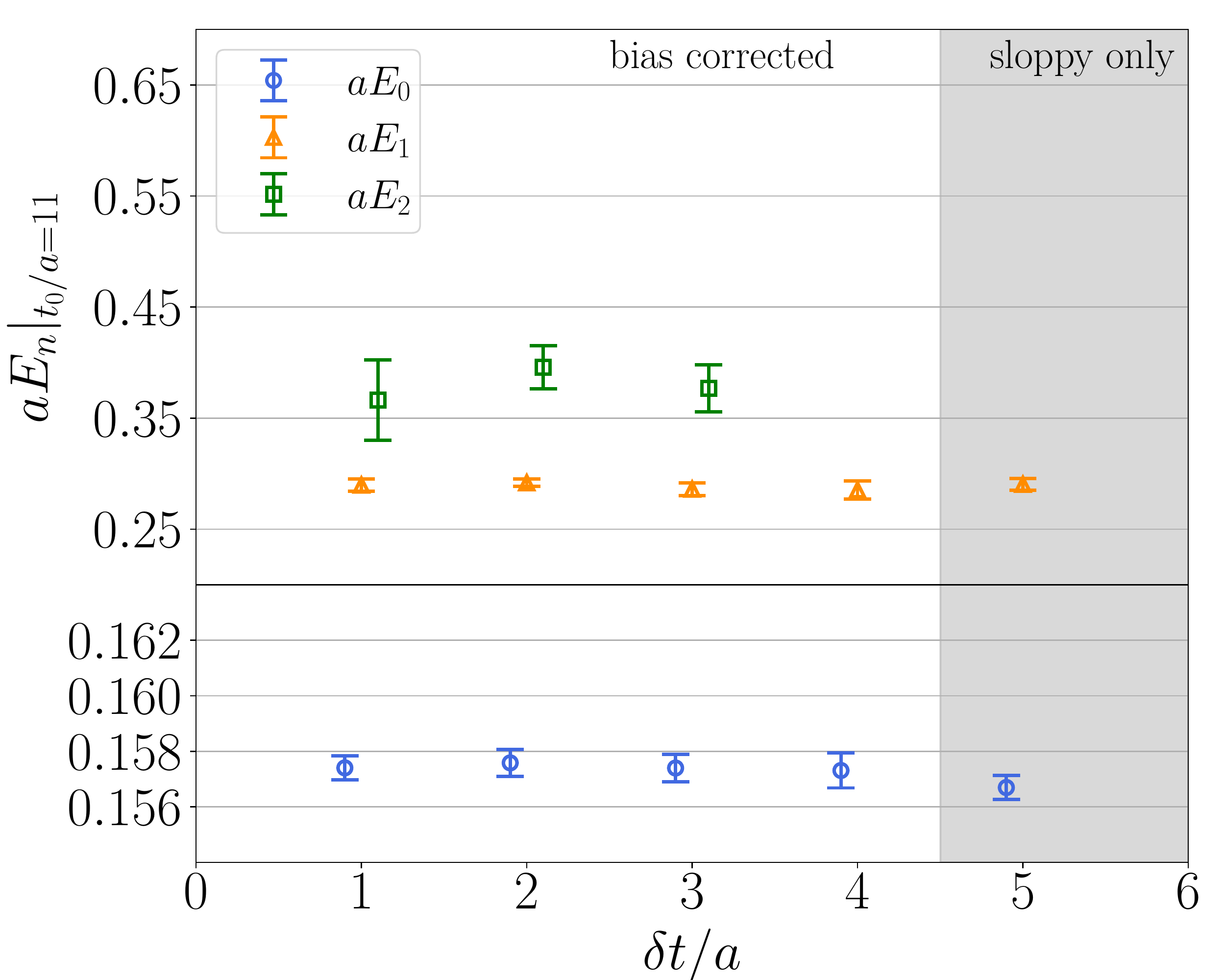} &
 \includegraphics[width=\figsize\textwidth]{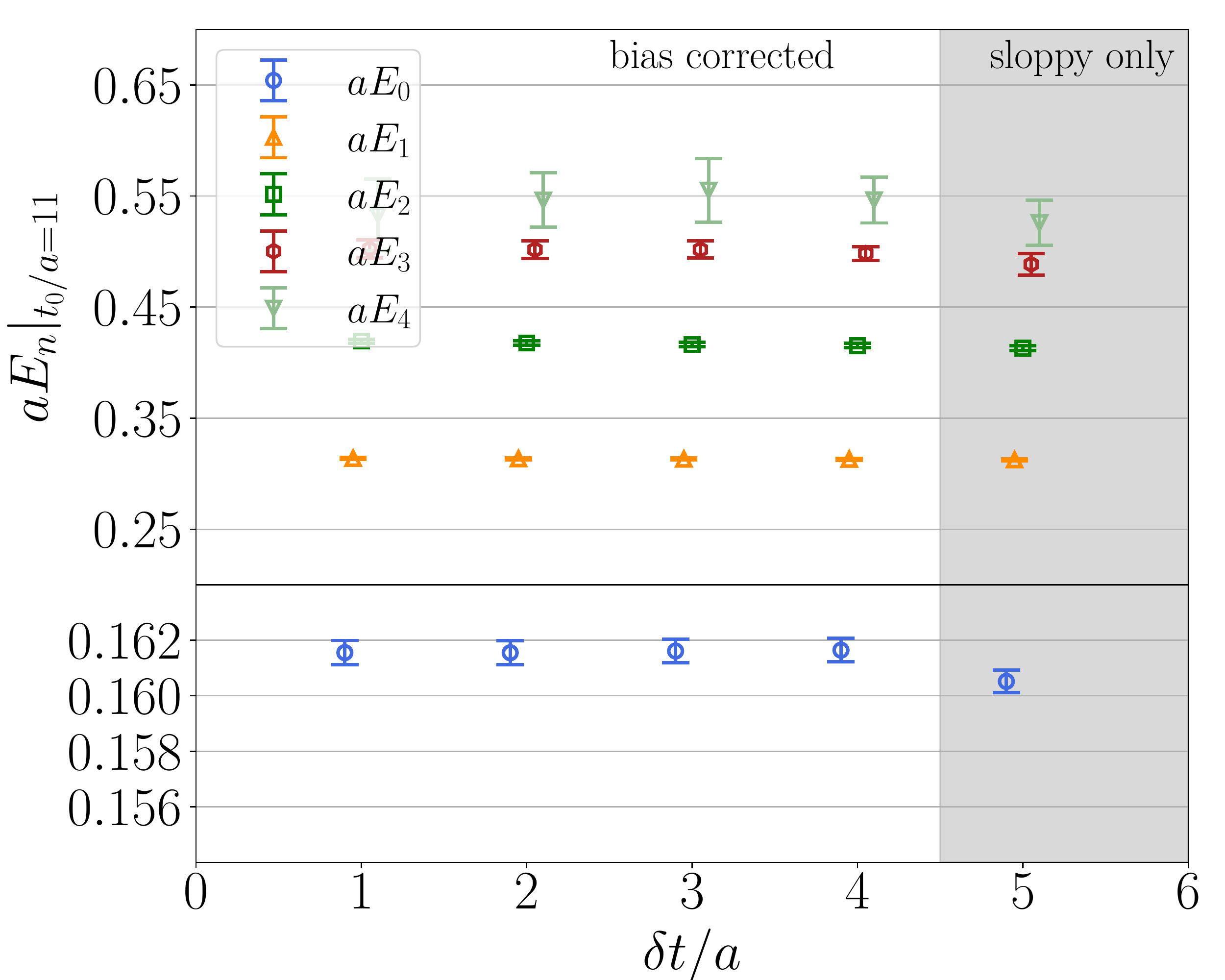}
 \\
 \end{tabular}
 \caption{
 Plots of the spectrum of states obtained from solving the GEVP on the 48I ensemble.
 The left (right) plot shows the results for the GEVP applied to the isospin 0 (isospin 2) channel.
 The top (bottom) panels show the state effective energies as a function of
 $t_0/a$ ($\delta t/a$) with fixed $\delta t/a$ ($t_0/a$).
 The vertical axis is the effective energy of the states implied by the GEVP.
 For each plot, multiple state progressions are shown with separate symbols and colors.
 Each panel has an inset to zoom in on the lowest state's effective energy.
 The unshaded region has the full AMA procedure applied to correct for bias,
 while the shaded region is only obtained from low-precision solves and
 does not correct for this bias.
 The bias can be significant on the level of the statistical precision,
 as is seen in the lowest effective energies (blue circles) in the upper-right panel.
 \label{fig:gevpspectra}
 }
\end{figure*}

Plots of the spectrum obtained from the GEVP as a function of $t_0$
 for both the matrix element subtraction and the time series subtraction schemes
 are shown in Fig.~\ref{fig:gevpspectra}.
The plotted data are the energy levels inferred from the eigenvalues
 after solving the GEVP equation~(\ref{eq:effectiveenergy}).
For each jackknife sample, the eigenvalues are sorted so that they
 optimize Eq.~(\ref{eq:eigsorting}) with $t'=t-1$ for $t>3$.
In the top row of plots,
 $\delta t$ is held fixed and $t_0$ is varied along the horizontal axis.
The bottom row shows instead with $t_0$ fixed and $\delta t$ varied.
For the isospin 2 channel, the thermal contributions have been subtracted away.
For the isospin 0 channel, both the (corrected) vacuum and thermal contributions
 have been subtracted.

For the isospin 2 channel, the states are well-separated over the entire
 range of $t_0$ and $\delta t$ studied for the 48I.
The 24ID ensemble exhibits a similar pattern for
 all states in the isospin 2 channel
 and up to and including the second excited state in the isospin 0 channel,
 although the precision for the highest excited states
 starts to degrade for times $t_0\gtrsim 8$.
 
For the isospin 0 channel on the 48I ensemble,
 a spurious state appears in the spectrum at small Euclidean times
 between the first and second excited states.
This spurious state is particularly noticeable in
 Fig.~\ref{fig:gevpspectra} for the state labeled $aE_2$ (green squares)
 in the range $t_0/a\in\{6,7,8\}$.
At $t_0/a = 8$, a partial level crossing takes place that causes
 a degeneracy betweeen $aE_1$ and $aE_2$,
 spoiling the identification of the state and producing overlapping error bands.
For $t_0/a\geq 9$,
 the spurious state effective energy becomes undefined
 and the second excited state takes its place as $aE_2$.
Investigation of this spurious state reveals that the only way to remove this
 state is to remove the scalar operator from the operator basis,
 which also shifts of the other effective energies outside of their
 statistical uncertainty.
This spurious state is most likely a fit artifact from a
 slight over- or under-subtraction of one of the contaminants
 that has been amplified by the GEVP.
No such spurious state appears in the spectrum when the time series subtraction scheme is used,
 backing up this claim.
To avoid this issue, results are taken for sufficiently large times
 such that no such spurious state is present and both the time series subtraction
 corrected vacuum subtraction schemes produce consistent results.

The final spectrum results are obtained by increasing
 $\delta t = t-t_0$ and $t_0$ until
 the GEVP eigenvalues plateau within statistics.
This strategy has the advantage of being sufficiently
 simple without the need to introduce correlator fitting
 and the corresponding systematics.
The data approach these plateau values exponentially,
 where the exponent is proportional to the energy gap between
 the largest state that is constrained by the GEVP and
 the next most significant state that is not sufficiently
 constrained by the basis of correlators.
The operator basis used in this study is composed of
 the quark bilinear operators and $\pi\pi$ operators
 with various back-to-back momenta,
 so states with similar particle content
 (namely the finite volume $\sigma$ state and
 the $\pi\pi$ states)
 are expected to be well-constrained by the basis choice.
The remaining poorly-constrained states are then expected
 to be made up of states that are not well represented
 by the operator basis,
 most likely consisting of states that involve
 two-particle scattering with an excited pion ($\pi\pi^\ast$).
Better overlap with these states could be achieved
 with an operator basis including derivatives
 within the quark bilinears
 that make up the two-particle interpolating operators~\cite{Dudek:2009qf,Dudek:2012gj}.

The GEVP parameters for each isospin channel and ensemble
 are given in Table~\ref{tab:2pigevpspec-fitparams}.
These values will have residual contamination from excited states that depends
 on $t_0$ and $\delta t$, which are estimated in two systematic uncertainties obtained
 by shifting either $t_0 \to t_0-1$ or $\delta t \to \delta t-1$
 and taking the difference of central values between this result and the nominal choice.

\begin{table}[t!]
\begin{tabular}{c|ccc|ccc||ccc|ccc}
 \multicolumn{1}{c}{}
 & \multicolumn{6}{|c||}{matrix element subtraction}
 & \multicolumn{6}{c}{time series subtraction} \\
 \multicolumn{1}{c}{}
 & \multicolumn{3}{|c}{24ID} & \multicolumn{3}{c||}{48I}
 & \multicolumn{3}{c}{24ID} & \multicolumn{3}{c}{48I} \\
Isospin
& $\delta t$ & $t_0$ & $N_{\text{op}}$ & $\delta t$ & $t_0$ & $N_{\text{op}}$ 
& $\delta t$ & $t_0$ & $N_{\text{op}}$ & $\delta t$ & $t_0$ & $N_{\text{op}}$ \\
\hline\hline
2 & 3 & 7 & 5 & 3 & 11 & 5
  & 3 & 6 & 5 & 3 &  8 & 5\\
0 & 3 & 7 & 6 & 3 & 11 & 6
  & 3 & 6 & 6 & 3 &  8 & 6
\end{tabular}
\caption{
 Parameter choices used in the GEVP analysis for this study.
 $t$ and $t_0$ are defined in Eq.~(\ref{eq:gevp}), with $\delta t = t-t_0$.
 $N_{\text{op}}$ is the size of the operator basis,
  where the set of operators used is described in the text.
 \label{tab:2pigevpspec-fitparams}
}
\end{table}

The noninteracting $\pi\pi$ correlators in this study were used in both the
 isospin 0 and isospin 2 channels.
The isospin 2 channel has no resonance states to spoil the
 identification of dispersion states.
We did not investigate correlating the interacting $\pi\pi$ spectrum
 with linear combinations of the energy levels in the noninteracting spectrum.
This strategy could be useful for the isospin 0,
 where a weak resonant state should be present,
 and might be able to amend this method for future use in the isospin 1 channel
 where there is a strong overlap with the $\rho$ resonance state.
In the isospin 0 channel, the improvement in statistics from applying this method is negligible
 and the statistical precision is not sufficient to identify any resonant behavior.
However, correlating the interacting correlators with their noninteracting counterparts
 is still advantageous because it reduces systematics from lattice cutoff effects.

The spectrum obtained by solving the GEVP for all of the isospin channels
 as well as the noninteracting values obtained from the process described
 in Sec.~\ref{sec:freepicorrelation} are given in
 Tables~\ref{tab:2pigevpspec-24id}~and~\ref{tab:2pigevpspec-48i}.
In addition to the raw values, the correlated energy differences between the
 interacting and noninteracting spectra are also provided.
Tables~\ref{tab:2pigevpspec-24id}~and~\ref{tab:2pigevpspec-48i} give
 the spectrum results for all isospin channels in the rest frame for the 24ID and 48I
 ensembles, respectively.

\begin{table*}[t!]
\begin{tabular}{c|ccc|cc}
    & \multicolumn{3}{c}{$aE_n$}  & \multicolumn{2}{c}{$a(E_n-E_n^{\text{NI}})$}   \\
\hline
$n$ & NI & I=0 & I=2 & I=0 & I=2 \\
\hline
0 & 0.28032(68)(01) & 0.2729(18)(03) & 0.28273(69)(06) & -0.0075(16)(03) & 0.002407(74)(57) \\
1 & 0.5917(12)(04) & 0.5520(94)(19) & 0.6061(15)(03) & -0.0398(91)(23) & 0.01440(55)(13) \\
2 & 0.7874(25)(14) & 0.764(55)(03) & 0.8141(42)(04) & -0.024(56)(02) & 0.0267(30)(11) \\
3 & 0.9434(55)(20) &  -- & 0.961(20)(09) &  -- & 0.018(17)(07) \\
4 & 1.063(15)(02) &  -- & 1.055(43)(24) &  -- & -0.008(43)(26) \\
\end{tabular}
\caption{
 A list of the $\pi\pi$ spectrum in each isospin channel for the 24ID ensemble.
 $n$ is an index identifying which state is listed.
 Columns under the $aE_n$ heading are the unmodified spectrum results in lattice units,
  and the columns under $a(E_n-E_n^{\text{NI}})$ are correlated differences between
  the interacting and noninteracting $\pi\pi$ energy levels.
 As a reference,
  the column labeled ``NI'' gives the noninteracting $\pi\pi$ energy levels obtained from
  a $1\times1$ GEVP with $\delta t/a=3$ and $t_0/a=4$.
 Note that the energy differences are computed using the same choice of
  $\delta t$ and $t_0$ for both the interacting and noninteracting energy levels,
  so the differences are not computed with the values in the ``NI'' column.
  All quantities include both statistical and systematic errors.
 \label{tab:2pigevpspec-24id}
}
\end{table*}

\begin{table*}[t!]
\begin{tabular}{c|ccc|cc}
    & \multicolumn{3}{c}{$aE_n$}  & \multicolumn{2}{c}{$a(E_n-E_n^{\text{NI}})$}   \\
\hline
$n$ & NI & I=0 & I=2 & I=0 & I=2 \\
\hline
0 & 0.16074(42)(03) & 0.15740(50)(33) & 0.16161(42)(06) & -0.00334(32)(32) & 0.000870(66)(62) \\
1 & 0.3075(08)(07) & 0.2860(58)(81) & 0.3131(09)(07) & -0.0215(59)(74) & 0.00562(20)(08) \\
2 & 0.4057(16)(07) & 0.377(21)(21) & 0.4163(19)(27) & -0.029(21)(21) & 0.0105(11)(21) \\
3 & 0.4854(40)(20) &  -- & 0.5018(77)(19) &  -- & 0.0163(59)(27) \\
4 & 0.5568(90)(26) &  -- & 0.555(29)(14) &  -- & -0.002(29)(16) \\
\end{tabular}
\caption{
 Same as Table~\ref{tab:2pigevpspec-24id}, but for the 48I ensemble.
 The ``NI'' column of data is taken at $\delta t/a=3$ and $t_0/a=10$.
 \label{tab:2pigevpspec-48i}
}
\end{table*}

\section{Scattering Length and Phase Shift Results}
 \label{sec:results}

The $\pi\pi$ scattering phase shifts are obtained by applying
 the L\"uscher formalism in Eqs.~(\ref{eq:2pimomentum})--(\ref{eq:Msecond})
 to the spectra that are obtained from the correlation function data.
The deviation of the measured spectra away from their noninteracting values
 encodes the strength of the interaction, and the sign indicates whether the interaction
 is attractive or repulsive.
There is no resonance present in the isospin 2 channel,
 so little deviation from the noninteracting spectrum is expected.
In the isospin 0 channel however, the presence of the $\sigma$ resonance
 will contribute to shifts away from the noninteracting spectrum,
 which yields a positive phase shift at energies above threshold.

The uncertainty due to discretization effects
 up to this point has not been considered because all comparisons
 have been performed on a single ensemble.
When comparing results to the continuum and to other ensembles,
 it is important to quantify this uncertainty.
Although this analysis uses results from two ensembles
 with different lattice spacings, these ensembles use different
 lattice actions and so cannot be combined together
 to come up with a direct estimate of discretization effects.
To circumvent this difficulty,
 we employ the same strategy as our sister paper~\cite{RBC:2021acc}
 of using the average of several other discretization error coefficients
\begin{align}
 |c^{X}_{\text{avg}}| = \frac{1}{4}
 \left(
  |c^{X}_{f}| 
 +|c^{X}_{f^{(K)}}| 
 +|c^{X}_{\sqrt{t_0},a}| 
 +|c^{X}_{w_0,a}| 
 \right)
\end{align}
 using the prefactor coefficients listed
 in Table~XVII of Ref.~\cite{Blum:2014tka}.
Here, the superscript $X$ is used to denote the action ID (I)
 for the 24ID (48I) ensemble.
From this quantity,
 a fractional discretization uncertainty is assigned for the energy
\begin{align}
 \frac{\delta E_{X}}{E_{X}} = |c_{\text{avg}}^{X}| a_{X}^2.
 \label{eq:discretizationuncertainty}
\end{align}
In the case that the energy difference $\Delta E_{\pi\pi}$ is used instead,
 as in Eq.~(\ref{eq:ppicomdifference}),
 the fractional discretization error is instead applied to 
 $\Delta E_{\pi\pi}$.
No discretization error is needed on $M_\pi$ since it is
 assumed to be free from discretization errors during the scale setting procedure.
For the 24ID (48I) ensemble,
 the fractional uncertainty for discretization effects
 $\delta E/E$ comes out to be about 2.5\% (1.1\%).

The phase shifts computed here assume that only
 the lowest partial wave allowed by symmetry contributes
 appreciably and all other higher partial waves are suppressed.
All of the data in this study transform under the $A_1$ octahedral irrep,
 for which the $\ell=4$ partial wave is the next-lowest partial wave
 with nonvanishing coupling.
Coupling of the higher partial wave is expected to
 be significantly suppressed compared to the lowest $\ell=0$ partial wave
 due to a coupling dependent on the momentum or angular momentum.
Under this assumption, no model of the mixing between partial waves as a
 function of energy is needed to compute the phase shifts
 and the values obtained from the lattice
 may be directly compared with phenomenological estimates.

Although the L\"uscher formalism works well for two-particle states,
 the formalism is expected to break down when thresholds for creation
 of more than two particles open up.
For the scattering channels in this study,
 the lowest of these thresholds is the four-pion threshold at
 $\sqrt{s} = 4M_\pi \approx 0.55$~GeV,
 which sits just below (above) the first excited state of the
 24ID (48I) ensemble in both isospin channels.
Although there is no strictly valid interpretation of the scattering phase shifts
 above this threshold, the data are still plotted so the breakdown
 of the data from more complicated channels may be qualitatively examined.
Three-pion states do not contribute to the correlation functions
 in this study because they are forbidden by $G$-parity symmetry.

\subsection{Scattering Phase Shifts}

\begin{figure*}[t!]
\begin{tabular}{cc}
 \includegraphics[width=\figsize\textwidth]{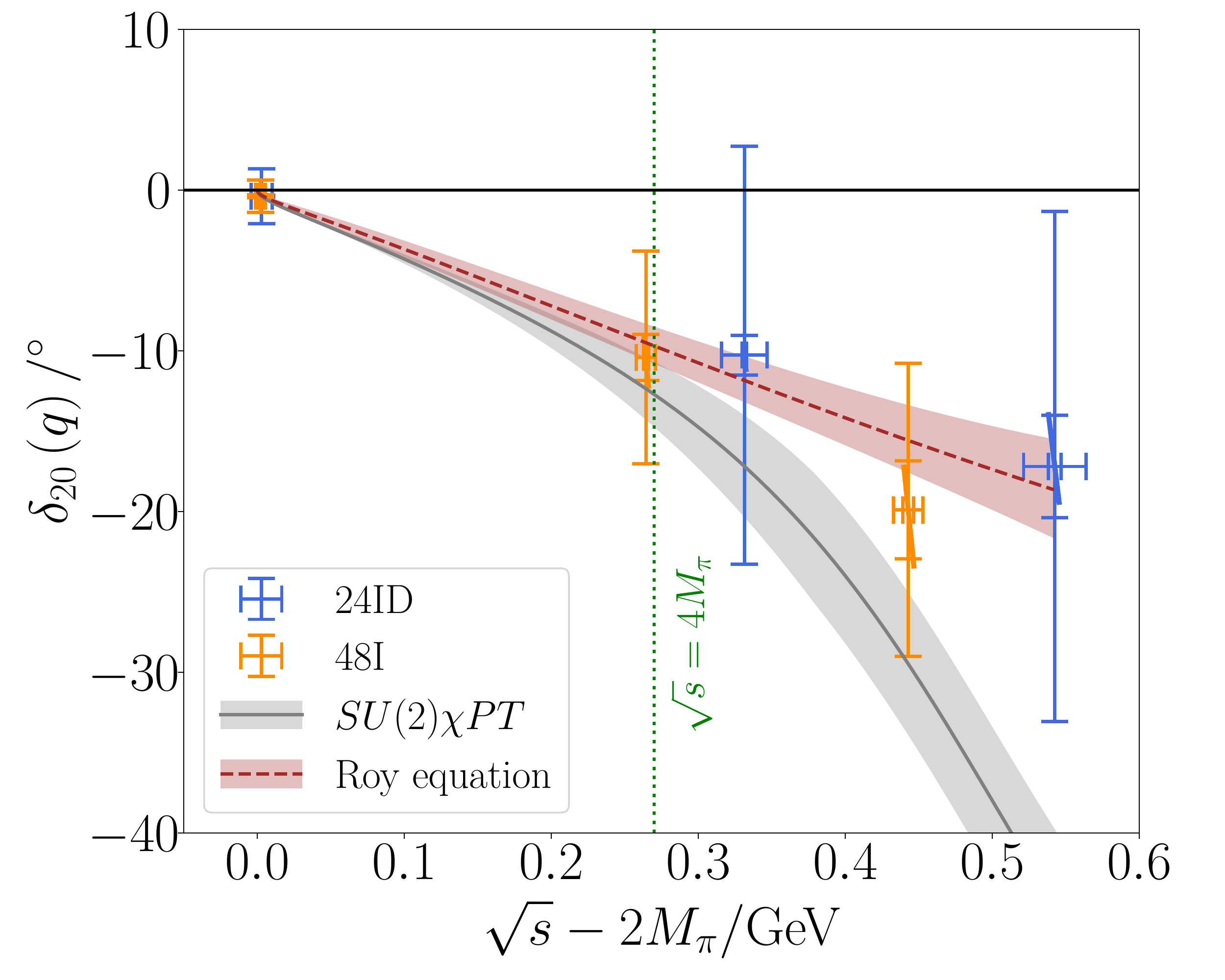} &
 \includegraphics[width=\figsize\textwidth]{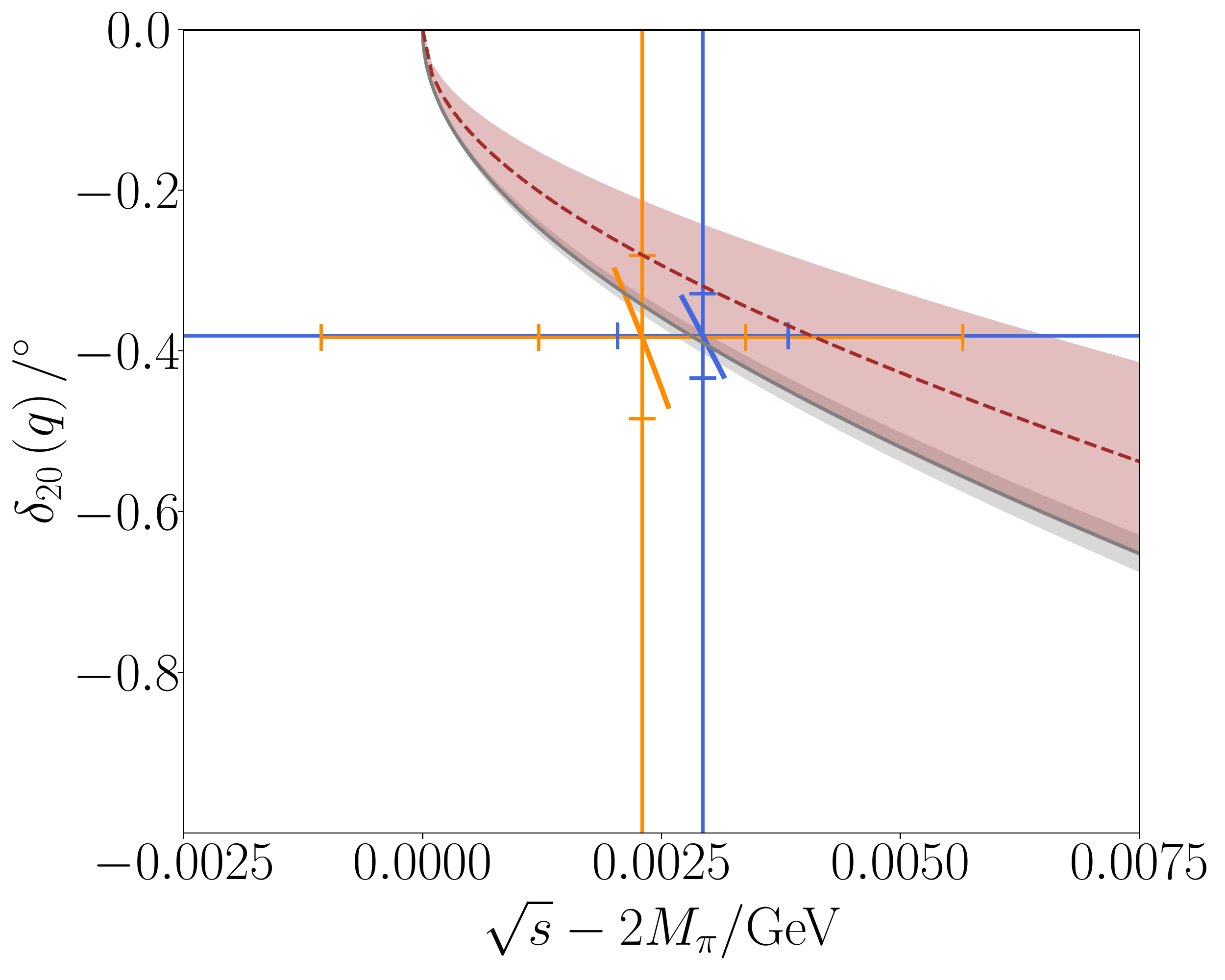} \\
 \includegraphics[width=\figsize\textwidth]{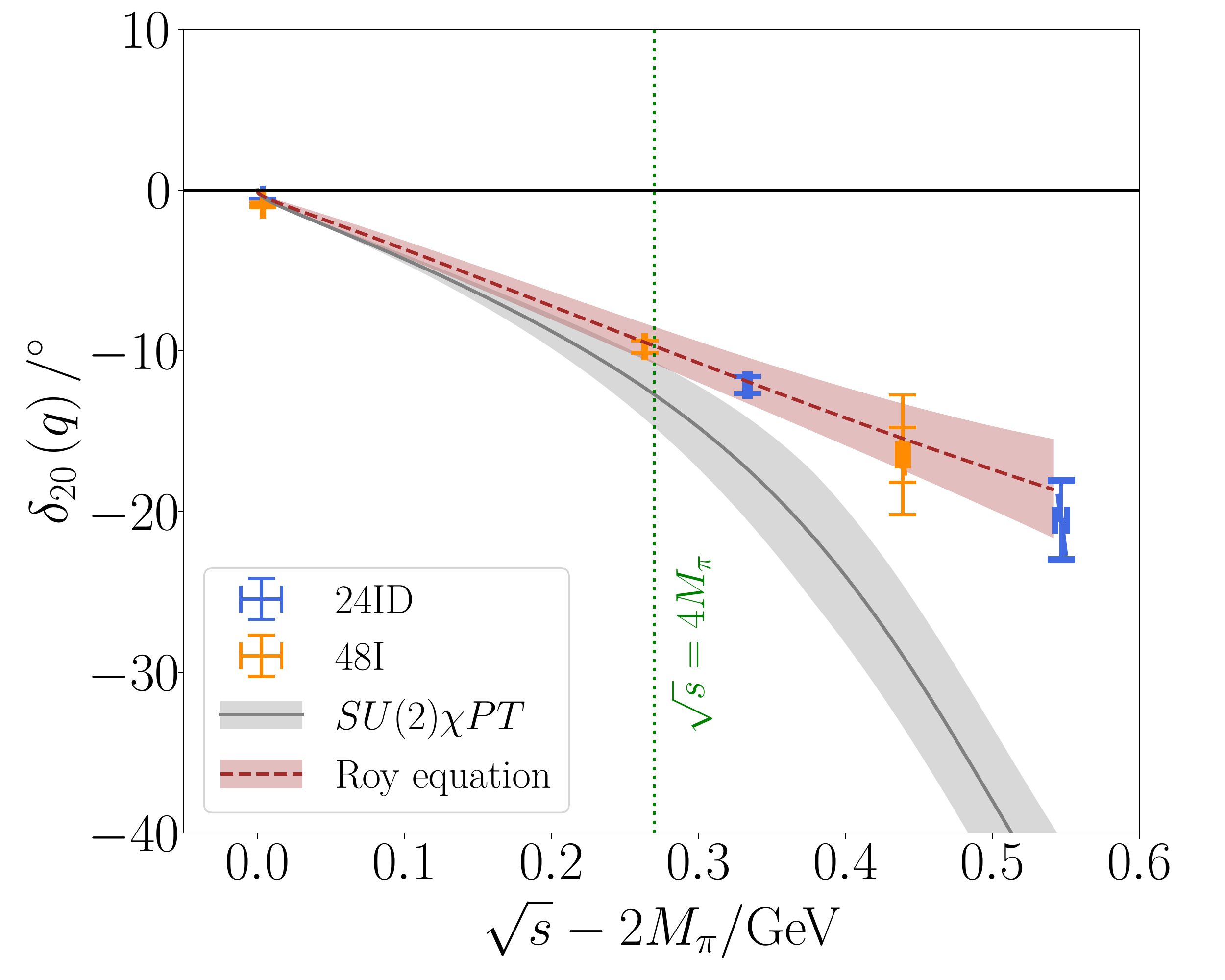} &
 \includegraphics[width=\figsize\textwidth]{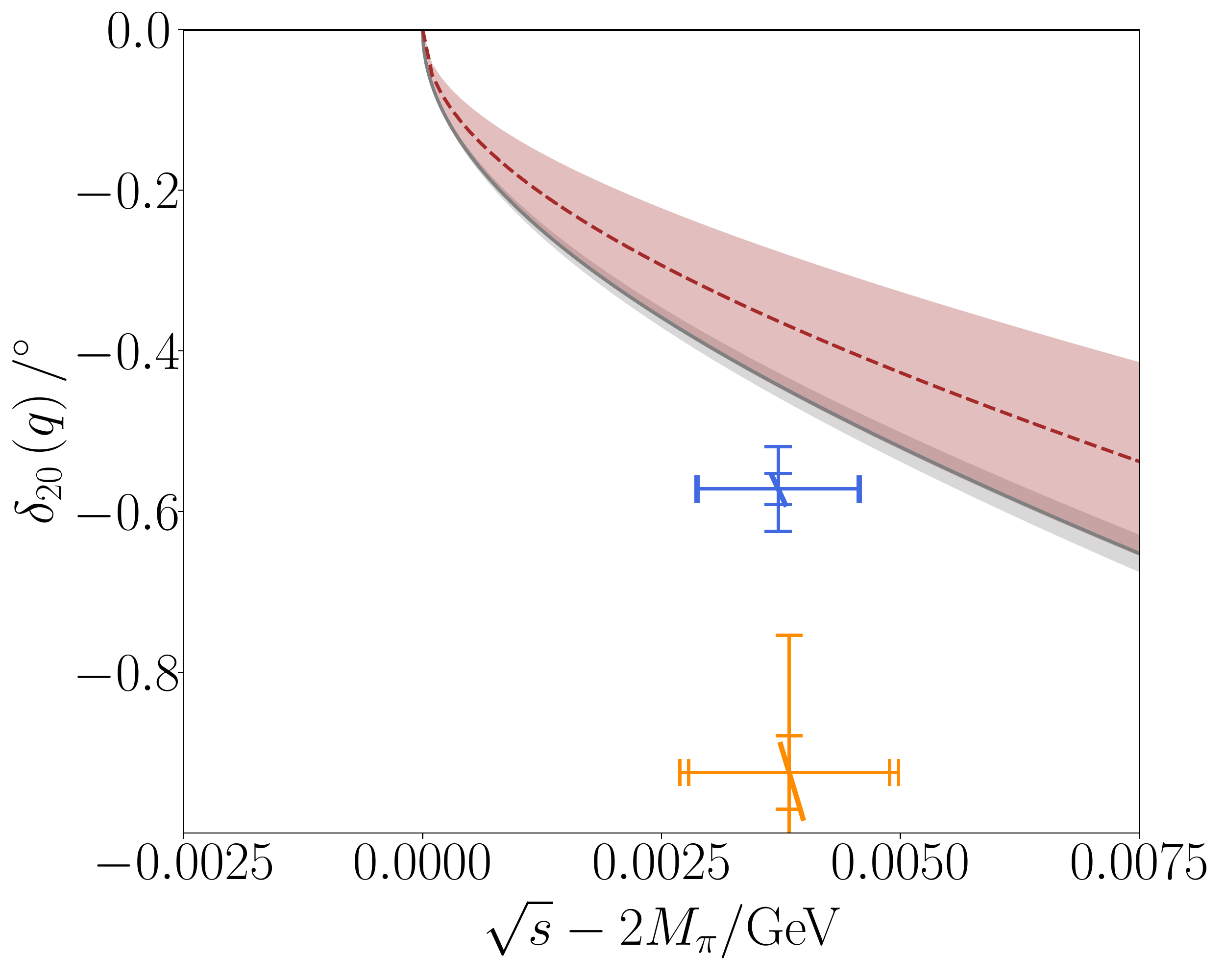} \\
 \includegraphics[width=\figsize\textwidth]{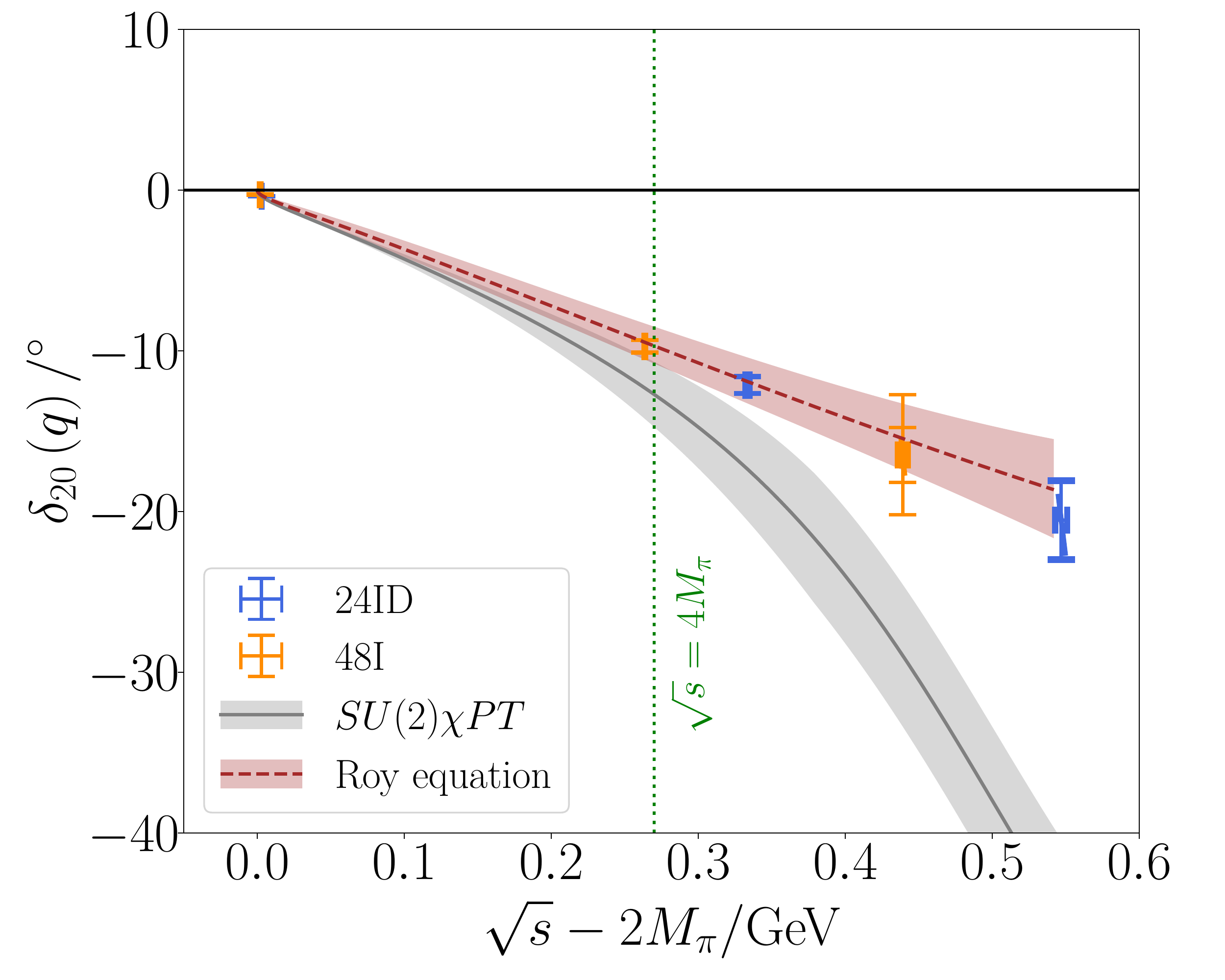} &
 \includegraphics[width=\figsize\textwidth]{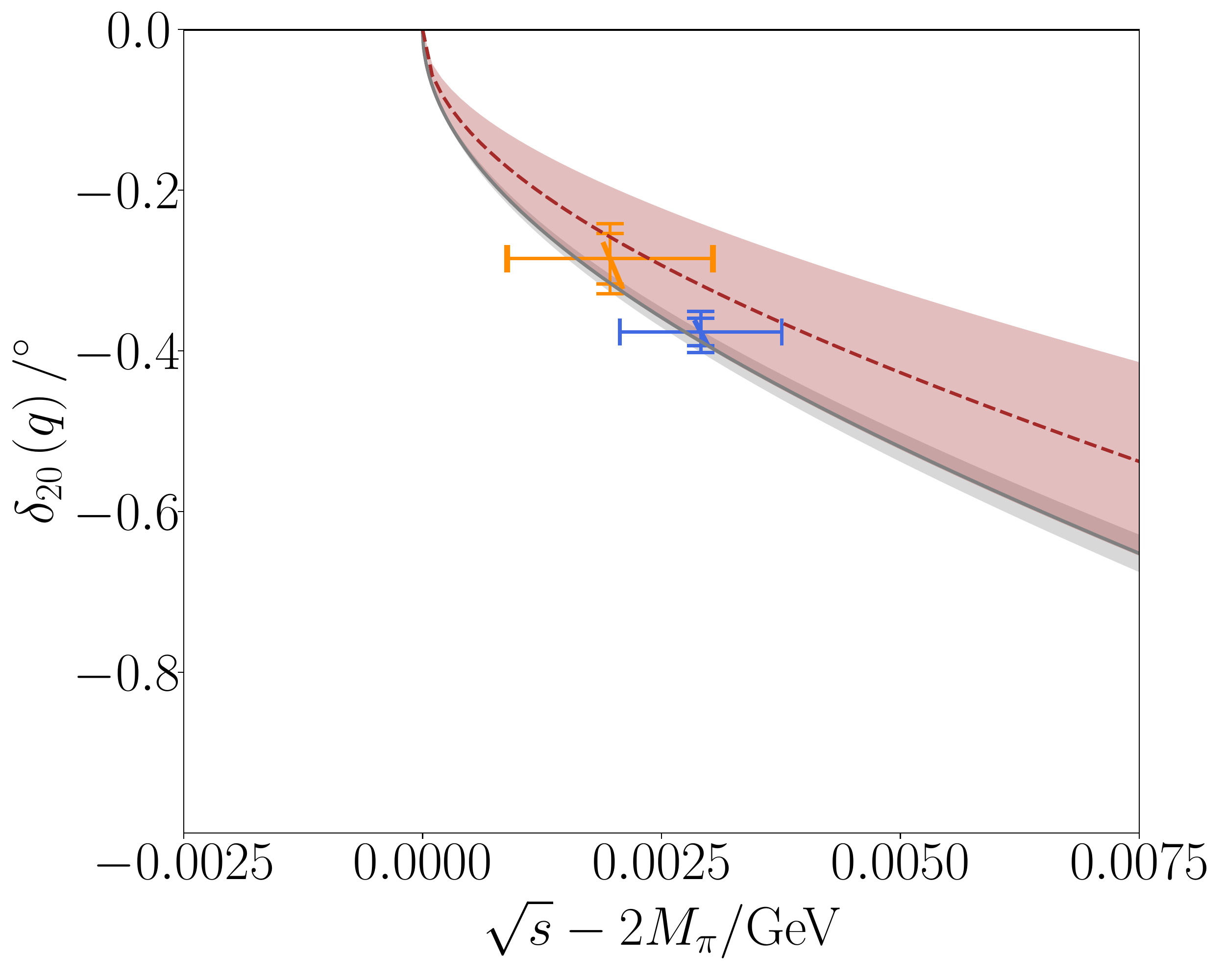} \\
\end{tabular}
\caption{
 Phase shift plots for the lattice data in the isospin 2 channel.
 The left column shows the phase shifts as a function of center of mass energy $\sqrt{s}$,
  and the right column is the same plot zoomed in to near the threshold value.
 The center of mass energy has been shifted by $2M_\pi$ to account for slight differences
  in the pion mass.
 The blue (orange) denote the 24ID (48I) ensemble results.
 In addition to the statistical and systematic error bars,
  a curve has been plotted on top of each scatter point that
  follows the L\"uscher quantization curve
  over the range of phase shifts covered by the middle 68\%
  of the jackknife samples at fixed pion mass.
 Phenomenological curves using Roy Equation and
  \sutwo~chiral perturbation theory have been plotted for comparison.
 The Roy equation uncertainty band
  from Ref.~\cite{Ananthanarayan:2000ht} is plotted to demonstrate the
  good agreement with the LQCD scattering data at low center of mass energies.
 A vertical dotted green line is included to show the $4\pi$ threshold cutoff,
  above which the L\"uscher quantization condition is expected to break down.
 The top row shows the results obtained using only the interacting energies
  obtained from the lattice simulation,
  reported under the I$=2$ column in the left half of
  Tables~\ref{tab:2pigevpspec-24id}~and~\ref{tab:2pigevpspec-48i}.
 The middle row shows the results using the spectrum obtained from the
  procedure using the difference between interacting and noninteracting energies
  described in Sec.~\ref{sec:freepicorrelation}.
 The bottom row shows again the results using the spectrum with the 
  difference between interacting and noninteracting energies,
  but additionally subtracts away the backward-propagating contributions
  in the noninteracting correlation functions,
  as described in Eq.~(\ref{eq:corrNIsub}).
 \label{fig:phaseshiftiso2}
}
\end{figure*}

Fig.~\ref{fig:phaseshiftiso2} shows the scattering phase shifts in the isospin 2 channel
 as a function of the center of mass energies of the $\pi\pi$ system
 for both the 24ID and 48I ensembles.
The data plotted here correspond to
 the ``thermal subtraction'' method of Eq.~(\ref{eq:2piexpectvalue}).
Two phenomenological curves are also plotted;
 the first was computed using next-to-next to leading order
 \sutwo~chiral perturbation theory~\cite{Mawhinney:2015sfj,Boyle:2015exm}.
The second uses integral equations for the scattering phase shifts first derived by
 Roy~\cite{Roy:1971tc,Ananthanarayan:2000ht,Colangelo:2001df}.
Specifically, we use the results of
 Ref.~\cite{Ananthanarayan:2000ht} to calculate an error band
 for a qualitative comparison against our results.
A comprehensive study involving the more accurate prediction in
 Ref.~\cite{Colangelo:2001df} is deferred to a future publication.
The top row of the figure shows the interacting $\pi\pi$ data using
 Eq.~(\ref{eq:ppicom}) to compute the center of mass momentum.
Considering only statistical uncertainties,
 the two ensembles have good agreement for the ground state and first excited state,
 and agreement up to systematic uncertainties for the second excited state at larger $\sqrt{s}$.
The systematic uncertainty due to discretization effects completely
 dominates the total uncertainty,
 leading to a result that is mostly consistent with zero.

In the middle row of Fig.~\ref{fig:phaseshiftiso2},
 Eq.~(\ref{eq:ppicomdifference}) is used together
 with the noninteracting $\pi\pi$ energies to compute the phase shift instead,
 which leads to a reduction in statistical error.
The dominant systematic uncertainty due to discretization effects
 is reduced dramatically, resulting in roughly comparable
 statistical and systematic uncertainties.
The excited states in the $\pi\pi$ correlation functions are
 also correlated between the interacting and noninteracting energies
 and their effects are reduced when taking the difference.
This accounts for the large shift in the second excited state energy,
 leading to better agreement between the two ensembles.

However, the subtraction between interacting and noninteracting energies also
 produces a large discrepancy between results for the ground state.
This discrepancy is due to the introduction of backward-propagating terms in the
 noninteracting correlation functions of
 Eq.~(\ref{eq:noninteractingcorrelator}),
\begin{align}
 &C^{\text{NI}}_{\,\pi\pi,\mathbf{P},-\mathbf{P}} (t) =
 C_{\pi,\mathbf{P}} (t) C_{\pi,-\mathbf{P}} (t)
 \nonumber\\
 &=\sum_{mn}
  \big|\langle0|{\cal O}_{\pi,\mathbf{P}}|\pi^0\rangle_m\big|^2
  \big|\langle0|{\cal O}_{\pi,\mathbf{P}}|\pi^0\rangle_n\big|^2
 \nonumber\\
 &\quad\times
  \Big( e^{-E_{\pi,m} t-E_{\pi,n} t} +e^{-E_{\pi,m} t-E_{\pi,n} (T-t)}
 \nonumber\\
 &\qquad+e^{-E_{\pi,m} (T-t)-E_{\pi,n} t} +e^{-E_{\pi,m} (T-t)-E_{\pi,n} (T-t)} \Big).
\end{align}
The terms suppressed by $e^{-ET}$ produce small but non negligible shifts in the
 noninteracting correlation function and a large shift in the correlated difference
 between interacting and noninteracting energies.
The solution is to replace the noninteracting correlation function with a thermally-subtracted
 correlator that removes the contributions including backward-propagating pions,
\begin{align}
 &C^{\text{NI,sub}}_{\pi\pi,\mathbf{P},-\mathbf{P}} (t)
 \nonumber\\
 &=
 C^{\text{NI}}_{\pi\pi,\mathbf{P},-\mathbf{P}} (t)
 \nonumber\\
 &\quad
 -\big|\langle0|{\cal O}_{\pi,\mathbf{P}}|\pi^0\rangle\big|^2
   \big|\langle0|{\cal O}_{\pi,\mathbf{P}}|\pi^0\rangle\big|^2
 \Big(2e^{-E_\pi T} +e^{-2E_\pi (T-t)} \Big),
 \label{eq:corrNIsub}
\end{align}
 where only the 0-momentum backward-propagating term is relevant.
The subtracted correlator from Eq.~(\ref{eq:corrNIsub})
 is used in place of the
 unsubtracted noninteracting correlator of
 Eq.~(\ref{eq:noninteractingcorrelator})
 to produce the bottom row of Fig.~\ref{fig:phaseshiftiso2}.
In this pair of plots,
 removal of the backward-propagating pion terms
 restores the agreement between the ground-state energies in the phase
 shift and has negligible impact on the higher excited states.
This subtraction also removes much of the systematic error,
 which is attributed to a mismatch in the energies of the
 interacting and noninteracting values.

The combination of thermal subtractions on the
 interacting and noninteracting correlation functions
 as well as the correlated subtraction between
 interacting and noninteracting energies
 produces a set of phase shifts for each ensemble that
 are consistent with each other and agree well with the
 result from the Roy Equation.
This remains true even after passing the threshold for
 production of four-pion states,
 where there is no expectation that the L\"uscher
 phase shift quantization condition still holds.
There is also good agreement with the \sutwo $\chi$PT
 expectation for the lowest energy state,
 as depicted in the lower-right panel of Fig.~\ref{fig:phaseshiftiso2}.
The validity of $\chi$PT is however expected to break down
 at large $\sqrt{s}$, which spoils the comparison with higher excited states.

\begin{figure}[t!]
\begin{tabular}{c}
 \includegraphics[width=\figsize\textwidth]{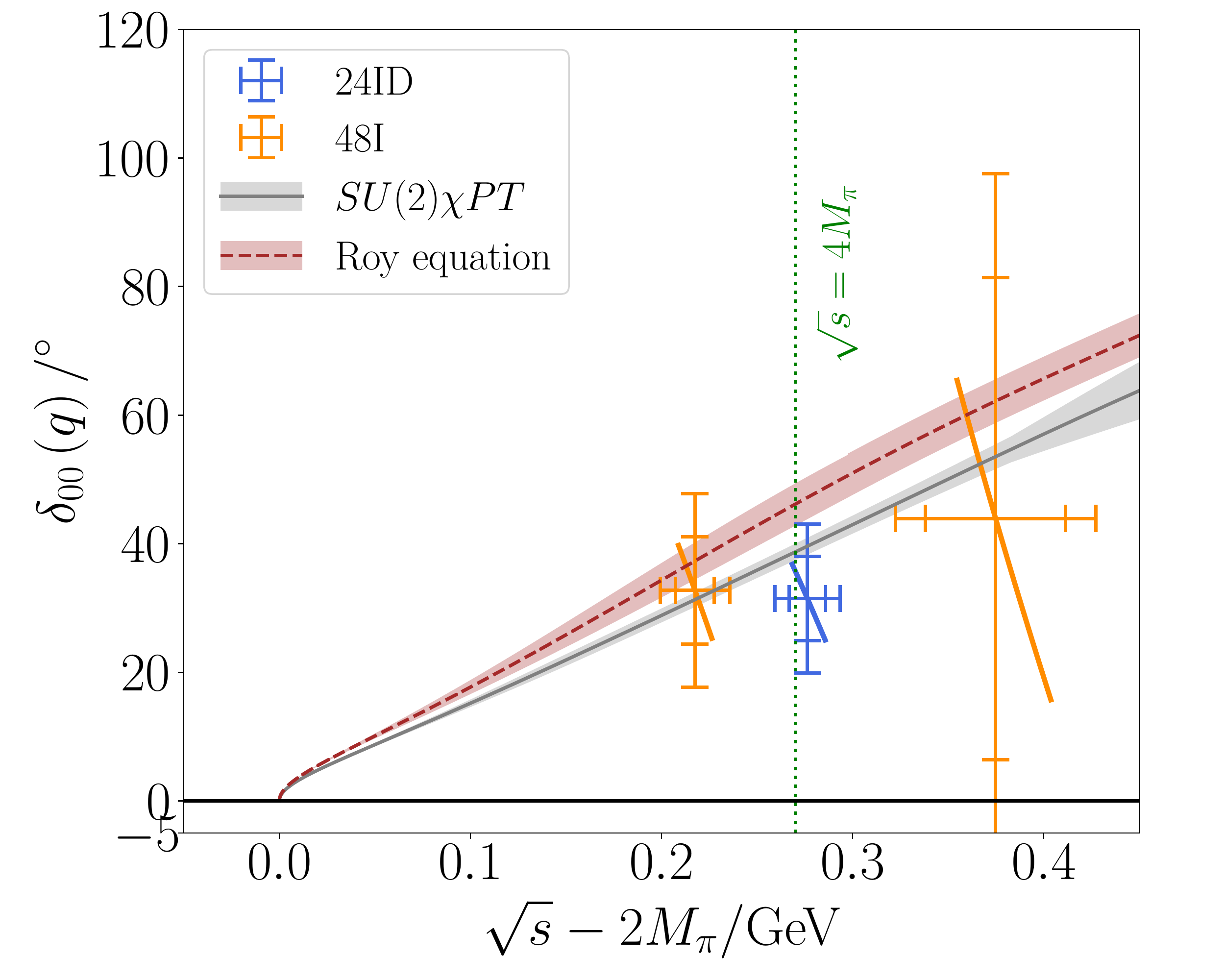} \\
 \includegraphics[width=\figsize\textwidth]{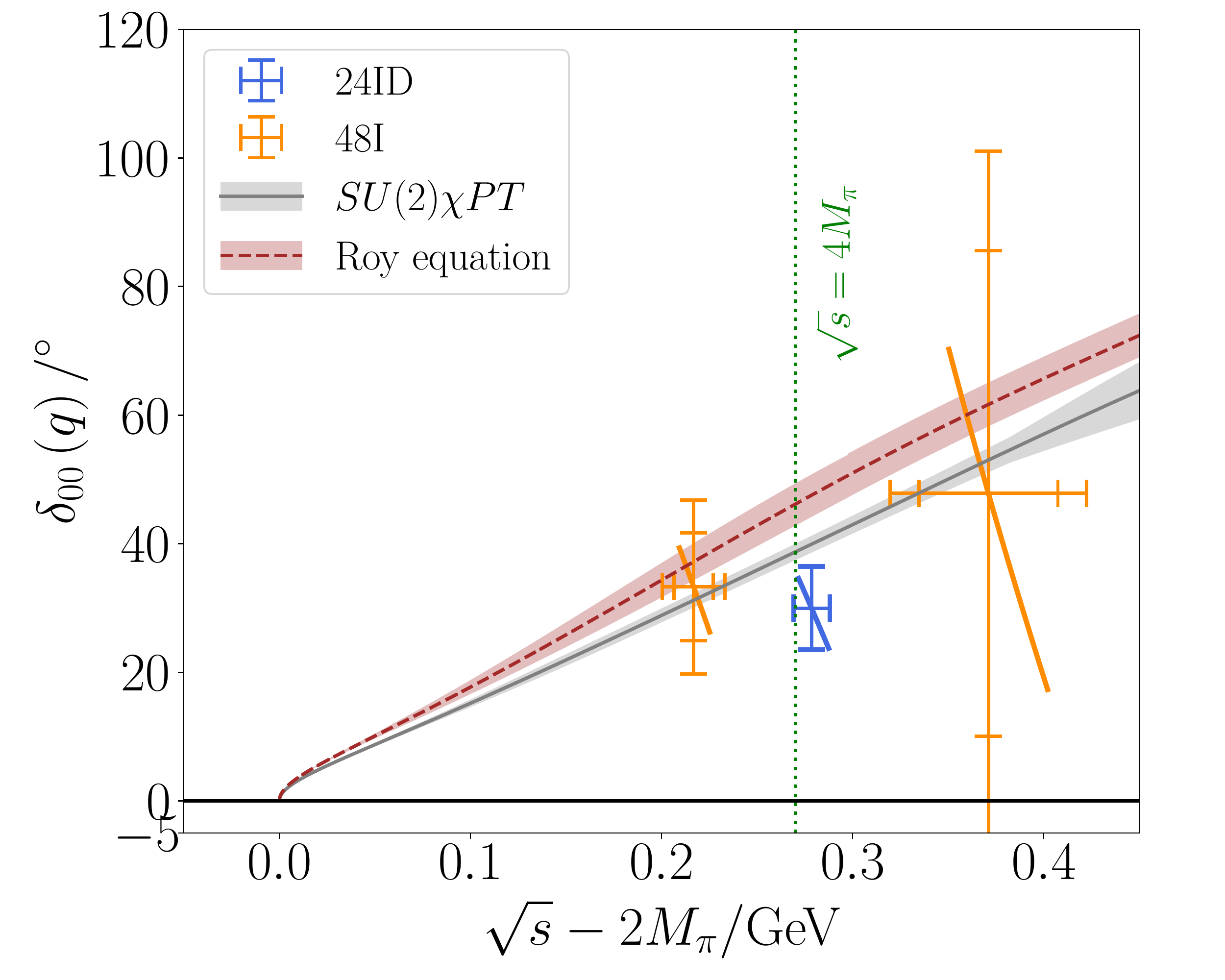} \\
\end{tabular}
\caption{
 Phase shift plots for the lattice data in the isospin 0 channel.
 The description for the top (bottom) plot is the same here
  as for the top (bottom) plot in the left column of
  Fig.~\ref{fig:phaseshiftiso2}.
 The energies plotted here correspond to $n=1$~and~2 in the I$=0$ columns of
  Tables~\ref{tab:2pigevpspec-24id}~and~\ref{tab:2pigevpspec-48i}.
 Note that the $n=0$ point close to threshold is not shown
  because $q^2<0$ cannot be mapped to a real value of $\sqrt{s}$.
 The Roy equation uncertainty band is calculated
 from Ref~\cite{Ananthanarayan:2000ht}.
 \label{fig:phaseshiftiso0}
}
\end{figure}

Fig.~\ref{fig:phaseshiftiso0} shows the scattering phase shifts
 in the isospin 0 channel as a function of the
 center of mass energies of the $\pi\pi$ system
 for both the 24ID and 48I ensembles.
All data here correspond to the  ``thermal+corrected vacuum''
 subtraction in Eqs.~(\ref{eq:vac})--(\ref{eq:iso0subtraction}).
The top plot shows the scattering phase shift computed from
 only the interacting  correlator data,
 which have large statistical uncertainties.
In the bottom plot,
 the difference between interacting and noninteracting energies
 are again applied, but the statistical fluctuations of the two spectra
 are not correlated enough to produce a noticeable reduction
 in statistical error.
The systematic error due to discretization uncertainty is reduced after the subtraction,
 resulting in a phase shift that is largely dominated by the statistical uncertainty
 and demonstrating that the lattice spacing uncertainty in the isospin 0 channel is not appreciable.
For the 48I ensemble, there is an approximate factor of 2 reduction
 in the statistical uncertainty on the ground state energy due to the subtraction,
 which is not visible in Fig.~\ref{fig:phaseshiftiso0}
 because the energy is below threshold and therefore 
 corresponds to imaginary $q$.

\subsection{Scattering Lengths}

The computed spectra on each ensemble close to threshold
 may also be used to produce estimates of the parameters
 associated with the effective range expansion.
The effective range expansion is expressed as the relation
\begin{align}
 p \,\text{cot}\delta_{I\ell}(pL/2\pi) =
 \frac{1}{a_{I\ell}}
 +\frac12 r_{I\ell} p^2 +O(p^4) ,
 \label{eq:ere}
\end{align}
 where $I$ and $\ell$ correspond to the isospin and spin channels,
 and $a_0$ and $r_0$ are the scattering length and effective range, respectively.
Taking $q = pL/2\pi$ close to threshold,
 the terms proportional to higher powers of $p^2$ can be neglected
 and the scattering length computed directly from the 
 ground state energy,
\begin{align}
 q \,\text{cot}\delta_{I\ell} (q) \Big|_{q \ll 1}
 =
 \frac{L}{2\pi a_{I\ell}} +O(q^2).
 \label{eq:eresmallq}
\end{align}
From the L\"uscher relation, the phase shift
 in the spin-0 channel with generic isospin $I$ is written
\begin{align}
 \frac{{\cal Z}_{00}(1;q^2)}{\pi^{\frac32}} =
 q \,\text{cot}\delta_{I0}(q),
 \label{eq:erezeta}
\end{align}
 which when combined with Eq.~(\ref{eq:eresmallq})
 may be used to estimate the scattering length.

\begin{table}[htb!]
\centering
\begin{tabular}{c|l}
Ens & \multicolumn{1}{c}{$M_\pi a_{00}$ (matrix element subtraction)} \\
\hline
24ID & $0.184(46)_{\text{stat}}(00)_{1\pi}(08)_{t_0}(04)_{\delta t}(05)_{a^2}(01)_{\text{3pt}}[47]_{\text{total}}$ \\
48I & $0.218(24)_{\text{stat}}(00)_{1\pi}(19)_{t_0}(15)_{\delta t}(03)_{a^2}(00)_{\text{3pt}}[34]_{\text{total}}$ \\
\hline \hline
Ens & \multicolumn{1}{c}{$M_\pi a_{00}$ (time series subtraction)} \\
\hline
24ID & $0.214(28)_{\text{stat}}(00)_{1\pi}(17)_{t_0}(12)_{\delta t}(06)_{a^2}[35]_{\text{total}}$ \\
48I & $0.125(41)_{\text{stat}}(00)_{1\pi}(53)_{t_0}(41)_{\delta t}(02)_{a^2}[79]_{\text{total}}$ \\
\end{tabular}
\caption{
 Results for the pion mass times the scattering length in the isospin 0 channel, $M_\pi a_{00}$,
  for each ensemble and subtraction scheme.
 The values in this table are computed assuming Eqs.~(\ref{eq:eresmallq})~and~(\ref{eq:erezeta})
  and using only the ground state energy.
 The full breakdown of uncertainties is included;
  stat denotes the statistical uncertainty,
  $1\pi$ denotes the uncertainty due to excited states
   in the $1\pi$ fits as described in Sec.~\ref{sec:onepion},
  $t_0$ and $\delta t$ denote the uncertainty from shifting the
   corresponding time variable in Eq.~(\ref{eq:effectiveenergy})
   by 1 timeslice,
  3pt denotes the uncertainty on the ratio with a
   pion transition matrix element in Eq.~(\ref{eq:c3ptratio})
   as described in Sec.~\ref{sec:piontransition},
  $a^2$ denotes the discretization uncertainty applied as described
   in Eq.~(\ref{eq:discretizationuncertainty}),
  and the number in square brackets is the total uncertainty on the quantity
   after adding all errors in quadrature.
 For the time series subtraction scheme, the pion transition matrix elements
  are unused and so their systematic uncertainty is not included in the error budget.
 Uncertainties are computed using standard error propagation
  and the uncertainties from similar sources are added in quadrature.
 \label{tab:scatlen_nor0}
}
\end{table}

\begin{table}[htb!]
\centering
\begin{tabular}{c|l}
Ens & \multicolumn{1}{c}{$M_\pi a_{20}$ (matrix element subtraction)} \\
\hline
24ID & $-0.0500(15)_{\text{stat}}(00)_{1\pi}(11)_{t_0}(03)_{\delta t}(12)_{a^2}(01)_{\text{3pt}}[22]_{\text{total}}$ \\
48I & $-0.0477(35)_{\text{stat}}(00)_{1\pi}(20)_{t_0}(26)_{\delta t}(05)_{a^2}(01)_{\text{3pt}}[48]_{\text{total}}$ \\
\hline \hline
Ens & \multicolumn{1}{c}{$M_\pi a_{20}$ (time series subtraction)} \\
\hline
24ID & $-0.0519(27)_{\text{stat}}(00)_{1\pi}(25)_{t_0}(14)_{\delta t}(12)_{a^2}[41]_{\text{total}}$ \\
48I & $-0.0450(53)_{\text{stat}}(00)_{1\pi}(28)_{t_0}(39)_{\delta t}(05)_{a^2}[71]_{\text{total}}$ \\
\end{tabular}
\caption{
 The same as Table~\ref{tab:scatlen_nor0}, but for the isospin 2 channel.
 \label{tab:scatlen_nor2}
}
\end{table}

The pion mass times scattering lengths obtained by applying the
 both the time series subtraction (Eq.~(\ref{eq:timeseriessubtraction}))
 and matrix element subtraction
 (the ``corrected vacuum+thermal subtraction'' for isospin 0
 and ``thermal subtraction'' for isospin 2)
 schemes combined with the difference between interacting and noninteracting
 effective energies (Eq.~(\ref{eq:niediff}))
 are given in Tables~\ref{tab:scatlen_nor0}~and~\ref{tab:scatlen_nor2}
 for the isospin 0 and isospin 2 channels, respectively.
These tables give the full breakdown of statistical and systematic uncertainties
 on the computed scattering lengths.
For the matrix element subtraction scheme on the 48I ensemble,
 the scattering lengths obtained for both the isospin 0 and 2 channels
 are in excellent agreement with phenomenology~\cite{Caprini:2011ky}.
These values will be taken as our final results when comparing to other literature.

We find reasonable agreement between the determinations of the scattering lengths
 on both ensembles using both subtraction schemes within their uncertainties,
 and with phenomenological determinations from
 Caprini \emph{et al.}~\cite{Caprini:2011ky}.
The largest disagreements appear for the 24ID ensemble in the isospin 2 channel.
This is likely due to
 1) the uncontrolled large discretization errors from the large lattice spacing,
 and 2) a statistical fluctuation to a low pion mass combined with
 an underestimated uncertainty from too few statistical measurements.
The low pion mass would cause the noninteracting effective energy to be underestimated
 and increase the interacting-noninteracting effective energy difference,
 which produces a larger negative scattering length in competition with the explicit
 pion mass dependence in the product $M_\pi a_{20}$.
This should also produce a smaller $a_{00}$ in the isospin 0 channel on the 24ID ensemble,
 although the uncertainties are large enough that the disagreement is not as obvious.

Another possible cause for concern is the isospin 0 scattering length for the matrix
 subtraction scheme on the 48I ensemble.
This estimate is quite low in comparison to the phenomenological value
 when only the statistical uncertainty is considered.
This is due to a large excited state contamination in the \emph{noninteracting}
 $\pi\pi$ correlators that is amplified by the time series subtraction scheme.
The contamination is strongly correlated with the isospin 2 channel interacting correlators
 but absent from the isospin 0 channel.
When the interacting-noninteracting effective energy difference is taken,
 the excited state contamination largely cancels in the isospin 2 channel
 but is introduced to the isospin 0 channel by the subtraction.
This large contamination is reflected in the large $t_0$ and $\delta t$ systematic uncertainties
 on $M_\pi a_{00}$ and the correspondingly large total uncertainty.

\begin{figure}[t!]
\begin{tabular}{c}
 \includegraphics[width=\figsize\textwidth]{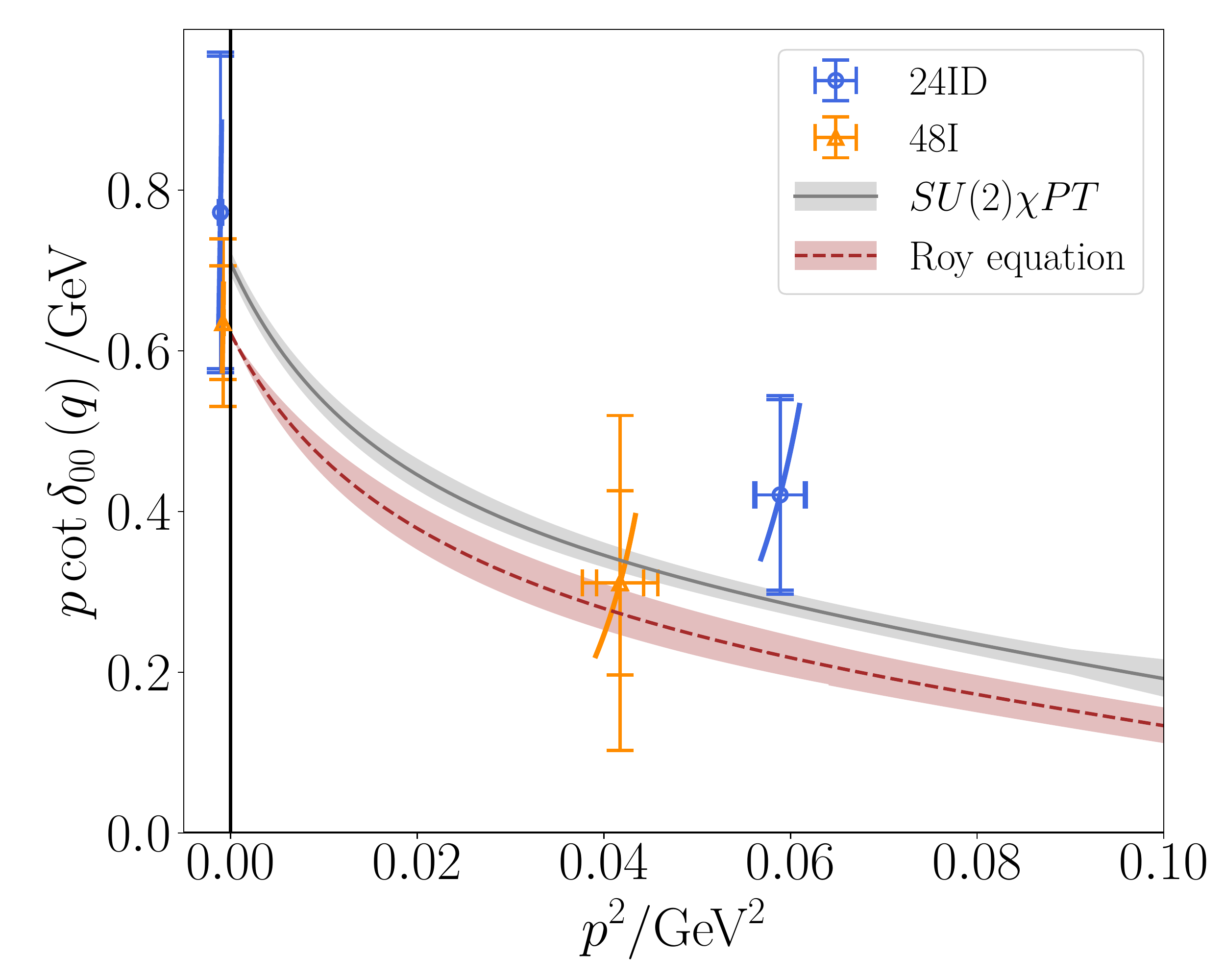} \\
 \includegraphics[width=\figsize\textwidth]{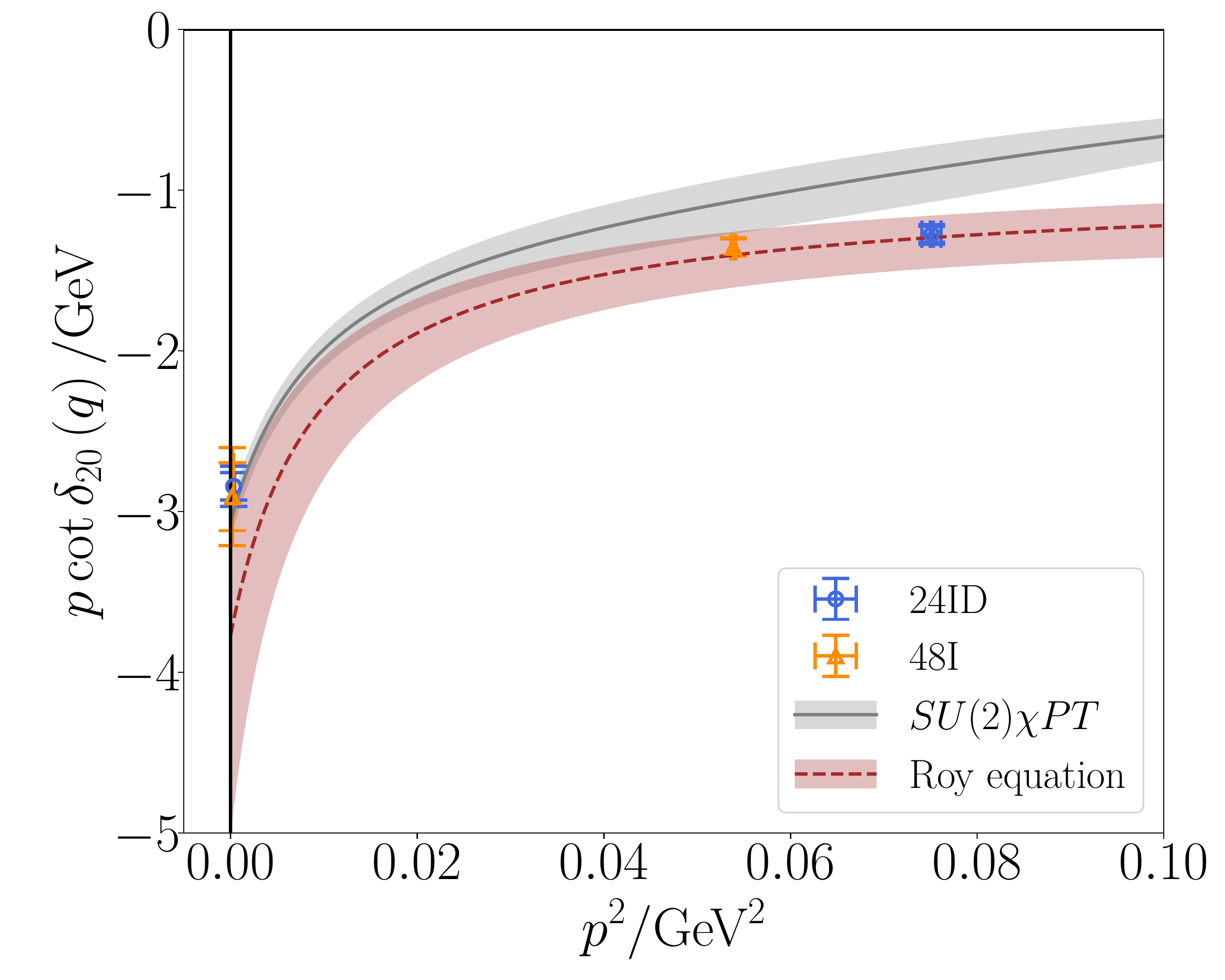} \\
\end{tabular}
\caption{
 Plot of the effective range expansion of Eq.~(\ref{eq:ere})
  as a function of $p^2$.
 Data for the ground state and first excited state of both
  ensembles are plotted as well as the predictions from
  \sutwo~chiral perturbation theory and the Roy equation.
 Like Fig.~\ref{fig:phaseshiftiso2},
  a curve has been plotted on top of each data point that
  follows the L\"uscher quantization curve at fixed pion mass
  over the range of phase shifts covered by the middle 68\%
  of the jackknife samples.
 \label{fig:ere}
}
\end{figure}

The effective range expansion of Eq.~(\ref{eq:ere})
 is shown in Fig.~\ref{fig:ere}.
In both isospin channels, the data close to threshold are in
 good agreement with the predictions from \sutwo~chiral perturbation theory
 and the Roy equation.
One might hope to obtain information about the effective range
 $r_{I\ell}$ in Eq.~(\ref{eq:ere}) by taking into account both
 the ground state and first excited state.
This would amount to fitting the $p^2$ dependence in Fig.~\ref{fig:ere}
 with the right-hand side of Eq.~(\ref{eq:ere}),
 specifically to access the information about the slope with respect
 to $p^2$ at threshold:
\begin{align}
 2\frac{\partial}{\partial p^2}
 \left[ p\cot{\delta_{I\ell}(q)} \right] = r_{I\ell}.
\end{align}
The nontrivial $p^2$ dependence of the \sutwo~chiral perturbation
 theory and Roy equation predictions over the range
 of $p^2$ between the ground state and first excited state
 suggest that the $p^2$ dependence must include higher-order corrections
 that are not properly captured by the formula in Eq.~(\ref{eq:ere}).
With the data available, the best that can be done is to fit the linear
 response with respect to $p^2$, which will not reproduce the correct slope.
For this reason, no attempt is made to determine the effective range
 from the lattice data and the first excited state is plotted in
 Fig.~\ref{fig:ere} only to show consistency with the phenomenological predictions.
It should be noted, however, that a linear extrapolation between
 the ground state and first excited state gives a
 scattering length that is consistent with the estimate from
 the ground state alone to well within $1\sigma$,
 indicating that the error from computing the scattering length
 at nonzero $q^2$ is negligible.

\section{Discussion \& Conclusions}
 \label{sec:discussion}

In this analysis, we solved the GEVP for matrices
 of correlation functions computed with large bases
 of one- and two-bilinear interpolating operators in LQCD,
 obtaining scattering lengths for both
 the isospin 0 and isospin 2 channels.
These results take advantage of the distillation framework
 to compute and subtract away contamination from
 unwanted thermal and vacuum contributions to the
 two-pion correlation functions.
By correlating the computed spectrum with another obtained
 from products of single pion correlation functions,
 we were able to improve the precision on the isospin-2
 channel scattering lengths.

Our analysis constitutes the first effort to obtain
 the isospin 0 scattering length from LQCD with both
 valence and sea quarks at physical pion mass
 and the second to achieve this claim in the isospin 2 channel.
Although our results use two ensembles of different lattice spacings,
 these ensembles have different quark actions and so lie
 on different continuum trajectories.
To attempt to account for the possible discretization errors,
 a fractional uncertainty is assigned to the $\pi\pi$ state energy
 based on $a^2$ scaling behavior of other parameters on the same lattice actions.
Despite the unknown discretization effects,
 the 48I ensemble produces scattering lengths for both the isospin 0 and isospin 2 channels
 that are consistent with phenomenology within their quoted statistical uncertainties.
This is the third work in a series of RBC+UKQCD collaboration sister 
 papers to understand $\pi\pi$ scattering~\cite{HoyingPipi,RBC:2021acc}.
 Comparisons between the results presented here and its sister works is deferred
 to a future paper, which will combine the results of these works with phenomenology.

\begin{figure}[htb!]
 \includegraphics[width=0.5\textwidth]{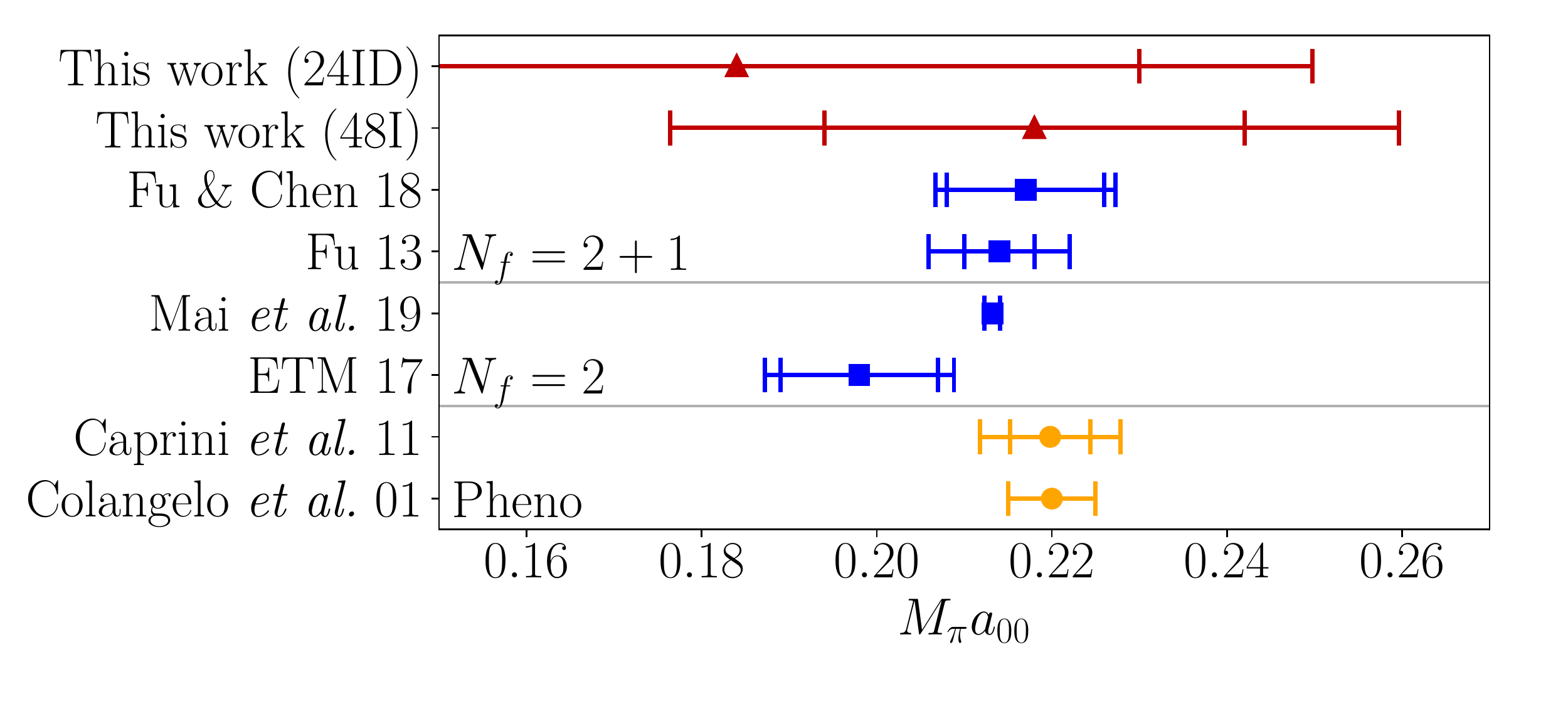}
 \includegraphics[width=0.5\textwidth]{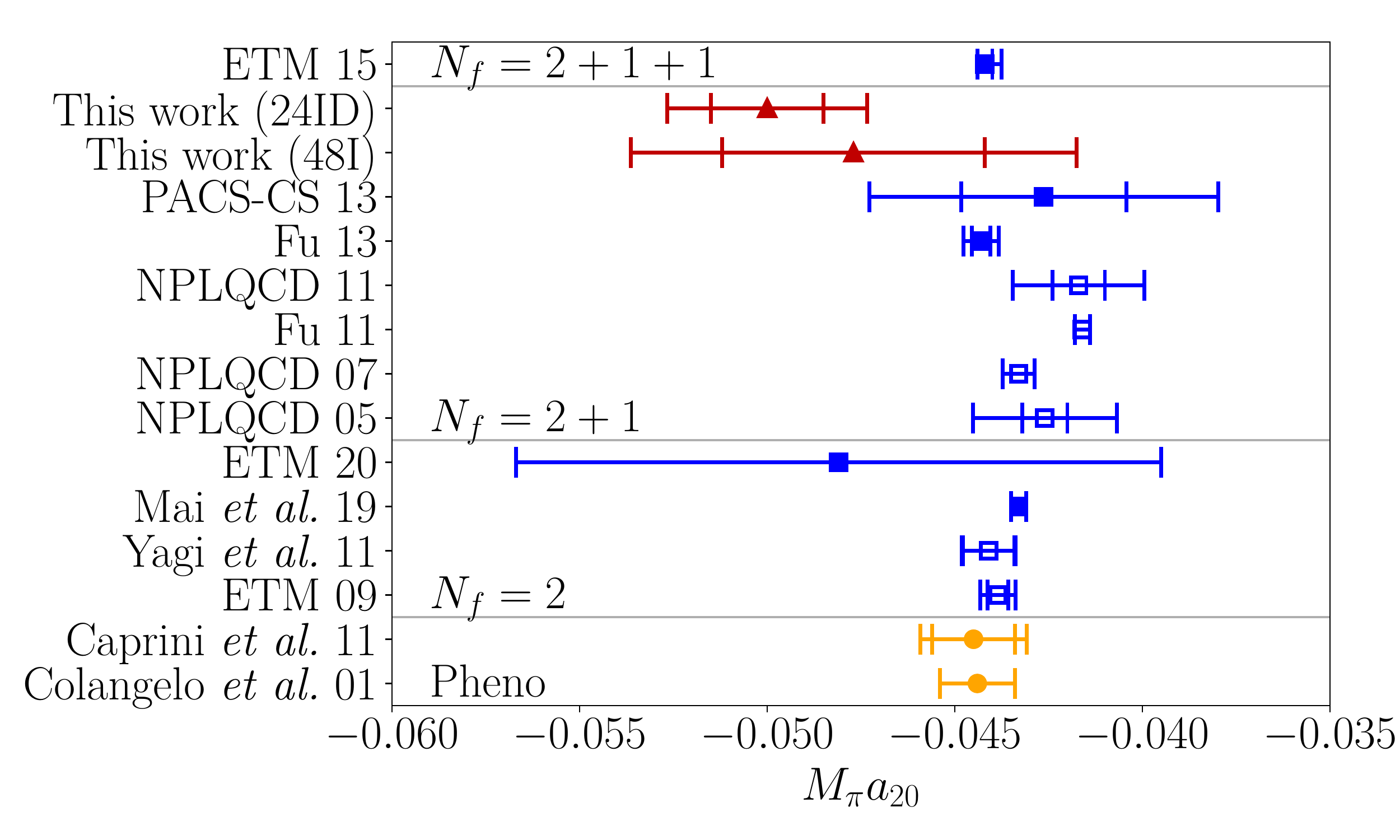}
 \caption{
  Summary plot of other $\pi\pi$ scattering length calculations
   from LQCD and phenomenology.
  Results plotted here are taken from the FLAG review~\cite{Aoki:2021kgd},
   with the addition of ETM~20~\cite{Fischer:2020jzp} and Mai~{\it et al.}~19.
  Results with a smallest $M_\pi \leq 250$~MeV are plotted with filled symbols,
   all others with open symbols.
  The full list of references is found in Table~\ref{tab:literature}.
  \label{fig:literature}
 }
\end{figure}

A comparison of these results to literature values for the scattering lengths
 from the FLAG review~\cite{Aoki:2021kgd} are given for both isospin channels
 in Table~\ref{tab:literature}.
The values from the tables are also given in a summary plot
 in Fig.~\ref{fig:literature}.
One other LQCD computation has been performed directly at physical $M_\pi$,
 labeled as ``ETM 20,'' which obtained the scattering phase shift
 in the isospin 2 channel~\cite{Fischer:2020jzp}.
This result is in good agreement with our isospin 2 result
 and with a comparable uncertainty to our extraction.
All of the other results are computed at heavier-than-physical pion masses
 and extrapolated down to the physical point,
 which all tend to have reasonable agreement with phenomenological estimates.
The majority of these have a lightest pion mass at the level of $M_\pi \approx 250$~MeV
 and several ensembles for extrapolations to lighter pion masses.

\begin{table*}[htb!]
\begin{tabular}{llllcc}
Authors & $N_f$ & Ref. & $M_\pi a_{00}$ & $M_\pi \geq$[MeV] & $N_{\rm ens}$ \\
\hline
This work (48I)           & 2+1 & \,---                   & 0.218(24)(24) & 139 & 1\\
Mai {\it et al.} 19       & 2   & \cite{Mai:2019pqr}      & 0.2132(9)     & 220 & 6\\ 
Fu \& Chen 18             & 2+1 & \cite{Fu:2017apw}       & 0.217(9)(5)   & 247 & 3\\ 
ETM 17                    & 2   & \cite{Liu:2016cba}      & 0.198(9)(6)   & 250 & 3\\ 
Fu 13                     & 2+1 & \cite{Fu:2013ffa}       & 0.214(4)(7)   & 240 & 6\\ 
\hline
Caprini {\it et al.} 11   & --- & \cite{Caprini:2011ky}   & 0.2198(46)(16)(64) & --- & ---\\
Colangelo {\it et al.} 01 & --- & \cite{Colangelo:2001df} & 0.220(5)           & --- & ---\\[2em]
Authors & $N_f$ & Ref. & $M_\pi a_{20}$ & $M_\pi \geq$[MeV] & $N_{\rm ens}$ \\
\hline
This work (48I)           & 2+1   & \,---                   & -0.0477(35)(33)     & 139 & 1\\
ETM 20                    & 2     & \cite{Fischer:2020jzp}  & -0.0481(86)         & 134 & 3\\ 
Mai {\it et al.} 19       & 2     & \cite{Mai:2019pqr}      & -0.0433(2)          & 220 & 6\\ 
ETM 15                    & 2+1+1 & \cite{Helmes:2015gla}   & -0.0442(2)($^4_0$)  & 250 &11\\ 
PACS-CS 13                & 2+1   & \cite{Sasaki:2013vxa}   & -0.04263(22)(41)    & 170 & $1^\ast$\\ 
Fu 13                     & 2+1   & \cite{Fu:2013ffa}       & -0.04430(25)(40)    & 240 & 6\\ 
NPLQCD 11                 & 2+1   & \cite{Beane:2011sc}     & -0.0417(07)(02)(16) & 390 & 4\\ 
Fu 11                     & 2+1   & \cite{Fu:2011bz}        & -0.0416(2)          & 330 & $1^\ast$\\ 
Yagi {\it et al.} 11      & 2     & \cite{Yagi:2011jn}      & -0.04410(69)(18)    & 290 & $1^\ast$\\ 
ETM 09                    & 2     & \cite{Feng:2009ij}      & -0.04385(28)(38)    & 270 & 6\\ 
NPLQCD 07                 & 2+1   & \cite{Beane:2007xs}     & -0.04330(42)        & 290 & 4\\ 
NPLQCD 05                 & 2+1   & \cite{Beane:2005rj}     & -0.0426(06)(03)(18) & 290 & 3\\ 
\hline
Caprini {\it et al.} 11   & ---  & \cite{Caprini:2011ky}   & -0.0445(11)(4)(8)    & --- & --- \\
Colangelo {\it et al.} 01 & ---  & \cite{Colangelo:2001df} & -0.0444(10)          & --- & --- \\
\end{tabular}
\caption{
 Most recent results for the isospin 0 and isospin 2 scattering lengths,
  compiled in the FLAG review~\cite{Aoki:2021kgd}.
 Additional references for ``ETM 20'' and ``Mai {\it et al.} 19''
  are listed which has appeared in the literature after the FLAG review.
 Minimum valence pion mass for each reference, in GeV,
  is listed in the fifth column
  and the number of ensembles that were used in each reference in the sixth column.
 For the three references where only one ensemble was used,
  multiple valence pion masses were computed.
 ``PACS-CS 13'' had a total of 5 valence pion masses,
  while both ``Fu 11'' and ``Yagi \emph{et al.} 11'' had 6 valence pion masses.
 \label{tab:literature}
}
\end{table*}

The distillation setup that was used in this analysis is of great interest
 to the RBC+UKQCD collaborations' effort to compute
 the anomalous magnetic moment of the muon~\cite{RBC:2018dos}.
For this purpose, distillation perambulators have been generated on several other ensembles,
 including 2 other ensembles at physical pion mass (correspondingly referred to as 64I and 96I)
 at finer lattice spacing using the same action as the 48I.
An identical analysis could be carried out on these ensembles to facilitate
 a proper continuum extrapolation entirely at the physical pion mass
 in order to assess discretization effects on the scattering lengths and phase shifts.
Correlation of these distillation data with the point vector current
 insertions needed for the $g-2$ analysis would provide additional constraints
 on the isospin 1 channel, which was ignored for the purposes of this analysis.

Distillation is also ideal for constructing additional correlation functions
 without the need for additional solutions of the Dirac matrix.
These advantages permit studies of moving frames~\cite{Thomas:2011rh},
 different operator smearings~\cite{Dudek:2009qf,Dudek:2012gj},
 and other cubic group irreducible representations~\cite{Dudek:2010wm}
 with little extra cost.
This would allow the phase shift to be probed at more center of mass energies
 below the $4\pi$ production threshold, allowing the phase shift curve
 and the effective range expansion to be mapped out in more detail.
This could also be used to constrain the strength of the D-wave
 $\pi\pi$ scattering channel.

For the moving frames, the subtraction techniques using other correlation functions
 that are described in this manuscript are especially powerful.
At nontrivial center-of-mass momentum,
 the appropriate transition matrix elements for assessing the thermal contamination
 to the $\pi\pi$ two-point correlation functions connect in states and out states
 with different momenta.
This produces a nontrivial Euclidean time dependence that does not exactly cancel
 with the time series subtraction method in Eq.~(\ref{eq:timeseriessubtraction}).
Other subtraction schemes have been adopted to circumvent this flaw~\cite{Thomas:2011rh},
 but these subtractions are only approximate and still suffer from the enhancement
 of excited state contamination that appeared in the 48I ensemble in this analysis.
It is also possible to fit this correlator contamination directly,
 but the exponential suppression with the temporal extent makes this prospect challenging.
Construction of alternative correlation functions provides a complimentary
 method for constraining the required matrix elements without
 enhancing excited state contamination or competing with the exponential suppression.

The phase shifts computed in this analysis are in excellent agreement with
 the scattering phase shift parameterization from Roy~\cite{Roy:1971tc}.
This is true even beyond the $4\pi$ threshold,
 where the L\"uscher quantization condition is not strictly valid.
This bodes well for future analyses,
 suggesting that the corrections to the spectrum from mixing with $4\pi$ states
 are small and that it might be possible to understand these contaminations
 as only a perturbation on elastic scattering phase shift results.

With correlation functions computed directly with physical pion mass,
 at several lattice spacings and volumes, and moving frames,
 LQCD will soon be able to make definitive claims about scattering
 in the $\pi\pi$ channels at various isospin.
These results could become competitive with phenomenological estimates
 in the near future, especially for higher partial wave contributions,
 and provide an alternative theoretical prediction
 of the low energy constants relevant to pion scattering.

\section{Acknowledgements}
We would like to thank our colleagues in the RBC and UKQCD collaborations
 for their interesting discussions.
The research of M.B. is funded through the MUR program
 for young researchers ``Rita Levi Montalcini''.
The work of D.H. is supported by the Swiss National Science Foundation (SNSF)
 through grant No. 200020\_208222.
T.I. and C.L. were supported in part by US DOE Contract DESC0012704(BNL),
 and T.I. was supported in part by the Scientific Discovery
 through Advanced Computing (SciDAC) program LAB 22-2580.
The work of A.S.M. was supported by the Department of Energy,
 Office of Nuclear Physics, under Contract No. DE-SC00046548.
The work of A.S.M. is also supported in part by
 Lawrence Livermore National Security, LLC
 under Contract No. DE-AC52-07NA27344 with the U.S. Department of Energy.
The work of M.T. was supported in part by US DOE awards
 DE-SC0010339 and DE-SC002114.
This work was supported by resources provided by the
 Scientific Data and Computing Center (SDCC)
 at Brookhaven National Laboratory (BNL),
 a DOE Office of Science User Facility supported by the
 Office of Science of the US Department of Energy.
The SDCC is a major component of the Computational Science Initiative at BNL.
We gratefully acknowledge computing resources provided through USQCD clusters
 at BNL and Jefferson Lab.

\appendix
\section{Scattering Phase Shift Formalism}
 \label{sec:formalism}

Due to the finite volume of calculations in LQCD,
 it is not possible to isolate an asymptotic state.
Particles in multiparticle states will travel through the periodic boundaries
 and reinteract with each other.
These interactions induce deviations in the energy spectrum
 as the states mix and undergo avoided level crossings.
As a result, the spectrum of the multiparticle states observed on the lattice
 cannot be directly deduced from the infinite volume continuum spectrum.

However, this problem may be turned on its head
 by using the finite volume as a probe of the physics of multiparticle scattering.
The deviations of the spectrum away from the 
 noninteracting limit are related to the infinite volume
 scattering phase shifts.
It is possible to vary the volume and use the deviations
 in the spectrum to deduce the scattering phase shifts.
This was first proposed by L\"uscher in a series of
 papers~\cite{Luscher:1986pf,Luscher:1990ux,Luscher:1991cf}, and was later
 generalized to moving frames in Ref.~\cite{Rummukainen:1995vs}.
The simplest system that exhibits this behavior with the volume is the
 $\pi\pi$ interaction channel, which is target of this study.
For the purposes of this appendix,
 we will discuss $\pi\pi$ scattering at general center-of-mass momentum,
 although the entire analysis in this manuscript was performed
 in the $\pi\pi$ center-of-mass rest frame only.

The matrix elements may be deduced by computing the shifts in the $\pi\pi$
 energy levels away from the expected infinite-volume continuum values,
\begin{equation}
    \mathbf{p}^2 = \frac{1}{4} \left( E_{\pi\pi}^2
    -(2 M_\pi)^2 -\mathbf{P}^2 \right) \,,
\end{equation}
 where the momentum of the moving frame is given by 
 $\mathbf{P} = {2\pi \mathbf{d}}/{L}$ for $\mathbf{d} \in \mathds{Z}^3$.
For convenience, we define the normalized momentum
\begin{align}
 \mathbf{q} = \mathbf{p}L/2\pi.
\end{align}
The scattering phase shifts may then be computed by solving
\begin{equation}
    {\rm det}\left[e^{2i\delta(q)} -U (q)\right] = 0
\end{equation}
 for a matrix $U$ computed from studying the mixing of states in the finite volume,
\begin{equation}
    U = ({\cal M} +i\mathds{1})({\cal M} -i\mathds{1})^{-1}
    \label{eq:umatrix}
\end{equation}
 with the scattering matrix ${\cal M}$ to be given below.
The expression for ${\cal M}$ is deduced from $M$, the continuous, finite-volume expression
 coupling the angular momentum eigenstates given by
\begin{widetext}
\begin{equation}
    M^{\mathbf{d}}_{\ell m, \ell^\prime m^\prime} (q) = -i\gamma^{-1}
    \frac{(-1)^\ell}{\pi^{3/2}}
    \sum_{j=|\ell-\ell^\prime|}^{\ell+\ell^\prime}
    \sum_{s=-j}^{j} \left( \frac{2\pi i}{p L} \right)^{j+1}
    {\cal Z}_{js}^{\mathbf{d}}\left(1;q\right)
    C_{\ell m,js, \ell^\prime m^\prime}.
    \label{eq:mmatrixcontinuum}
\end{equation}
The term $C$ is proportional to the typical \sutwo~Clebsch-Gordan coefficients,
\begin{align}
    C_{\ell m,js, \ell^\prime m^\prime}
    &= \left[ i^{\ell-j+\ell'} \sqrt{\frac{(2\ell+1)(2j+1)}{(2\ell'+1)}}
    \braket{ \ell,0;j,0|\ell',0} \right]
    \braket{ \ell,m;j,s|\ell',m'} \,,
    \label{eq:matrixcgfactor}
\end{align}
\end{widetext}
 where the term in square brackets is independent of the angular momentum
 $\mathbf{z}$ components, $m$, $s$, and $m'$.
The Lorentz factor $\gamma$ is computed from the moving frame momentum,
\begin{equation}
    \gamma = \frac{1}{\sqrt{ 1 - \frac{\mathbf{P}^2}{E^2_{\pi\pi}} }} \,,
\end{equation}
 and ${\cal Z}^{\mathbf{d}}$ is a spherical derivative of the zeta function,
\begin{equation}
    {\cal Z}_{\ell m}^{\mathbf{d}}\left( s; q \right)
    = \sum_{\mathbf{r}\in P_{\mathbf{d}}} {\cal Y}_{\ell m} (\mathbf{r})
    \left(\mathbf{r}^2-q^2 \right)^{-s} \,,
\end{equation}
 with ${\cal Y}$ proportional to a spherical harmonic,
\begin{equation}
    {\cal Y}_{\ell m}(\mathbf{r}) = r^{\ell} Y_{\ell m}(\hat{r}) \,.
\end{equation}
The domain of the sum in the expression for ${\cal Z}$ is
\begin{equation}
    P_{\mathbf{d}} = \left\{ \mathbf{r}\in \mathds{R}^3 \,\Big|\, \mathbf{r}
    = \hat{\gamma}^{-1} \left(\mathbf{n}+\frac{\mathbf{d}}{2} \right)
    \,, \, \mathbf{n} \in \mathds{Z}^3 \right\} \,,
\end{equation}
 with the operator $\hat{\gamma}^{-1}$ generating a boost of the momentum
 along the direction of $\mathbf{d}$,
\begin{equation}
    \hat{\gamma}^{-1} \mathbf{k}
    = \gamma^{-1} \mathbf{k}_{\parallel} + \mathbf{k}_{\perp}
\end{equation}
 for
\begin{equation}
    \mathbf{k}_{\parallel} = \frac{\mathbf{k}\cdot\mathbf{d}}{d^2}\mathbf{d}
    \,, \quad \mathbf{k}_{\perp} = \mathbf{k} - \mathbf{k}_{\parallel} \,.
\end{equation}

The expression in Eq.~(\ref{eq:mmatrixcontinuum}) is derived for a theory
 that is continuous but restricted to finite volume, but the matrix element
 in Eq.~(\ref{eq:matrixcgfactor}) is expressed in terms of the infinite volume
 angular momentum eigenstates.
These continuum angular momentum eigenstates are considered in the center of mass
 frame of the $\pi\pi$ system, which is boosted relative to the lab frame when
 $\mathbf{P}\neq\mathbf{0}$.
Another Clebsch-Gordan coefficient must be included to project the system to the
 irreducible representations of the moving $\pi\pi$ system.

The center of mass states belong to a reducible representation
 and are conventionally referred to as canonical states,
 given simply by a boost of a rest-frame angular momentum eigenstate,
\begin{equation}
     \hat{L}(\mathbf{P}) \ket{\mathbf{0},\ell,m}_{\cal C} =
     \ket{\mathbf{P},\ell,m}_{\cal C} \,.
     \label{eq:canonicalboost}
\end{equation}
The canonical states are denoted with a subscript ${\cal C}$ and satisfy
 the action under rotation given by
\begin{equation}
     \hat{R} \ket{\mathbf{P},\ell,m}_{\cal C} =
     \sum_{m'} {\cal D}^{\ell}_{m'm}(\hat{R}) \ket{\hat{R}
     \mathbf{P},\ell,m'}_{\cal C} \,.
     \label{eq:canonicalrotation}
\end{equation}
The irreducible representations of boosted systems are helicity irreps,
 with states that satisfy the action under rotation
\begin{equation}
     \hat{R} \ket{\mathbf{P},\ell,\lambda} =
     \ket{\hat{R} \mathbf{P},\ell,\lambda}
     \label{eq:helicityrotation}
\end{equation}
 with $\lambda=\pm|\lambda|$ denoting the helicity.
Combining Eqs.~(\ref{eq:canonicalboost})--(\ref{eq:helicityrotation})
 results in the desired relation between canonical and helicity states,
\begin{equation}
     \braket{\mathbf{P},\ell,\lambda | \mathbf{P},\ell,m}_{\cal C} =
     {\cal D}^{\ell\;\ast}_{m\lambda}(\hat{R}_0) \,,
     \label{eq:canonicaltohelicity}
\end{equation}
 for $\mathbf{P} = \hat{R}_0 P \hat{z}$ and
 noting that ${\cal D}^{\ell\;\ast}_{m \lambda}(\hat{R}_0)
 = {\cal D}^{\ell}_{\lambda m}(\hat{R}^{-1}_0)$.

There is an additional Clebsch-Gordan coefficient that is needed because the finite volume
 breaks the continuous rotational symmetry down into the finite octahedral rotation group.
The restriction to the octahedral group is imposed by contracting
 Eq.~(\ref{eq:mmatrixcontinuum}) with subduction coefficients, 
 which for properly normalized states may be written as
\begin{equation}
     S^{\mathbf{P},\ell,\lambda}_{\Lambda^{(n)},\mu}
     = \braket{ \Lambda^{(n)},\mu | \mathbf{P},\ell,\lambda } \,.
     \label{eq:subductioncoefficient}
\end{equation}
The index $\lambda$ denotes helicity when $\mathbf{P}\neq\mathbf{0}$ or otherwise
 the $\mathbf{z}$ component of angular momentum, and ${\cal P}$ denotes the parity
 of the state.
The octahedral group representations $\Lambda$ have their own set of indices $\mu$,
 which implicitly encode the momentum in the case of moving frames.
In some cases, the dimension of the angular momentum irrep may be large enough to
 permit several embeddings of the same irrep, for which the embedding is distinguished
 by the additional index $n$.
These embeddings are unique only up to a unitary change of basis.
From sets of known subduction coefficients, other representations can be built up by
 imposing a tensor product relation,
\begin{widetext}
\begin{equation}
     S^{\mathbf{P},\ell,\lambda}_{\Lambda^{(n)},\mu}
     = N \sum_{\mu_1,\mu_2} \sum_{m_1,m_2}
     \braket{ \Lambda, \mu | \Lambda_1, \mu_1; \Lambda_2, \mu_2 }
     S^{\mathbf{P}_1,\ell_1,\lambda_1}_{\Lambda^{(n_1)}_1,\mu_1}
     S^{\mathbf{P}_2,\ell_2,\lambda_2}_{\Lambda^{(n_2)}_2,\mu_2} 
     \braket{\mathbf{P}_1,\ell_1,\lambda_1;
     \mathbf{P}_2,\ell_2,\lambda_2 | \mathbf{P},\ell,\lambda}
     \label{eq:subductionladder} \,.
\end{equation}
\end{widetext}
This equation may be used to derive the higher-spin subduction coefficients from those
 of lower-spin.

With the expressions in
 Eqns.~(\ref{eq:canonicaltohelicity})~and~(\ref{eq:subductioncoefficient}),
 it is now possible to write the full matrix element needed for Eq.~(\ref{eq:umatrix}),
\begin{align}
     {\cal M}^{\mathbf{P},\Lambda}_{\ell,n;\ell',n'}(q) &=
     \frac{1}{{\rm dim}\Lambda_{\mathbf{P}}}
     \sum_{\mu}
     \sum_{\lambda,\lambda'}
     \sum_{mm'} \nonumber\\
     &\times
     S^{\mathbf{P},\ell,\lambda\;\ast}_{\Lambda^{(n )},\mu}
     S^{\mathbf{P},\ell',\lambda'}_{\Lambda^{(n')},\mu}
     {\cal D}^{\ell\;\ast}_{m \lambda }(\hat{R}_0)
     {\cal D}^{\ell'     }_{m'\lambda'}(\hat{R}_0) \nonumber\\
     &\times
     M^{\mathbf{P}L/2\pi}_{\ell,m;\ell',m'} (q) \,,
     \label{eq:mmatrix}
\end{align}
 where ${\rm dim}\Lambda_{\mathbf{P}}$ gives the dimension of only the elements
 of $\Lambda$ with total momentum $\mathbf{P}$.
Tracing over all irrep indices $\mu$ makes this expression manifestly
 invariant under a change of basis of the lattice
 irreducible representations since under a unitary change of basis,
\begin{equation}
    \sum_\mu S^{\ast}_{\mu} S_{\mu} \to
    \sum_{\mu\bar{\mu}\bar{\mu}'} S^{\ast}_{\bar{\mu}}
    U^{\ast}_{\bar{\mu}\mu} U_{\mu\bar{\mu}'}
    S_{\bar{\mu}'}
    = \sum_\mu S^{\ast}_{\mu} S_{\mu} \,.
\end{equation}
Eq.~(\ref{eq:mmatrix}) respects rotational symmetry,
\begin{align}
    &\sum_{\bar{m}\bar{s}\bar{m}'}
    {\cal D}^{\ell\;\ast}_{\bar{m}m}(\hat{R})
    {\cal D}^{j\;\ast}_{\bar{s}s}(\hat{R})
    {\cal D}^{\ell'}_{\bar{m}'m'}(\hat{R})
    C_{\ell \bar{m},j\bar{s}, \ell^\prime \bar{m}^\prime} \nonumber\\
    &= C_{\ell m,js, \ell^\prime m^\prime}
\end{align}
 and
\begin{equation}
    \sum_{\bar{m}} {\cal D}^{\ell}_{\bar{m}m}(\hat{R})
    {\cal Z}_{\ell \bar{m}}^{\mathbf{d}}\left( s; q \right)
    = {\cal Z}_{\ell m}^{\hat{R}\mathbf{d}}\left( s; q \right)
    = {\cal Z}_{\ell m}^{\mathbf{d}}\left( s; q \right)
\end{equation}
 imply the symmetry
\begin{align}
    {\cal M}^{\mathbf{P},\Lambda}_{\ell,n;\ell',n'}(q)
    &= {\cal M}^{\hat{R}\mathbf{P},\Lambda}_{\ell,n;\ell',n'}(q) \nonumber\\
    =& \frac{1}{{\rm dim}\Lambda_{\mathbf{P}}}
    \sum_{\mu}
    \sum_{\lambda,\lambda'}
    \sum_{mm'} \nonumber\\
    &\times
    S^{\hat{R}\mathbf{P},\ell,\lambda\;\ast}_{\Lambda^{(n )},\mu}
    S^{\hat{R}\mathbf{P},\ell',\lambda'}_{\Lambda^{(n')},\mu} \nonumber\\
    &\times
    {\cal D}^{\ell\;\ast}_{m \lambda }(\hat{R}\hat{R}_0)
    {\cal D}^{\ell'     }_{m'\lambda'}(\hat{R}\hat{R}_0) \nonumber\\
    &\times
    M^{\hat{R}\mathbf{P}L/2\pi}_{\ell,m;\ell',m'} (q) \,,
    \label{eq:rotationalsymmetrycalM}
\end{align}
 with $\hat{R}$ in the octahedral rotation group $O_h$.
This symmetry relates momenta that are connected by lattice rotations,
 so it is not necessary to compute more than one choice of momentum.
In general, the momentum $\mathbf{P}$ cannot be rotated to be parallel to the
 $\hat{z}$ axis, so the expression still depends
 on the direction that $\mathbf{P}$ is oriented with respect to the lattice axes.

The rotational symmetry of Eq.~(\ref{eq:rotationalsymmetrycalM})
 implies that the sum over the irrep row is also not necessary
 to produce a basis-independent result.
This is easily seen by noting that, when the sum over $\mu$ is absent,
 the expression in Eq.~(\ref{eq:mmatrix}) describes a tensor product
 in the octahedral representation,
\begin{align}
  &\sum_{\lambda,\lambda'} \sum_{mm'}
  S^{\mathbf{P},\ell,\lambda\;\ast}_{\Lambda^{(n )},\mu}
  S^{\mathbf{P},\ell',\lambda'}_{\Lambda^{(n')},\mu'} \nonumber\\
  &\times
  {\cal D}^{\ell\;\ast}_{m \lambda }(\hat{R}_0)
  {\cal D}^{\ell'     }_{m'\lambda'}(\hat{R}_0)
  M^{\mathbf{P}L/2\pi}_{\ell,m;\ell',m'} (q)
  \nonumber\\
  =&
  \sum_\sigma \sum_{\nu_\sigma} {\cal M}^{\bar\Lambda\otimes\Lambda\to\sigma}_{\nu_\sigma}
  \langle \Lambda^{(n')},\mu' | \Lambda^{(n)},\mu ; \sigma, \nu_\sigma \rangle
  \nonumber\\
  =&
  {\cal M}^{\bar\Lambda\otimes\Lambda\to A_1}_{0} \delta_{\mu\mu'} +(...),
\end{align}
 where
\begin{align}
 {\cal M}^{\bar\Lambda\otimes\Lambda\to\sigma}_{\nu_\sigma}
 =&\sum_{\mu\mu'} \sum_{\lambda,\lambda'} \sum_{mm'}
  S^{\mathbf{P},\ell,\lambda\;\ast}_{\Lambda^{(n )},\mu}
  S^{\mathbf{P},\ell',\lambda'}_{\Lambda^{(n')},\mu'} \nonumber\\
  &\times
  {\cal D}^{\ell\;\ast}_{m \lambda }(\hat{R}_0)
  {\cal D}^{\ell'     }_{m'\lambda'}(\hat{R}_0)
  \nonumber\\
  &\times
  M^{\mathbf{P}L/2\pi}_{\ell,m;\ell',m'} (q)
  \langle \Lambda^{(n)},\mu ; \sigma, \nu_\sigma | \Lambda^{(n')},\mu' \rangle
\end{align}
 denotes the contribution to ${\cal M}$ in Eq.~(\ref{eq:mmatrix}) from irrep $\sigma$ and
 the ellipses indicate other nontrivial irrep contributions that all vanish in the gauge average.
The trivial $A_1$ irrep has only one index $\nu_{A_1}=0$.
From this point, it is easy to verify that
\begin{align}
  {\cal M}^{\mathbf{P},\Lambda}_{\ell,n;\ell',n'}(q)
  =
  &\sum_{\lambda,\lambda'} \sum_{mm'}
  S^{\mathbf{P},\ell,\lambda\;\ast}_{\Lambda^{(n )},\mu}
  S^{\mathbf{P},\ell',\lambda'}_{\Lambda^{(n')},\mu} \nonumber\\
  &\times
  {\cal D}^{\ell\;\ast}_{m \lambda }(\hat{R}_0)
  {\cal D}^{\ell'     }_{m'\lambda'}(\hat{R}_0)
  M^{\mathbf{P}L/2\pi}_{\ell,m;\ell',m'} (q)
\end{align}
 independent of $\mu$.

When the angular momentum and embedding are fixed with $\ell=\ell'$ and $n=n'$,
 the sum over the product of subduction coefficients forms a projection matrix.
For the cases considered in this manuscript, namely where $\ell=0$
 in the rest frame, this projection is either the identity matrix or zero
 with only one possible embedding $n=n'=0$.
For these cases, the expression in Eq.~(\ref{eq:mmatrix}) simplifies.
Assuming
\begin{equation}
    \sum_{\mu}
    S^{\mathbf{P},\ell,\lambda\;\ast}_{\Lambda,\mu}
    S^{\mathbf{P},\ell,\lambda'}_{\Lambda,\mu}
    = \delta^{\lambda\lambda'} \,,
\end{equation}
 the simplified expression is
\begin{align}
     {\cal M}^{\mathbf{P},\Lambda}_{\ell,0;\ell,0}(q) =&
     \sum_{\lambda}
     \sum_{mm'}
     {\cal D}^{\ell\;\ast}_{m \lambda}(\hat{R}_0)
     {\cal D}^{\ell      }_{m'\lambda}(\hat{R}_0)
     M^{\mathbf{d}}_{\ell,m;\ell,m'} (q) \nonumber\\
     =& \sum_m
     M^{\mathbf{P}L/2\pi}_{\ell,m;\ell,m} (q) \,.
\end{align}

\bibliographystyle{apsrev4-1}
\bibliography{main}

\begin{thebibliography}{68}%
\makeatletter
\providecommand \@ifxundefined [1]{%
 \@ifx{#1\undefined}
}%
\providecommand \@ifnum [1]{%
 \ifnum #1\expandafter \@firstoftwo
 \else \expandafter \@secondoftwo
 \fi
}%
\providecommand \@ifx [1]{%
 \ifx #1\expandafter \@firstoftwo
 \else \expandafter \@secondoftwo
 \fi
}%
\providecommand \natexlab [1]{#1}%
\providecommand \enquote  [1]{``#1''}%
\providecommand \bibnamefont  [1]{#1}%
\providecommand \bibfnamefont [1]{#1}%
\providecommand \citenamefont [1]{#1}%
\providecommand \href@noop [0]{\@secondoftwo}%
\providecommand \href [0]{\begingroup \@sanitize@url \@href}%
\providecommand \@href[1]{\@@startlink{#1}\@@href}%
\providecommand \@@href[1]{\endgroup#1\@@endlink}%
\providecommand \@sanitize@url [0]{\catcode `\\12\catcode `\$12\catcode
  `\&12\catcode `\#12\catcode `\^12\catcode `\_12\catcode `\%12\relax}%
\providecommand \@@startlink[1]{}%
\providecommand \@@endlink[0]{}%
\providecommand \url  [0]{\begingroup\@sanitize@url \@url }%
\providecommand \@url [1]{\endgroup\@href {#1}{\urlprefix }}%
\providecommand \urlprefix  [0]{URL }%
\providecommand \Eprint [0]{\href }%
\providecommand \doibase [0]{http://dx.doi.org/}%
\providecommand \selectlanguage [0]{\@gobble}%
\providecommand \bibinfo  [0]{\@secondoftwo}%
\providecommand \bibfield  [0]{\@secondoftwo}%
\providecommand \translation [1]{[#1]}%
\providecommand \BibitemOpen [0]{}%
\providecommand \bibitemStop [0]{}%
\providecommand \bibitemNoStop [0]{.\EOS\space}%
\providecommand \EOS [0]{\spacefactor3000\relax}%
\providecommand \BibitemShut  [1]{\csname bibitem#1\endcsname}%
\let\auto@bib@innerbib\@empty
\bibitem [{\citenamefont {Batley}\ \emph {et~al.}(2002)\citenamefont {Batley}
  \emph {et~al.}}]{Batley:2002gn}%
  \BibitemOpen
  \bibfield  {author} {\bibinfo {author} {\bibfnamefont {J.~R.}\ \bibnamefont
  {Batley}} \emph {et~al.} (\bibinfo {collaboration} {NA48}),\ }\href {\doibase
  10.1016/S0370-2693(02)02476-0} {\bibfield  {journal} {\bibinfo  {journal}
  {Phys. Lett. B}\ }\textbf {\bibinfo {volume} {544}},\ \bibinfo {pages} {97}
  (\bibinfo {year} {2002})},\ \Eprint {http://arxiv.org/abs/hep-ex/0208009}
  {arXiv:hep-ex/0208009} \BibitemShut {NoStop}%
\bibitem [{\citenamefont {Abouzaid}\ \emph {et~al.}(2011)\citenamefont
  {Abouzaid} \emph {et~al.}}]{Abouzaid:2010ny}%
  \BibitemOpen
  \bibfield  {author} {\bibinfo {author} {\bibfnamefont {E.}~\bibnamefont
  {Abouzaid}} \emph {et~al.} (\bibinfo {collaboration} {KTeV}),\ }\href
  {\doibase 10.1103/PhysRevD.83.092001} {\bibfield  {journal} {\bibinfo
  {journal} {Phys. Rev. D}\ }\textbf {\bibinfo {volume} {83}},\ \bibinfo
  {pages} {092001} (\bibinfo {year} {2011})},\ \Eprint
  {http://arxiv.org/abs/1011.0127} {arXiv:1011.0127 [hep-ex]} \BibitemShut
  {NoStop}%
\bibitem [{\citenamefont {Bai}\ \emph {et~al.}(2015)\citenamefont {Bai} \emph
  {et~al.}}]{Bai:2015nea}%
  \BibitemOpen
  \bibfield  {author} {\bibinfo {author} {\bibfnamefont {Z.}~\bibnamefont
  {Bai}} \emph {et~al.} (\bibinfo {collaboration} {RBC, UKQCD}),\ }\href
  {\doibase 10.1103/PhysRevLett.115.212001} {\bibfield  {journal} {\bibinfo
  {journal} {Phys. Rev. Lett.}\ }\textbf {\bibinfo {volume} {115}},\ \bibinfo
  {pages} {212001} (\bibinfo {year} {2015})},\ \Eprint
  {http://arxiv.org/abs/1505.07863} {arXiv:1505.07863 [hep-lat]} \BibitemShut
  {NoStop}%
\bibitem [{\citenamefont {Abbott}\ \emph {et~al.}(2020)\citenamefont {Abbott}
  \emph {et~al.}}]{Abbott:2020hxn}%
  \BibitemOpen
  \bibfield  {author} {\bibinfo {author} {\bibfnamefont {R.}~\bibnamefont
  {Abbott}} \emph {et~al.} (\bibinfo {collaboration} {RBC, UKQCD}),\ }\href
  {\doibase 10.1103/PhysRevD.102.054509} {\bibfield  {journal} {\bibinfo
  {journal} {Phys. Rev. D}\ }\textbf {\bibinfo {volume} {102}},\ \bibinfo
  {pages} {054509} (\bibinfo {year} {2020})},\ \Eprint
  {http://arxiv.org/abs/2004.09440} {arXiv:2004.09440 [hep-lat]} \BibitemShut
  {NoStop}%
\bibitem [{\citenamefont {Blum}\ \emph {et~al.}(2021)\citenamefont {Blum} \emph
  {et~al.}}]{RBC:2021acc}%
  \BibitemOpen
  \bibfield  {author} {\bibinfo {author} {\bibfnamefont {T.}~\bibnamefont
  {Blum}} \emph {et~al.} (\bibinfo {collaboration} {RBC, UKQCD}),\ }\href
  {\doibase 10.1103/PhysRevD.104.114506} {\bibfield  {journal} {\bibinfo
  {journal} {Phys. Rev. D}\ }\textbf {\bibinfo {volume} {104}},\ \bibinfo
  {pages} {114506} (\bibinfo {year} {2021})},\ \Eprint
  {http://arxiv.org/abs/2103.15131} {arXiv:2103.15131 [hep-lat]} \BibitemShut
  {NoStop}%
\bibitem [{\citenamefont {Lehner}(2016)}]{LehnerTalkLGT16}%
  \BibitemOpen
  \bibfield  {author} {\bibinfo {author} {\bibfnamefont {C.}~\bibnamefont
  {Lehner}},\ }\href
  {https://indico.bnl.gov/getFile.py/access?contribId=1&sessionId=0&resId=0&materialId=slides&confId=1628}
  {\enquote {\bibinfo {title} {The hadronic vacuum polarization contribution to
  the muon anomalous magnetic moment},}\ } (\bibinfo {year} {2016}),\ \bibinfo
  {note} {{RBRC Workshop on Lattice Gauge Theories}}\BibitemShut {NoStop}%
\bibitem [{\citenamefont {Borsanyi}\ \emph {et~al.}(2017)\citenamefont
  {Borsanyi}, \citenamefont {Fodor}, \citenamefont {Kawanai}, \citenamefont
  {Krieg}, \citenamefont {Lellouch}, \citenamefont {Malak}, \citenamefont
  {Miura}, \citenamefont {Szabo}, \citenamefont {Torrero},\ and\ \citenamefont
  {Toth}}]{Borsanyi:2016lpl}%
  \BibitemOpen
  \bibfield  {author} {\bibinfo {author} {\bibfnamefont {S.}~\bibnamefont
  {Borsanyi}}, \bibinfo {author} {\bibfnamefont {Z.}~\bibnamefont {Fodor}},
  \bibinfo {author} {\bibfnamefont {T.}~\bibnamefont {Kawanai}}, \bibinfo
  {author} {\bibfnamefont {S.}~\bibnamefont {Krieg}}, \bibinfo {author}
  {\bibfnamefont {L.}~\bibnamefont {Lellouch}}, \bibinfo {author}
  {\bibfnamefont {R.}~\bibnamefont {Malak}}, \bibinfo {author} {\bibfnamefont
  {K.}~\bibnamefont {Miura}}, \bibinfo {author} {\bibfnamefont {K.~K.}\
  \bibnamefont {Szabo}}, \bibinfo {author} {\bibfnamefont {C.}~\bibnamefont
  {Torrero}}, \ and\ \bibinfo {author} {\bibfnamefont {B.}~\bibnamefont
  {Toth}},\ }\href {\doibase 10.1103/PhysRevD.96.074507} {\bibfield  {journal}
  {\bibinfo  {journal} {Phys. Rev. D}\ }\textbf {\bibinfo {volume} {96}},\
  \bibinfo {pages} {074507} (\bibinfo {year} {2017})},\ \Eprint
  {http://arxiv.org/abs/1612.02364} {arXiv:1612.02364 [hep-lat]} \BibitemShut
  {NoStop}%
\bibitem [{\citenamefont {Bruno}\ \emph {et~al.}(2019)\citenamefont {Bruno},
  \citenamefont {Izubuchi}, \citenamefont {Lehner},\ and\ \citenamefont
  {Meyer}}]{Bruno:2019nzm}%
  \BibitemOpen
  \bibfield  {author} {\bibinfo {author} {\bibfnamefont {M.}~\bibnamefont
  {Bruno}}, \bibinfo {author} {\bibfnamefont {T.}~\bibnamefont {Izubuchi}},
  \bibinfo {author} {\bibfnamefont {C.}~\bibnamefont {Lehner}}, \ and\ \bibinfo
  {author} {\bibfnamefont {A.~S.}\ \bibnamefont {Meyer}},\ }\href {\doibase
  10.22323/1.363.0239} {\bibfield  {journal} {\bibinfo  {journal} {PoS}\
  }\textbf {\bibinfo {volume} {LATTICE2019}},\ \bibinfo {pages} {239} (\bibinfo
  {year} {2019})},\ \Eprint {http://arxiv.org/abs/1910.11745} {arXiv:1910.11745
  [hep-lat]} \BibitemShut {NoStop}%
\bibitem [{\citenamefont {Aoyama}\ \emph {et~al.}(2020)\citenamefont {Aoyama}
  \emph {et~al.}}]{Aoyama:2020ynm}%
  \BibitemOpen
  \bibfield  {author} {\bibinfo {author} {\bibfnamefont {T.}~\bibnamefont
  {Aoyama}} \emph {et~al.},\ }\href {\doibase 10.1016/j.physrep.2020.07.006}
  {\bibfield  {journal} {\bibinfo  {journal} {Phys. Rept.}\ }\textbf {\bibinfo
  {volume} {887}},\ \bibinfo {pages} {1} (\bibinfo {year} {2020})},\ \Eprint
  {http://arxiv.org/abs/2006.04822} {arXiv:2006.04822 [hep-ph]} \BibitemShut
  {NoStop}%
\bibitem [{\citenamefont {Kronfeld}\ \emph {et~al.}(2019)\citenamefont
  {Kronfeld}, \citenamefont {Richards}, \citenamefont {Detmold}, \citenamefont
  {Gupta}, \citenamefont {Lin}, \citenamefont {Liu}, \citenamefont {Meyer},
  \citenamefont {Sufian},\ and\ \citenamefont {Syritsyn}}]{Kronfeld:2019nfb}%
  \BibitemOpen
  \bibfield  {author} {\bibinfo {author} {\bibfnamefont {A.~S.}\ \bibnamefont
  {Kronfeld}}, \bibinfo {author} {\bibfnamefont {D.~G.}\ \bibnamefont
  {Richards}}, \bibinfo {author} {\bibfnamefont {W.}~\bibnamefont {Detmold}},
  \bibinfo {author} {\bibfnamefont {R.}~\bibnamefont {Gupta}}, \bibinfo
  {author} {\bibfnamefont {H.-W.}\ \bibnamefont {Lin}}, \bibinfo {author}
  {\bibfnamefont {K.-F.}\ \bibnamefont {Liu}}, \bibinfo {author} {\bibfnamefont
  {A.~S.}\ \bibnamefont {Meyer}}, \bibinfo {author} {\bibfnamefont
  {R.}~\bibnamefont {Sufian}}, \ and\ \bibinfo {author} {\bibfnamefont
  {S.}~\bibnamefont {Syritsyn}} (\bibinfo {collaboration} {USQCD}),\ }\href
  {\doibase 10.1140/epja/i2019-12916-x} {\bibfield  {journal} {\bibinfo
  {journal} {Eur. Phys. J. A}\ }\textbf {\bibinfo {volume} {55}},\ \bibinfo
  {pages} {196} (\bibinfo {year} {2019})},\ \Eprint
  {http://arxiv.org/abs/1904.09931} {arXiv:1904.09931 [hep-lat]} \BibitemShut
  {NoStop}%
\bibitem [{\citenamefont {Luscher}(1986)}]{Luscher:1986pf}%
  \BibitemOpen
  \bibfield  {author} {\bibinfo {author} {\bibfnamefont {M.}~\bibnamefont
  {Luscher}},\ }\href {\doibase 10.1007/BF01211097} {\bibfield  {journal}
  {\bibinfo  {journal} {Commun. Math. Phys.}\ }\textbf {\bibinfo {volume}
  {105}},\ \bibinfo {pages} {153} (\bibinfo {year} {1986})}\BibitemShut
  {NoStop}%
\bibitem [{\citenamefont {Luscher}(1991{\natexlab{a}})}]{Luscher:1990ux}%
  \BibitemOpen
  \bibfield  {author} {\bibinfo {author} {\bibfnamefont {M.}~\bibnamefont
  {Luscher}},\ }\href {\doibase 10.1016/0550-3213(91)90366-6} {\bibfield
  {journal} {\bibinfo  {journal} {Nucl. Phys.}\ }\textbf {\bibinfo {volume}
  {B354}},\ \bibinfo {pages} {531} (\bibinfo {year}
  {1991}{\natexlab{a}})}\BibitemShut {NoStop}%
\bibitem [{\citenamefont {Luscher}(1991{\natexlab{b}})}]{Luscher:1991cf}%
  \BibitemOpen
  \bibfield  {author} {\bibinfo {author} {\bibfnamefont {M.}~\bibnamefont
  {Luscher}},\ }\href {\doibase 10.1016/0550-3213(91)90584-K} {\bibfield
  {journal} {\bibinfo  {journal} {Nucl. Phys.}\ }\textbf {\bibinfo {volume}
  {B364}},\ \bibinfo {pages} {237} (\bibinfo {year}
  {1991}{\natexlab{b}})}\BibitemShut {NoStop}%
\bibitem [{\citenamefont {Rummukainen}\ and\ \citenamefont
  {Gottlieb}(1995)}]{Rummukainen:1995vs}%
  \BibitemOpen
  \bibfield  {author} {\bibinfo {author} {\bibfnamefont {K.}~\bibnamefont
  {Rummukainen}}\ and\ \bibinfo {author} {\bibfnamefont {S.~A.}\ \bibnamefont
  {Gottlieb}},\ }\href {\doibase 10.1016/0550-3213(95)00313-H} {\bibfield
  {journal} {\bibinfo  {journal} {Nucl. Phys.}\ }\textbf {\bibinfo {volume}
  {B450}},\ \bibinfo {pages} {397} (\bibinfo {year} {1995})},\ \Eprint
  {http://arxiv.org/abs/hep-lat/9503028} {arXiv:hep-lat/9503028 [hep-lat]}
  \BibitemShut {NoStop}%
\bibitem [{\citenamefont {Leskovec}\ and\ \citenamefont
  {Prelovsek}(2012)}]{Leskovec:2012gb}%
  \BibitemOpen
  \bibfield  {author} {\bibinfo {author} {\bibfnamefont {L.}~\bibnamefont
  {Leskovec}}\ and\ \bibinfo {author} {\bibfnamefont {S.}~\bibnamefont
  {Prelovsek}},\ }\href {\doibase 10.1103/PhysRevD.85.114507} {\bibfield
  {journal} {\bibinfo  {journal} {Phys. Rev. D}\ }\textbf {\bibinfo {volume}
  {85}},\ \bibinfo {pages} {114507} (\bibinfo {year} {2012})},\ \Eprint
  {http://arxiv.org/abs/1202.2145} {arXiv:1202.2145 [hep-lat]} \BibitemShut
  {NoStop}%
\bibitem [{\citenamefont {Guo}\ \emph {et~al.}(2013)\citenamefont {Guo},
  \citenamefont {Dudek}, \citenamefont {Edwards},\ and\ \citenamefont
  {Szczepaniak}}]{Guo:2012hv}%
  \BibitemOpen
  \bibfield  {author} {\bibinfo {author} {\bibfnamefont {P.}~\bibnamefont
  {Guo}}, \bibinfo {author} {\bibfnamefont {J.}~\bibnamefont {Dudek}}, \bibinfo
  {author} {\bibfnamefont {R.}~\bibnamefont {Edwards}}, \ and\ \bibinfo
  {author} {\bibfnamefont {A.~P.}\ \bibnamefont {Szczepaniak}},\ }\href
  {\doibase 10.1103/PhysRevD.88.014501} {\bibfield  {journal} {\bibinfo
  {journal} {Phys. Rev. D}\ }\textbf {\bibinfo {volume} {88}},\ \bibinfo
  {pages} {014501} (\bibinfo {year} {2013})},\ \Eprint
  {http://arxiv.org/abs/1211.0929} {arXiv:1211.0929 [hep-lat]} \BibitemShut
  {NoStop}%
\bibitem [{\citenamefont {Woss}\ \emph {et~al.}(2018)\citenamefont {Woss},
  \citenamefont {Thomas}, \citenamefont {Dudek}, \citenamefont {Edwards},\ and\
  \citenamefont {Wilson}}]{Woss:2018irj}%
  \BibitemOpen
  \bibfield  {author} {\bibinfo {author} {\bibfnamefont {A.}~\bibnamefont
  {Woss}}, \bibinfo {author} {\bibfnamefont {C.~E.}\ \bibnamefont {Thomas}},
  \bibinfo {author} {\bibfnamefont {J.~J.}\ \bibnamefont {Dudek}}, \bibinfo
  {author} {\bibfnamefont {R.~G.}\ \bibnamefont {Edwards}}, \ and\ \bibinfo
  {author} {\bibfnamefont {D.~J.}\ \bibnamefont {Wilson}},\ }\href {\doibase
  10.1007/JHEP07(2018)043} {\bibfield  {journal} {\bibinfo  {journal} {JHEP}\
  }\textbf {\bibinfo {volume} {07}},\ \bibinfo {pages} {043} (\bibinfo {year}
  {2018})},\ \Eprint {http://arxiv.org/abs/1802.05580} {arXiv:1802.05580
  [hep-lat]} \BibitemShut {NoStop}%
\bibitem [{\citenamefont {Woss}(2019)}]{Woss:2019dnv}%
  \BibitemOpen
  \bibfield  {author} {\bibinfo {author} {\bibfnamefont {A.~J.}\ \bibnamefont
  {Woss}},\ }\emph {\bibinfo {title} {{The scattering of spinning hadrons from
  lattice QCD}}},\ \href {\doibase 10.17863/CAM.45966} {Ph.D. thesis},\
  \bibinfo  {school} {Cambridge U., DAMTP} (\bibinfo {year} {2019})\BibitemShut
  {NoStop}%
\bibitem [{\citenamefont {Hansen}\ and\ \citenamefont
  {Sharpe}(2019)}]{Hansen:2019nir}%
  \BibitemOpen
  \bibfield  {author} {\bibinfo {author} {\bibfnamefont {M.~T.}\ \bibnamefont
  {Hansen}}\ and\ \bibinfo {author} {\bibfnamefont {S.~R.}\ \bibnamefont
  {Sharpe}},\ }\href {\doibase 10.1146/annurev-nucl-101918-023723} {\bibfield
  {journal} {\bibinfo  {journal} {Ann. Rev. Nucl. Part. Sci.}\ }\textbf
  {\bibinfo {volume} {69}},\ \bibinfo {pages} {65} (\bibinfo {year} {2019})},\
  \Eprint {http://arxiv.org/abs/1901.00483} {arXiv:1901.00483 [hep-lat]}
  \BibitemShut {NoStop}%
\bibitem [{\citenamefont {Rusetsky}(2019)}]{Rusetsky:2019gyk}%
  \BibitemOpen
  \bibfield  {author} {\bibinfo {author} {\bibfnamefont {A.}~\bibnamefont
  {Rusetsky}},\ }\href {\doibase 10.22323/1.363.0281} {\bibfield  {journal}
  {\bibinfo  {journal} {PoS}\ }\textbf {\bibinfo {volume} {LATTICE2019}},\
  \bibinfo {pages} {281} (\bibinfo {year} {2019})},\ \Eprint
  {http://arxiv.org/abs/1911.01253} {arXiv:1911.01253 [hep-lat]} \BibitemShut
  {NoStop}%
\bibitem [{\citenamefont {Mai}\ \emph {et~al.}(2021)\citenamefont {Mai},
  \citenamefont {D\"oring},\ and\ \citenamefont {Rusetsky}}]{Mai:2021lwb}%
  \BibitemOpen
  \bibfield  {author} {\bibinfo {author} {\bibfnamefont {M.}~\bibnamefont
  {Mai}}, \bibinfo {author} {\bibfnamefont {M.}~\bibnamefont {D\"oring}}, \
  and\ \bibinfo {author} {\bibfnamefont {A.}~\bibnamefont {Rusetsky}},\ }\href
  {\doibase 10.1140/epjs/s11734-021-00146-5} {\bibfield  {journal} {\bibinfo
  {journal} {Eur. Phys. J. ST}\ }\textbf {\bibinfo {volume} {230}},\ \bibinfo
  {pages} {1623} (\bibinfo {year} {2021})},\ \Eprint
  {http://arxiv.org/abs/2103.00577} {arXiv:2103.00577 [hep-lat]} \BibitemShut
  {NoStop}%
\bibitem [{\citenamefont {H\"orz}\ and\ \citenamefont
  {Hanlon}(2019)}]{Horz:2019rrn}%
  \BibitemOpen
  \bibfield  {author} {\bibinfo {author} {\bibfnamefont {B.}~\bibnamefont
  {H\"orz}}\ and\ \bibinfo {author} {\bibfnamefont {A.}~\bibnamefont
  {Hanlon}},\ }\href {\doibase 10.1103/PhysRevLett.123.142002} {\bibfield
  {journal} {\bibinfo  {journal} {Phys. Rev. Lett.}\ }\textbf {\bibinfo
  {volume} {123}},\ \bibinfo {pages} {142002} (\bibinfo {year} {2019})},\
  \Eprint {http://arxiv.org/abs/1905.04277} {arXiv:1905.04277 [hep-lat]}
  \BibitemShut {NoStop}%
\bibitem [{\citenamefont {Fischer}\ \emph
  {et~al.}(2020{\natexlab{a}})\citenamefont {Fischer}, \citenamefont
  {Kostrzewa}, \citenamefont {Liu}, \citenamefont {Romero-L\'opez},
  \citenamefont {Ueding},\ and\ \citenamefont {Urbach}}]{Fischer:2020jzp}%
  \BibitemOpen
  \bibfield  {author} {\bibinfo {author} {\bibfnamefont {M.}~\bibnamefont
  {Fischer}}, \bibinfo {author} {\bibfnamefont {B.}~\bibnamefont {Kostrzewa}},
  \bibinfo {author} {\bibfnamefont {L.}~\bibnamefont {Liu}}, \bibinfo {author}
  {\bibfnamefont {F.}~\bibnamefont {Romero-L\'opez}}, \bibinfo {author}
  {\bibfnamefont {M.}~\bibnamefont {Ueding}}, \ and\ \bibinfo {author}
  {\bibfnamefont {C.}~\bibnamefont {Urbach}},\ }\href@noop {} {\  (\bibinfo
  {year} {2020}{\natexlab{a}})},\ \Eprint {http://arxiv.org/abs/2008.03035}
  {arXiv:2008.03035 [hep-lat]} \BibitemShut {NoStop}%
\bibitem [{\citenamefont {Ishii}\ \emph {et~al.}(2007)\citenamefont {Ishii},
  \citenamefont {Aoki},\ and\ \citenamefont {Hatsuda}}]{Ishii:2006ec}%
  \BibitemOpen
  \bibfield  {author} {\bibinfo {author} {\bibfnamefont {N.}~\bibnamefont
  {Ishii}}, \bibinfo {author} {\bibfnamefont {S.}~\bibnamefont {Aoki}}, \ and\
  \bibinfo {author} {\bibfnamefont {T.}~\bibnamefont {Hatsuda}},\ }\href
  {\doibase 10.1103/PhysRevLett.99.022001} {\bibfield  {journal} {\bibinfo
  {journal} {Phys. Rev. Lett.}\ }\textbf {\bibinfo {volume} {99}},\ \bibinfo
  {pages} {022001} (\bibinfo {year} {2007})},\ \Eprint
  {http://arxiv.org/abs/nucl-th/0611096} {arXiv:nucl-th/0611096} \BibitemShut
  {NoStop}%
\bibitem [{\citenamefont {Aoki}\ \emph {et~al.}(2010)\citenamefont {Aoki},
  \citenamefont {Hatsuda},\ and\ \citenamefont {Ishii}}]{Aoki:2009ji}%
  \BibitemOpen
  \bibfield  {author} {\bibinfo {author} {\bibfnamefont {S.}~\bibnamefont
  {Aoki}}, \bibinfo {author} {\bibfnamefont {T.}~\bibnamefont {Hatsuda}}, \
  and\ \bibinfo {author} {\bibfnamefont {N.}~\bibnamefont {Ishii}},\ }\href
  {\doibase 10.1143/PTP.123.89} {\bibfield  {journal} {\bibinfo  {journal}
  {Prog. Theor. Phys.}\ }\textbf {\bibinfo {volume} {123}},\ \bibinfo {pages}
  {89} (\bibinfo {year} {2010})},\ \Eprint {http://arxiv.org/abs/0909.5585}
  {arXiv:0909.5585 [hep-lat]} \BibitemShut {NoStop}%
\bibitem [{\citenamefont {Ishii}\ \emph {et~al.}(2012)\citenamefont {Ishii},
  \citenamefont {Aoki}, \citenamefont {Doi}, \citenamefont {Hatsuda},
  \citenamefont {Ikeda}, \citenamefont {Inoue}, \citenamefont {Murano},
  \citenamefont {Nemura},\ and\ \citenamefont {Sasaki}}]{HALQCD:2012aa}%
  \BibitemOpen
  \bibfield  {author} {\bibinfo {author} {\bibfnamefont {N.}~\bibnamefont
  {Ishii}}, \bibinfo {author} {\bibfnamefont {S.}~\bibnamefont {Aoki}},
  \bibinfo {author} {\bibfnamefont {T.}~\bibnamefont {Doi}}, \bibinfo {author}
  {\bibfnamefont {T.}~\bibnamefont {Hatsuda}}, \bibinfo {author} {\bibfnamefont
  {Y.}~\bibnamefont {Ikeda}}, \bibinfo {author} {\bibfnamefont
  {T.}~\bibnamefont {Inoue}}, \bibinfo {author} {\bibfnamefont
  {K.}~\bibnamefont {Murano}}, \bibinfo {author} {\bibfnamefont
  {H.}~\bibnamefont {Nemura}}, \ and\ \bibinfo {author} {\bibfnamefont
  {K.}~\bibnamefont {Sasaki}} (\bibinfo {collaboration} {HAL QCD}),\ }\href
  {\doibase 10.1016/j.physletb.2012.04.076} {\bibfield  {journal} {\bibinfo
  {journal} {Phys. Lett. B}\ }\textbf {\bibinfo {volume} {712}},\ \bibinfo
  {pages} {437} (\bibinfo {year} {2012})},\ \Eprint
  {http://arxiv.org/abs/1203.3642} {arXiv:1203.3642 [hep-lat]} \BibitemShut
  {NoStop}%
\bibitem [{\citenamefont {Kawai}\ \emph {et~al.}(2018)\citenamefont {Kawai},
  \citenamefont {Aoki}, \citenamefont {Doi}, \citenamefont {Ikeda},
  \citenamefont {Inoue}, \citenamefont {Iritani}, \citenamefont {Ishii},
  \citenamefont {Miyamoto}, \citenamefont {Nemura},\ and\ \citenamefont
  {Sasaki}}]{Kawai:2017goq}%
  \BibitemOpen
  \bibfield  {author} {\bibinfo {author} {\bibfnamefont {D.}~\bibnamefont
  {Kawai}}, \bibinfo {author} {\bibfnamefont {S.}~\bibnamefont {Aoki}},
  \bibinfo {author} {\bibfnamefont {T.}~\bibnamefont {Doi}}, \bibinfo {author}
  {\bibfnamefont {Y.}~\bibnamefont {Ikeda}}, \bibinfo {author} {\bibfnamefont
  {T.}~\bibnamefont {Inoue}}, \bibinfo {author} {\bibfnamefont
  {T.}~\bibnamefont {Iritani}}, \bibinfo {author} {\bibfnamefont
  {N.}~\bibnamefont {Ishii}}, \bibinfo {author} {\bibfnamefont
  {T.}~\bibnamefont {Miyamoto}}, \bibinfo {author} {\bibfnamefont
  {H.}~\bibnamefont {Nemura}}, \ and\ \bibinfo {author} {\bibfnamefont
  {K.}~\bibnamefont {Sasaki}} (\bibinfo {collaboration} {HAL QCD}),\ }\href
  {\doibase 10.1093/ptep/pty032} {\bibfield  {journal} {\bibinfo  {journal}
  {PTEP}\ }\textbf {\bibinfo {volume} {2018}},\ \bibinfo {pages} {043B04}
  (\bibinfo {year} {2018})},\ \Eprint {http://arxiv.org/abs/1711.01883}
  {arXiv:1711.01883 [hep-lat]} \BibitemShut {NoStop}%
\bibitem [{\citenamefont {Peardon}\ \emph {et~al.}(2009)\citenamefont
  {Peardon}, \citenamefont {Bulava}, \citenamefont {Foley}, \citenamefont
  {Morningstar}, \citenamefont {Dudek}, \citenamefont {Edwards}, \citenamefont
  {Joo}, \citenamefont {Lin}, \citenamefont {Richards},\ and\ \citenamefont
  {Juge}}]{Peardon:2009gh}%
  \BibitemOpen
  \bibfield  {author} {\bibinfo {author} {\bibfnamefont {M.}~\bibnamefont
  {Peardon}}, \bibinfo {author} {\bibfnamefont {J.}~\bibnamefont {Bulava}},
  \bibinfo {author} {\bibfnamefont {J.}~\bibnamefont {Foley}}, \bibinfo
  {author} {\bibfnamefont {C.}~\bibnamefont {Morningstar}}, \bibinfo {author}
  {\bibfnamefont {J.}~\bibnamefont {Dudek}}, \bibinfo {author} {\bibfnamefont
  {R.~G.}\ \bibnamefont {Edwards}}, \bibinfo {author} {\bibfnamefont
  {B.}~\bibnamefont {Joo}}, \bibinfo {author} {\bibfnamefont {H.-W.}\
  \bibnamefont {Lin}}, \bibinfo {author} {\bibfnamefont {D.~G.}\ \bibnamefont
  {Richards}}, \ and\ \bibinfo {author} {\bibfnamefont {K.~J.}\ \bibnamefont
  {Juge}} (\bibinfo {collaboration} {Hadron Spectrum}),\ }\href {\doibase
  10.1103/PhysRevD.80.054506} {\bibfield  {journal} {\bibinfo  {journal} {Phys.
  Rev.}\ }\textbf {\bibinfo {volume} {D80}},\ \bibinfo {pages} {054506}
  (\bibinfo {year} {2009})},\ \Eprint {http://arxiv.org/abs/0905.2160}
  {arXiv:0905.2160 [hep-lat]} \BibitemShut {NoStop}%
\bibitem [{\citenamefont {Blossier}\ \emph {et~al.}(2009)\citenamefont
  {Blossier}, \citenamefont {Della~Morte}, \citenamefont {von Hippel},
  \citenamefont {Mendes},\ and\ \citenamefont {Sommer}}]{Blossier:2009kd}%
  \BibitemOpen
  \bibfield  {author} {\bibinfo {author} {\bibfnamefont {B.}~\bibnamefont
  {Blossier}}, \bibinfo {author} {\bibfnamefont {M.}~\bibnamefont
  {Della~Morte}}, \bibinfo {author} {\bibfnamefont {G.}~\bibnamefont {von
  Hippel}}, \bibinfo {author} {\bibfnamefont {T.}~\bibnamefont {Mendes}}, \
  and\ \bibinfo {author} {\bibfnamefont {R.}~\bibnamefont {Sommer}},\ }\href
  {\doibase 10.1088/1126-6708/2009/04/094} {\bibfield  {journal} {\bibinfo
  {journal} {JHEP}\ }\textbf {\bibinfo {volume} {04}},\ \bibinfo {pages} {094}
  (\bibinfo {year} {2009})},\ \Eprint {http://arxiv.org/abs/0902.1265}
  {arXiv:0902.1265 [hep-lat]} \BibitemShut {NoStop}%
\bibitem [{\citenamefont {Bulava}\ \emph {et~al.}(2014)\citenamefont {Bulava},
  \citenamefont {Fahy}, \citenamefont {Foley}, \citenamefont {Jhang},
  \citenamefont {Juge}, \citenamefont {Lenkner}, \citenamefont {Morningstar},\
  and\ \citenamefont {Wong}}]{Bulava:2014vua}%
  \BibitemOpen
  \bibfield  {author} {\bibinfo {author} {\bibfnamefont {J.}~\bibnamefont
  {Bulava}}, \bibinfo {author} {\bibfnamefont {B.}~\bibnamefont {Fahy}},
  \bibinfo {author} {\bibfnamefont {J.}~\bibnamefont {Foley}}, \bibinfo
  {author} {\bibfnamefont {Y.-C.}\ \bibnamefont {Jhang}}, \bibinfo {author}
  {\bibfnamefont {K.~J.}\ \bibnamefont {Juge}}, \bibinfo {author}
  {\bibfnamefont {D.}~\bibnamefont {Lenkner}}, \bibinfo {author} {\bibfnamefont
  {C.}~\bibnamefont {Morningstar}}, \ and\ \bibinfo {author} {\bibfnamefont
  {C.~H.}\ \bibnamefont {Wong}},\ }\href {\doibase 10.1051/epjconf/20147303018}
  {\bibfield  {journal} {\bibinfo  {journal} {EPJ Web Conf.}\ }\textbf
  {\bibinfo {volume} {73}},\ \bibinfo {pages} {03018} (\bibinfo {year}
  {2014})},\ \Eprint {http://arxiv.org/abs/1401.2544} {arXiv:1401.2544
  [hep-lat]} \BibitemShut {NoStop}%
\bibitem [{\citenamefont {Fischer}\ \emph
  {et~al.}(2020{\natexlab{b}})\citenamefont {Fischer}, \citenamefont
  {Kostrzewa}, \citenamefont {Ostmeyer}, \citenamefont {Ottnad}, \citenamefont
  {Ueding},\ and\ \citenamefont {Urbach}}]{Fischer:2020bgv}%
  \BibitemOpen
  \bibfield  {author} {\bibinfo {author} {\bibfnamefont {M.}~\bibnamefont
  {Fischer}}, \bibinfo {author} {\bibfnamefont {B.}~\bibnamefont {Kostrzewa}},
  \bibinfo {author} {\bibfnamefont {J.}~\bibnamefont {Ostmeyer}}, \bibinfo
  {author} {\bibfnamefont {K.}~\bibnamefont {Ottnad}}, \bibinfo {author}
  {\bibfnamefont {M.}~\bibnamefont {Ueding}}, \ and\ \bibinfo {author}
  {\bibfnamefont {C.}~\bibnamefont {Urbach}},\ }\href {\doibase
  10.1140/epja/s10050-020-00205-w} {\bibfield  {journal} {\bibinfo  {journal}
  {Eur. Phys. J. A}\ }\textbf {\bibinfo {volume} {56}},\ \bibinfo {pages} {206}
  (\bibinfo {year} {2020}{\natexlab{b}})},\ \Eprint
  {http://arxiv.org/abs/2004.10472} {arXiv:2004.10472 [hep-lat]} \BibitemShut
  {NoStop}%
\bibitem [{\citenamefont {Dudek}\ \emph {et~al.}(2012)\citenamefont {Dudek},
  \citenamefont {Edwards},\ and\ \citenamefont {Thomas}}]{Dudek:2012gj}%
  \BibitemOpen
  \bibfield  {author} {\bibinfo {author} {\bibfnamefont {J.~J.}\ \bibnamefont
  {Dudek}}, \bibinfo {author} {\bibfnamefont {R.~G.}\ \bibnamefont {Edwards}},
  \ and\ \bibinfo {author} {\bibfnamefont {C.~E.}\ \bibnamefont {Thomas}},\
  }\href {\doibase 10.1103/PhysRevD.86.034031} {\bibfield  {journal} {\bibinfo
  {journal} {Phys. Rev.}\ }\textbf {\bibinfo {volume} {D86}},\ \bibinfo {pages}
  {034031} (\bibinfo {year} {2012})},\ \Eprint {http://arxiv.org/abs/1203.6041}
  {arXiv:1203.6041 [hep-ph]} \BibitemShut {NoStop}%
\bibitem [{\citenamefont {Sharpe}\ \emph {et~al.}(1992)\citenamefont {Sharpe},
  \citenamefont {Gupta},\ and\ \citenamefont {Kilcup}}]{Sharpe:1992pp}%
  \BibitemOpen
  \bibfield  {author} {\bibinfo {author} {\bibfnamefont {S.~R.}\ \bibnamefont
  {Sharpe}}, \bibinfo {author} {\bibfnamefont {R.}~\bibnamefont {Gupta}}, \
  and\ \bibinfo {author} {\bibfnamefont {G.~W.}\ \bibnamefont {Kilcup}},\
  }\href {\doibase 10.1016/0550-3213(92)90681-Z} {\bibfield  {journal}
  {\bibinfo  {journal} {Nucl. Phys. B}\ }\textbf {\bibinfo {volume} {383}},\
  \bibinfo {pages} {309} (\bibinfo {year} {1992})}\BibitemShut {NoStop}%
\bibitem [{\citenamefont {Gupta}\ \emph {et~al.}(1993)\citenamefont {Gupta},
  \citenamefont {Patel},\ and\ \citenamefont {Sharpe}}]{Gupta:1993rn}%
  \BibitemOpen
  \bibfield  {author} {\bibinfo {author} {\bibfnamefont {R.}~\bibnamefont
  {Gupta}}, \bibinfo {author} {\bibfnamefont {A.}~\bibnamefont {Patel}}, \ and\
  \bibinfo {author} {\bibfnamefont {S.~R.}\ \bibnamefont {Sharpe}},\ }\href
  {\doibase 10.1103/PhysRevD.48.388} {\bibfield  {journal} {\bibinfo  {journal}
  {Phys. Rev. D}\ }\textbf {\bibinfo {volume} {48}},\ \bibinfo {pages} {388}
  (\bibinfo {year} {1993})},\ \Eprint {http://arxiv.org/abs/hep-lat/9301016}
  {arXiv:hep-lat/9301016} \BibitemShut {NoStop}%
\bibitem [{\citenamefont {Kuramashi}\ \emph {et~al.}(1993)\citenamefont
  {Kuramashi}, \citenamefont {Fukugita}, \citenamefont {Mino}, \citenamefont
  {Okawa},\ and\ \citenamefont {Ukawa}}]{Kuramashi:1993ka}%
  \BibitemOpen
  \bibfield  {author} {\bibinfo {author} {\bibfnamefont {Y.}~\bibnamefont
  {Kuramashi}}, \bibinfo {author} {\bibfnamefont {M.}~\bibnamefont {Fukugita}},
  \bibinfo {author} {\bibfnamefont {H.}~\bibnamefont {Mino}}, \bibinfo {author}
  {\bibfnamefont {M.}~\bibnamefont {Okawa}}, \ and\ \bibinfo {author}
  {\bibfnamefont {A.}~\bibnamefont {Ukawa}},\ }\href {\doibase
  10.1103/PhysRevLett.71.2387} {\bibfield  {journal} {\bibinfo  {journal}
  {Phys. Rev. Lett.}\ }\textbf {\bibinfo {volume} {71}},\ \bibinfo {pages}
  {2387} (\bibinfo {year} {1993})}\BibitemShut {NoStop}%
\bibitem [{\citenamefont {Tu}(2020)}]{Tu:2020vpn}%
  \BibitemOpen
  \bibfield  {author} {\bibinfo {author} {\bibfnamefont {J.}~\bibnamefont
  {Tu}},\ }\emph {\bibinfo {title} {{Lattice QCD Simulations towards Strong and
  Weak Coupling Limits}}},\ \href {\doibase 10.7916/d8-74hj-ak60} {Ph.D.
  thesis},\ \bibinfo  {school} {Columbia U.} (\bibinfo {year}
  {2020})\BibitemShut {NoStop}%
\bibitem [{\citenamefont {Blum}\ \emph
  {et~al.}(2016{\natexlab{a}})\citenamefont {Blum} \emph
  {et~al.}}]{RBC:2014ntl}%
  \BibitemOpen
  \bibfield  {author} {\bibinfo {author} {\bibfnamefont {T.}~\bibnamefont
  {Blum}} \emph {et~al.} (\bibinfo {collaboration} {RBC, UKQCD}),\ }\href
  {\doibase 10.1103/PhysRevD.93.074505} {\bibfield  {journal} {\bibinfo
  {journal} {Phys. Rev. D}\ }\textbf {\bibinfo {volume} {93}},\ \bibinfo
  {pages} {074505} (\bibinfo {year} {2016}{\natexlab{a}})},\ \Eprint
  {http://arxiv.org/abs/1411.7017} {arXiv:1411.7017 [hep-lat]} \BibitemShut
  {NoStop}%
\bibitem [{\citenamefont {Vranas}(2000)}]{Vranas:1999rz}%
  \BibitemOpen
  \bibfield  {author} {\bibinfo {author} {\bibfnamefont {P.~M.}\ \bibnamefont
  {Vranas}},\ }\href@noop {} {\bibfield  {journal} {\bibinfo  {journal} {NATO
  Sci. Ser. C}\ }\textbf {\bibinfo {volume} {553}},\ \bibinfo {pages} {11}
  (\bibinfo {year} {2000})},\ \Eprint {http://arxiv.org/abs/hep-lat/0001006}
  {arXiv:hep-lat/0001006} \BibitemShut {NoStop}%
\bibitem [{\citenamefont {Vranas}(2006)}]{Vranas:2006zk}%
  \BibitemOpen
  \bibfield  {author} {\bibinfo {author} {\bibfnamefont {P.~M.}\ \bibnamefont
  {Vranas}},\ }\href {\doibase 10.1103/PhysRevD.74.034512} {\bibfield
  {journal} {\bibinfo  {journal} {Phys. Rev. D}\ }\textbf {\bibinfo {volume}
  {74}},\ \bibinfo {pages} {034512} (\bibinfo {year} {2006})},\ \Eprint
  {http://arxiv.org/abs/hep-lat/0606014} {arXiv:hep-lat/0606014} \BibitemShut
  {NoStop}%
\bibitem [{\citenamefont {Fukaya}\ \emph {et~al.}(2006)\citenamefont {Fukaya},
  \citenamefont {Hashimoto}, \citenamefont {Ishikawa}, \citenamefont {Kaneko},
  \citenamefont {Matsufuru}, \citenamefont {Onogi},\ and\ \citenamefont
  {Yamada}}]{Fukaya:2006vs}%
  \BibitemOpen
  \bibfield  {author} {\bibinfo {author} {\bibfnamefont {H.}~\bibnamefont
  {Fukaya}}, \bibinfo {author} {\bibfnamefont {S.}~\bibnamefont {Hashimoto}},
  \bibinfo {author} {\bibfnamefont {K.-I.}\ \bibnamefont {Ishikawa}}, \bibinfo
  {author} {\bibfnamefont {T.}~\bibnamefont {Kaneko}}, \bibinfo {author}
  {\bibfnamefont {H.}~\bibnamefont {Matsufuru}}, \bibinfo {author}
  {\bibfnamefont {T.}~\bibnamefont {Onogi}}, \ and\ \bibinfo {author}
  {\bibfnamefont {N.}~\bibnamefont {Yamada}} (\bibinfo {collaboration}
  {JLQCD}),\ }\href {\doibase 10.1103/PhysRevD.74.094505} {\bibfield  {journal}
  {\bibinfo  {journal} {Phys. Rev. D}\ }\textbf {\bibinfo {volume} {74}},\
  \bibinfo {pages} {094505} (\bibinfo {year} {2006})},\ \Eprint
  {http://arxiv.org/abs/hep-lat/0607020} {arXiv:hep-lat/0607020} \BibitemShut
  {NoStop}%
\bibitem [{\citenamefont {Renfrew}\ \emph {et~al.}(2008)\citenamefont
  {Renfrew}, \citenamefont {Blum}, \citenamefont {Christ}, \citenamefont
  {Mawhinney},\ and\ \citenamefont {Vranas}}]{Renfrew:2008zfx}%
  \BibitemOpen
  \bibfield  {author} {\bibinfo {author} {\bibfnamefont {D.}~\bibnamefont
  {Renfrew}}, \bibinfo {author} {\bibfnamefont {T.}~\bibnamefont {Blum}},
  \bibinfo {author} {\bibfnamefont {N.}~\bibnamefont {Christ}}, \bibinfo
  {author} {\bibfnamefont {R.}~\bibnamefont {Mawhinney}}, \ and\ \bibinfo
  {author} {\bibfnamefont {P.}~\bibnamefont {Vranas}},\ }\href {\doibase
  10.22323/1.066.0048} {\bibfield  {journal} {\bibinfo  {journal} {PoS}\
  }\textbf {\bibinfo {volume} {LATTICE2008}},\ \bibinfo {pages} {048} (\bibinfo
  {year} {2008})},\ \Eprint {http://arxiv.org/abs/0902.2587} {arXiv:0902.2587
  [hep-lat]} \BibitemShut {NoStop}%
\bibitem [{\citenamefont {Shintani}\ \emph {et~al.}(2015)\citenamefont
  {Shintani}, \citenamefont {Arthur}, \citenamefont {Blum}, \citenamefont
  {Izubuchi}, \citenamefont {Jung},\ and\ \citenamefont
  {Lehner}}]{Shintani:2014vja}%
  \BibitemOpen
  \bibfield  {author} {\bibinfo {author} {\bibfnamefont {E.}~\bibnamefont
  {Shintani}}, \bibinfo {author} {\bibfnamefont {R.}~\bibnamefont {Arthur}},
  \bibinfo {author} {\bibfnamefont {T.}~\bibnamefont {Blum}}, \bibinfo {author}
  {\bibfnamefont {T.}~\bibnamefont {Izubuchi}}, \bibinfo {author}
  {\bibfnamefont {C.}~\bibnamefont {Jung}}, \ and\ \bibinfo {author}
  {\bibfnamefont {C.}~\bibnamefont {Lehner}},\ }\href {\doibase
  10.1103/PhysRevD.91.114511} {\bibfield  {journal} {\bibinfo  {journal} {Phys.
  Rev.}\ }\textbf {\bibinfo {volume} {D91}},\ \bibinfo {pages} {114511}
  (\bibinfo {year} {2015})},\ \Eprint {http://arxiv.org/abs/1402.0244}
  {arXiv:1402.0244 [hep-lat]} \BibitemShut {NoStop}%
\bibitem [{\citenamefont {Tanabashi}\ \emph {et~al.}(2018)\citenamefont
  {Tanabashi} \emph {et~al.}}]{Tanabashi:2018oca}%
  \BibitemOpen
  \bibfield  {author} {\bibinfo {author} {\bibfnamefont {M.}~\bibnamefont
  {Tanabashi}} \emph {et~al.} (\bibinfo {collaboration} {Particle Data
  Group}),\ }\href {\doibase 10.1103/PhysRevD.98.030001} {\bibfield  {journal}
  {\bibinfo  {journal} {Phys. Rev.}\ }\textbf {\bibinfo {volume} {D98}},\
  \bibinfo {pages} {030001} (\bibinfo {year} {2018})}\BibitemShut {NoStop}%
\bibitem [{\citenamefont {Dudek}\ \emph {et~al.}(2009)\citenamefont {Dudek},
  \citenamefont {Edwards}, \citenamefont {Peardon}, \citenamefont {Richards},\
  and\ \citenamefont {Thomas}}]{Dudek:2009qf}%
  \BibitemOpen
  \bibfield  {author} {\bibinfo {author} {\bibfnamefont {J.~J.}\ \bibnamefont
  {Dudek}}, \bibinfo {author} {\bibfnamefont {R.~G.}\ \bibnamefont {Edwards}},
  \bibinfo {author} {\bibfnamefont {M.~J.}\ \bibnamefont {Peardon}}, \bibinfo
  {author} {\bibfnamefont {D.~G.}\ \bibnamefont {Richards}}, \ and\ \bibinfo
  {author} {\bibfnamefont {C.~E.}\ \bibnamefont {Thomas}},\ }\href {\doibase
  10.1103/PhysRevLett.103.262001} {\bibfield  {journal} {\bibinfo  {journal}
  {Phys. Rev. Lett.}\ }\textbf {\bibinfo {volume} {103}},\ \bibinfo {pages}
  {262001} (\bibinfo {year} {2009})},\ \Eprint {http://arxiv.org/abs/0909.0200}
  {arXiv:0909.0200 [hep-ph]} \BibitemShut {NoStop}%
\bibitem [{\citenamefont {Blum}\ \emph
  {et~al.}(2016{\natexlab{b}})\citenamefont {Blum} \emph
  {et~al.}}]{Blum:2014tka}%
  \BibitemOpen
  \bibfield  {author} {\bibinfo {author} {\bibfnamefont {T.}~\bibnamefont
  {Blum}} \emph {et~al.} (\bibinfo {collaboration} {RBC, UKQCD}),\ }\href
  {\doibase 10.1103/PhysRevD.93.074505} {\bibfield  {journal} {\bibinfo
  {journal} {Phys. Rev. D}\ }\textbf {\bibinfo {volume} {93}},\ \bibinfo
  {pages} {074505} (\bibinfo {year} {2016}{\natexlab{b}})},\ \Eprint
  {http://arxiv.org/abs/1411.7017} {arXiv:1411.7017 [hep-lat]} \BibitemShut
  {NoStop}%
\bibitem [{\citenamefont {Mawhinney}\ and\ \citenamefont
  {Murphy}(2016)}]{Mawhinney:2015sfj}%
  \BibitemOpen
  \bibfield  {author} {\bibinfo {author} {\bibfnamefont {R.~D.}\ \bibnamefont
  {Mawhinney}}\ and\ \bibinfo {author} {\bibfnamefont {D.~J.}\ \bibnamefont
  {Murphy}} (\bibinfo {collaboration} {RBC-UKQCD}),\ }\href {\doibase
  10.22323/1.251.0061} {\bibfield  {journal} {\bibinfo  {journal} {PoS}\
  }\textbf {\bibinfo {volume} {LATTICE2015}},\ \bibinfo {pages} {061} (\bibinfo
  {year} {2016})},\ \Eprint {http://arxiv.org/abs/1511.04419} {arXiv:1511.04419
  [hep-lat]} \BibitemShut {NoStop}%
\bibitem [{\citenamefont {Boyle}\ \emph {et~al.}(2016)\citenamefont {Boyle}
  \emph {et~al.}}]{Boyle:2015exm}%
  \BibitemOpen
  \bibfield  {author} {\bibinfo {author} {\bibfnamefont {P.~A.}\ \bibnamefont
  {Boyle}} \emph {et~al.},\ }\href {\doibase 10.1103/PhysRevD.93.054502}
  {\bibfield  {journal} {\bibinfo  {journal} {Phys. Rev. D}\ }\textbf {\bibinfo
  {volume} {93}},\ \bibinfo {pages} {054502} (\bibinfo {year} {2016})},\
  \Eprint {http://arxiv.org/abs/1511.01950} {arXiv:1511.01950 [hep-lat]}
  \BibitemShut {NoStop}%
\bibitem [{\citenamefont {Roy}(1971)}]{Roy:1971tc}%
  \BibitemOpen
  \bibfield  {author} {\bibinfo {author} {\bibfnamefont {S.~M.}\ \bibnamefont
  {Roy}},\ }\href {\doibase 10.1016/0370-2693(71)90724-6} {\bibfield  {journal}
  {\bibinfo  {journal} {Phys. Lett. B}\ }\textbf {\bibinfo {volume} {36}},\
  \bibinfo {pages} {353} (\bibinfo {year} {1971})}\BibitemShut {NoStop}%
\bibitem [{\citenamefont {Ananthanarayan}\ \emph {et~al.}(2001)\citenamefont
  {Ananthanarayan}, \citenamefont {Colangelo}, \citenamefont {Gasser},\ and\
  \citenamefont {Leutwyler}}]{Ananthanarayan:2000ht}%
  \BibitemOpen
  \bibfield  {author} {\bibinfo {author} {\bibfnamefont {B.}~\bibnamefont
  {Ananthanarayan}}, \bibinfo {author} {\bibfnamefont {G.}~\bibnamefont
  {Colangelo}}, \bibinfo {author} {\bibfnamefont {J.}~\bibnamefont {Gasser}}, \
  and\ \bibinfo {author} {\bibfnamefont {H.}~\bibnamefont {Leutwyler}},\ }\href
  {\doibase 10.1016/S0370-1573(01)00009-6} {\bibfield  {journal} {\bibinfo
  {journal} {Phys. Rept.}\ }\textbf {\bibinfo {volume} {353}},\ \bibinfo
  {pages} {207} (\bibinfo {year} {2001})},\ \Eprint
  {http://arxiv.org/abs/hep-ph/0005297} {arXiv:hep-ph/0005297} \BibitemShut
  {NoStop}%
\bibitem [{\citenamefont {Colangelo}\ \emph {et~al.}(2001)\citenamefont
  {Colangelo}, \citenamefont {Gasser},\ and\ \citenamefont
  {Leutwyler}}]{Colangelo:2001df}%
  \BibitemOpen
  \bibfield  {author} {\bibinfo {author} {\bibfnamefont {G.}~\bibnamefont
  {Colangelo}}, \bibinfo {author} {\bibfnamefont {J.}~\bibnamefont {Gasser}}, \
  and\ \bibinfo {author} {\bibfnamefont {H.}~\bibnamefont {Leutwyler}},\ }\href
  {\doibase 10.1016/S0550-3213(01)00147-X} {\bibfield  {journal} {\bibinfo
  {journal} {Nucl. Phys. B}\ }\textbf {\bibinfo {volume} {603}},\ \bibinfo
  {pages} {125} (\bibinfo {year} {2001})},\ \Eprint
  {http://arxiv.org/abs/hep-ph/0103088} {arXiv:hep-ph/0103088} \BibitemShut
  {NoStop}%
\bibitem [{\citenamefont {Caprini}\ \emph {et~al.}(2012)\citenamefont
  {Caprini}, \citenamefont {Colangelo},\ and\ \citenamefont
  {Leutwyler}}]{Caprini:2011ky}%
  \BibitemOpen
  \bibfield  {author} {\bibinfo {author} {\bibfnamefont {I.}~\bibnamefont
  {Caprini}}, \bibinfo {author} {\bibfnamefont {G.}~\bibnamefont {Colangelo}},
  \ and\ \bibinfo {author} {\bibfnamefont {H.}~\bibnamefont {Leutwyler}},\
  }\href {\doibase 10.1140/epjc/s10052-012-1860-1} {\bibfield  {journal}
  {\bibinfo  {journal} {Eur. Phys. J. C}\ }\textbf {\bibinfo {volume} {72}},\
  \bibinfo {pages} {1860} (\bibinfo {year} {2012})},\ \Eprint
  {http://arxiv.org/abs/1111.7160} {arXiv:1111.7160 [hep-ph]} \BibitemShut
  {NoStop}%
\bibitem [{\citenamefont {{}}()}]{HoyingPipi}%
  \BibitemOpen
  \bibfield  {author} {\bibinfo {author} {\bibnamefont {{}}} (\bibinfo
  {collaboration} {RBC, UKQCD}),\ }\href@noop {} {\ }\bibinfo {note} {In
  preparation}\BibitemShut {NoStop}%
\bibitem [{\citenamefont {Aoki}\ \emph {et~al.}(2021)\citenamefont {Aoki} \emph
  {et~al.}}]{Aoki:2021kgd}%
  \BibitemOpen
  \bibfield  {author} {\bibinfo {author} {\bibfnamefont {Y.}~\bibnamefont
  {Aoki}} \emph {et~al.},\ }\href@noop {} {\  (\bibinfo {year} {2021})},\
  \Eprint {http://arxiv.org/abs/2111.09849} {arXiv:2111.09849 [hep-lat]}
  \BibitemShut {NoStop}%
\bibitem [{\citenamefont {Mai}\ \emph {et~al.}(2019)\citenamefont {Mai},
  \citenamefont {Culver}, \citenamefont {Alexandru}, \citenamefont {D\"oring},\
  and\ \citenamefont {Lee}}]{Mai:2019pqr}%
  \BibitemOpen
  \bibfield  {author} {\bibinfo {author} {\bibfnamefont {M.}~\bibnamefont
  {Mai}}, \bibinfo {author} {\bibfnamefont {C.}~\bibnamefont {Culver}},
  \bibinfo {author} {\bibfnamefont {A.}~\bibnamefont {Alexandru}}, \bibinfo
  {author} {\bibfnamefont {M.}~\bibnamefont {D\"oring}}, \ and\ \bibinfo
  {author} {\bibfnamefont {F.~X.}\ \bibnamefont {Lee}},\ }\href {\doibase
  10.1103/PhysRevD.100.114514} {\bibfield  {journal} {\bibinfo  {journal}
  {Phys. Rev. D}\ }\textbf {\bibinfo {volume} {100}},\ \bibinfo {pages}
  {114514} (\bibinfo {year} {2019})},\ \Eprint
  {http://arxiv.org/abs/1908.01847} {arXiv:1908.01847 [hep-lat]} \BibitemShut
  {NoStop}%
\bibitem [{\citenamefont {Fu}\ and\ \citenamefont {Chen}(2018)}]{Fu:2017apw}%
  \BibitemOpen
  \bibfield  {author} {\bibinfo {author} {\bibfnamefont {Z.}~\bibnamefont
  {Fu}}\ and\ \bibinfo {author} {\bibfnamefont {X.}~\bibnamefont {Chen}},\
  }\href {\doibase 10.1103/PhysRevD.98.014514} {\bibfield  {journal} {\bibinfo
  {journal} {Phys. Rev. D}\ }\textbf {\bibinfo {volume} {98}},\ \bibinfo
  {pages} {014514} (\bibinfo {year} {2018})},\ \Eprint
  {http://arxiv.org/abs/1712.02219} {arXiv:1712.02219 [hep-lat]} \BibitemShut
  {NoStop}%
\bibitem [{\citenamefont {Liu}\ \emph {et~al.}(2017)\citenamefont {Liu} \emph
  {et~al.}}]{Liu:2016cba}%
  \BibitemOpen
  \bibfield  {author} {\bibinfo {author} {\bibfnamefont {L.}~\bibnamefont
  {Liu}} \emph {et~al.},\ }\href {\doibase 10.1103/PhysRevD.96.054516}
  {\bibfield  {journal} {\bibinfo  {journal} {Phys. Rev. D}\ }\textbf {\bibinfo
  {volume} {96}},\ \bibinfo {pages} {054516} (\bibinfo {year} {2017})},\
  \Eprint {http://arxiv.org/abs/1612.02061} {arXiv:1612.02061 [hep-lat]}
  \BibitemShut {NoStop}%
\bibitem [{\citenamefont {Fu}(2013)}]{Fu:2013ffa}%
  \BibitemOpen
  \bibfield  {author} {\bibinfo {author} {\bibfnamefont {Z.}~\bibnamefont
  {Fu}},\ }\href {\doibase 10.1103/PhysRevD.87.074501} {\bibfield  {journal}
  {\bibinfo  {journal} {Phys. Rev. D}\ }\textbf {\bibinfo {volume} {87}},\
  \bibinfo {pages} {074501} (\bibinfo {year} {2013})},\ \Eprint
  {http://arxiv.org/abs/1303.0517} {arXiv:1303.0517 [hep-lat]} \BibitemShut
  {NoStop}%
\bibitem [{\citenamefont {Helmes}\ \emph {et~al.}(2015)\citenamefont {Helmes},
  \citenamefont {Jost}, \citenamefont {Knippschild}, \citenamefont {Liu},
  \citenamefont {Liu}, \citenamefont {Liu}, \citenamefont {Urbach},
  \citenamefont {Ueding}, \citenamefont {Wang},\ and\ \citenamefont
  {Werner}}]{Helmes:2015gla}%
  \BibitemOpen
  \bibfield  {author} {\bibinfo {author} {\bibfnamefont {C.}~\bibnamefont
  {Helmes}}, \bibinfo {author} {\bibfnamefont {C.}~\bibnamefont {Jost}},
  \bibinfo {author} {\bibfnamefont {B.}~\bibnamefont {Knippschild}}, \bibinfo
  {author} {\bibfnamefont {C.}~\bibnamefont {Liu}}, \bibinfo {author}
  {\bibfnamefont {J.}~\bibnamefont {Liu}}, \bibinfo {author} {\bibfnamefont
  {L.}~\bibnamefont {Liu}}, \bibinfo {author} {\bibfnamefont {C.}~\bibnamefont
  {Urbach}}, \bibinfo {author} {\bibfnamefont {M.}~\bibnamefont {Ueding}},
  \bibinfo {author} {\bibfnamefont {Z.}~\bibnamefont {Wang}}, \ and\ \bibinfo
  {author} {\bibfnamefont {M.}~\bibnamefont {Werner}} (\bibinfo {collaboration}
  {ETM}),\ }\href {\doibase 10.1007/JHEP09(2015)109} {\bibfield  {journal}
  {\bibinfo  {journal} {JHEP}\ }\textbf {\bibinfo {volume} {09}},\ \bibinfo
  {pages} {109} (\bibinfo {year} {2015})},\ \Eprint
  {http://arxiv.org/abs/1506.00408} {arXiv:1506.00408 [hep-lat]} \BibitemShut
  {NoStop}%
\bibitem [{\citenamefont {Sasaki}\ \emph {et~al.}(2014)\citenamefont {Sasaki},
  \citenamefont {Ishizuka}, \citenamefont {Oka},\ and\ \citenamefont
  {Yamazaki}}]{Sasaki:2013vxa}%
  \BibitemOpen
  \bibfield  {author} {\bibinfo {author} {\bibfnamefont {K.}~\bibnamefont
  {Sasaki}}, \bibinfo {author} {\bibfnamefont {N.}~\bibnamefont {Ishizuka}},
  \bibinfo {author} {\bibfnamefont {M.}~\bibnamefont {Oka}}, \ and\ \bibinfo
  {author} {\bibfnamefont {T.}~\bibnamefont {Yamazaki}} (\bibinfo
  {collaboration} {PACS-CS}),\ }\href {\doibase 10.1103/PhysRevD.89.054502}
  {\bibfield  {journal} {\bibinfo  {journal} {Phys. Rev. D}\ }\textbf {\bibinfo
  {volume} {89}},\ \bibinfo {pages} {054502} (\bibinfo {year} {2014})},\
  \Eprint {http://arxiv.org/abs/1311.7226} {arXiv:1311.7226 [hep-lat]}
  \BibitemShut {NoStop}%
\bibitem [{\citenamefont {Beane}\ \emph {et~al.}(2012)\citenamefont {Beane},
  \citenamefont {Chang}, \citenamefont {Detmold}, \citenamefont {Lin},
  \citenamefont {Luu}, \citenamefont {Orginos}, \citenamefont {Parreno},
  \citenamefont {Savage}, \citenamefont {Torok},\ and\ \citenamefont
  {Walker-Loud}}]{Beane:2011sc}%
  \BibitemOpen
  \bibfield  {author} {\bibinfo {author} {\bibfnamefont {S.~R.}\ \bibnamefont
  {Beane}}, \bibinfo {author} {\bibfnamefont {E.}~\bibnamefont {Chang}},
  \bibinfo {author} {\bibfnamefont {W.}~\bibnamefont {Detmold}}, \bibinfo
  {author} {\bibfnamefont {H.~W.}\ \bibnamefont {Lin}}, \bibinfo {author}
  {\bibfnamefont {T.~C.}\ \bibnamefont {Luu}}, \bibinfo {author} {\bibfnamefont
  {K.}~\bibnamefont {Orginos}}, \bibinfo {author} {\bibfnamefont
  {A.}~\bibnamefont {Parreno}}, \bibinfo {author} {\bibfnamefont {M.~J.}\
  \bibnamefont {Savage}}, \bibinfo {author} {\bibfnamefont {A.}~\bibnamefont
  {Torok}}, \ and\ \bibinfo {author} {\bibfnamefont {A.}~\bibnamefont
  {Walker-Loud}} (\bibinfo {collaboration} {NPLQCD}),\ }\href {\doibase
  10.1103/PhysRevD.85.034505} {\bibfield  {journal} {\bibinfo  {journal} {Phys.
  Rev. D}\ }\textbf {\bibinfo {volume} {85}},\ \bibinfo {pages} {034505}
  (\bibinfo {year} {2012})},\ \Eprint {http://arxiv.org/abs/1107.5023}
  {arXiv:1107.5023 [hep-lat]} \BibitemShut {NoStop}%
\bibitem [{\citenamefont {Fu}(2012)}]{Fu:2011bz}%
  \BibitemOpen
  \bibfield  {author} {\bibinfo {author} {\bibfnamefont {Z.}~\bibnamefont
  {Fu}},\ }\href {\doibase 10.1088/0253-6102/57/1/13} {\bibfield  {journal}
  {\bibinfo  {journal} {Commun. Theor. Phys.}\ }\textbf {\bibinfo {volume}
  {57}},\ \bibinfo {pages} {78} (\bibinfo {year} {2012})},\ \Eprint
  {http://arxiv.org/abs/1110.3918} {arXiv:1110.3918 [hep-lat]} \BibitemShut
  {NoStop}%
\bibitem [{\citenamefont {Yagi}\ \emph {et~al.}(2011)\citenamefont {Yagi},
  \citenamefont {Hashimoto}, \citenamefont {Morimatsu},\ and\ \citenamefont
  {Ohtani}}]{Yagi:2011jn}%
  \BibitemOpen
  \bibfield  {author} {\bibinfo {author} {\bibfnamefont {T.}~\bibnamefont
  {Yagi}}, \bibinfo {author} {\bibfnamefont {S.}~\bibnamefont {Hashimoto}},
  \bibinfo {author} {\bibfnamefont {O.}~\bibnamefont {Morimatsu}}, \ and\
  \bibinfo {author} {\bibfnamefont {M.}~\bibnamefont {Ohtani}},\ }\href@noop {}
  {\  (\bibinfo {year} {2011})},\ \Eprint {http://arxiv.org/abs/1108.2970}
  {arXiv:1108.2970 [hep-lat]} \BibitemShut {NoStop}%
\bibitem [{\citenamefont {Feng}\ \emph {et~al.}(2010)\citenamefont {Feng},
  \citenamefont {Jansen},\ and\ \citenamefont {Renner}}]{Feng:2009ij}%
  \BibitemOpen
  \bibfield  {author} {\bibinfo {author} {\bibfnamefont {X.}~\bibnamefont
  {Feng}}, \bibinfo {author} {\bibfnamefont {K.}~\bibnamefont {Jansen}}, \ and\
  \bibinfo {author} {\bibfnamefont {D.~B.}\ \bibnamefont {Renner}},\ }\href
  {\doibase 10.1016/j.physletb.2010.01.018} {\bibfield  {journal} {\bibinfo
  {journal} {Phys. Lett. B}\ }\textbf {\bibinfo {volume} {684}},\ \bibinfo
  {pages} {268} (\bibinfo {year} {2010})},\ \Eprint
  {http://arxiv.org/abs/0909.3255} {arXiv:0909.3255 [hep-lat]} \BibitemShut
  {NoStop}%
\bibitem [{\citenamefont {Beane}\ \emph {et~al.}(2008)\citenamefont {Beane},
  \citenamefont {Luu}, \citenamefont {Orginos}, \citenamefont {Parreno},
  \citenamefont {Savage}, \citenamefont {Torok},\ and\ \citenamefont
  {Walker-Loud}}]{Beane:2007xs}%
  \BibitemOpen
  \bibfield  {author} {\bibinfo {author} {\bibfnamefont {S.~R.}\ \bibnamefont
  {Beane}}, \bibinfo {author} {\bibfnamefont {T.~C.}\ \bibnamefont {Luu}},
  \bibinfo {author} {\bibfnamefont {K.}~\bibnamefont {Orginos}}, \bibinfo
  {author} {\bibfnamefont {A.}~\bibnamefont {Parreno}}, \bibinfo {author}
  {\bibfnamefont {M.~J.}\ \bibnamefont {Savage}}, \bibinfo {author}
  {\bibfnamefont {A.}~\bibnamefont {Torok}}, \ and\ \bibinfo {author}
  {\bibfnamefont {A.}~\bibnamefont {Walker-Loud}},\ }\href {\doibase
  10.1103/PhysRevD.77.014505} {\bibfield  {journal} {\bibinfo  {journal} {Phys.
  Rev. D}\ }\textbf {\bibinfo {volume} {77}},\ \bibinfo {pages} {014505}
  (\bibinfo {year} {2008})},\ \Eprint {http://arxiv.org/abs/0706.3026}
  {arXiv:0706.3026 [hep-lat]} \BibitemShut {NoStop}%
\bibitem [{\citenamefont {Beane}\ \emph {et~al.}(2006)\citenamefont {Beane},
  \citenamefont {Bedaque}, \citenamefont {Orginos},\ and\ \citenamefont
  {Savage}}]{Beane:2005rj}%
  \BibitemOpen
  \bibfield  {author} {\bibinfo {author} {\bibfnamefont {S.~R.}\ \bibnamefont
  {Beane}}, \bibinfo {author} {\bibfnamefont {P.~F.}\ \bibnamefont {Bedaque}},
  \bibinfo {author} {\bibfnamefont {K.}~\bibnamefont {Orginos}}, \ and\
  \bibinfo {author} {\bibfnamefont {M.~J.}\ \bibnamefont {Savage}} (\bibinfo
  {collaboration} {NPLQCD}),\ }\href {\doibase 10.1103/PhysRevD.73.054503}
  {\bibfield  {journal} {\bibinfo  {journal} {Phys. Rev. D}\ }\textbf {\bibinfo
  {volume} {73}},\ \bibinfo {pages} {054503} (\bibinfo {year} {2006})},\
  \Eprint {http://arxiv.org/abs/hep-lat/0506013} {arXiv:hep-lat/0506013}
  \BibitemShut {NoStop}%
\bibitem [{\citenamefont {Blum}\ \emph {et~al.}(2018)\citenamefont {Blum},
  \citenamefont {Boyle}, \citenamefont {G\"ulpers}, \citenamefont {Izubuchi},
  \citenamefont {Jin}, \citenamefont {Jung}, \citenamefont {J\"uttner},
  \citenamefont {Lehner}, \citenamefont {Portelli},\ and\ \citenamefont
  {Tsang}}]{RBC:2018dos}%
  \BibitemOpen
  \bibfield  {author} {\bibinfo {author} {\bibfnamefont {T.}~\bibnamefont
  {Blum}}, \bibinfo {author} {\bibfnamefont {P.~A.}\ \bibnamefont {Boyle}},
  \bibinfo {author} {\bibfnamefont {V.}~\bibnamefont {G\"ulpers}}, \bibinfo
  {author} {\bibfnamefont {T.}~\bibnamefont {Izubuchi}}, \bibinfo {author}
  {\bibfnamefont {L.}~\bibnamefont {Jin}}, \bibinfo {author} {\bibfnamefont
  {C.}~\bibnamefont {Jung}}, \bibinfo {author} {\bibfnamefont {A.}~\bibnamefont
  {J\"uttner}}, \bibinfo {author} {\bibfnamefont {C.}~\bibnamefont {Lehner}},
  \bibinfo {author} {\bibfnamefont {A.}~\bibnamefont {Portelli}}, \ and\
  \bibinfo {author} {\bibfnamefont {J.~T.}\ \bibnamefont {Tsang}} (\bibinfo
  {collaboration} {RBC, UKQCD}),\ }\href {\doibase
  10.1103/PhysRevLett.121.022003} {\bibfield  {journal} {\bibinfo  {journal}
  {Phys. Rev. Lett.}\ }\textbf {\bibinfo {volume} {121}},\ \bibinfo {pages}
  {022003} (\bibinfo {year} {2018})},\ \Eprint
  {http://arxiv.org/abs/1801.07224} {arXiv:1801.07224 [hep-lat]} \BibitemShut
  {NoStop}%
\bibitem [{\citenamefont {Thomas}\ \emph {et~al.}(2012)\citenamefont {Thomas},
  \citenamefont {Edwards},\ and\ \citenamefont {Dudek}}]{Thomas:2011rh}%
  \BibitemOpen
  \bibfield  {author} {\bibinfo {author} {\bibfnamefont {C.~E.}\ \bibnamefont
  {Thomas}}, \bibinfo {author} {\bibfnamefont {R.~G.}\ \bibnamefont {Edwards}},
  \ and\ \bibinfo {author} {\bibfnamefont {J.~J.}\ \bibnamefont {Dudek}},\
  }\href {\doibase 10.1103/PhysRevD.85.014507, 10.1103/PhysRevD.85.039901}
  {\bibfield  {journal} {\bibinfo  {journal} {Phys. Rev.}\ }\textbf {\bibinfo
  {volume} {D85}},\ \bibinfo {pages} {014507} (\bibinfo {year} {2012})},\
  \Eprint {http://arxiv.org/abs/1107.1930} {arXiv:1107.1930 [hep-lat]}
  \BibitemShut {NoStop}%
\bibitem [{\citenamefont {Dudek}\ \emph {et~al.}(2010)\citenamefont {Dudek},
  \citenamefont {Edwards}, \citenamefont {Peardon}, \citenamefont {Richards},\
  and\ \citenamefont {Thomas}}]{Dudek:2010wm}%
  \BibitemOpen
  \bibfield  {author} {\bibinfo {author} {\bibfnamefont {J.~J.}\ \bibnamefont
  {Dudek}}, \bibinfo {author} {\bibfnamefont {R.~G.}\ \bibnamefont {Edwards}},
  \bibinfo {author} {\bibfnamefont {M.~J.}\ \bibnamefont {Peardon}}, \bibinfo
  {author} {\bibfnamefont {D.~G.}\ \bibnamefont {Richards}}, \ and\ \bibinfo
  {author} {\bibfnamefont {C.~E.}\ \bibnamefont {Thomas}},\ }\href {\doibase
  10.1103/PhysRevD.82.034508} {\bibfield  {journal} {\bibinfo  {journal} {Phys.
  Rev.}\ }\textbf {\bibinfo {volume} {D82}},\ \bibinfo {pages} {034508}
  (\bibinfo {year} {2010})},\ \Eprint {http://arxiv.org/abs/1004.4930}
  {arXiv:1004.4930 [hep-ph]} \BibitemShut {NoStop}%
\end{thebibliography}%

\end{document}